\documentclass[hyper,a4paper,notoc]{JHEP}
\pdfoutput=1

\usepackage{graphicx}
\usepackage{graphics}
\usepackage[mathscr]{eucal}
\usepackage{amssymb}
\usepackage{amsmath}
\usepackage{xspace}
\usepackage{listings}
\usepackage{ulem}

\newcommand{\SARAH}{{\tt SARAH}\xspace}
\newcommand{\SPheno}{{\tt SPheno}\xspace}
\newcommand{\MO}{{\tt MicrOmegas}\xspace}
\newcommand{\CalcHep}{{\tt CalcHep}\xspace}

\newcommand{\sign}{\text{sign}\xspace}

\newcommand{\BL}{{\ensuremath{B-L}}\xspace}
\newcommand{\UBL}{{\ensuremath{U(1)_{B-L}}}\xspace}
\newcommand{\BLSSM}{BLSSM\xspace}
\newcommand{\CBLSSM}{CBLSSM\xspace}

\newcommand{\vevs}{\textit{vev}s\xspace}
\newcommand{\msugra}{mSUGRA\xspace}
\newcommand{\blino}{BLino\xspace}
\newcommand{\EQ}{eq.\xspace}
\newcommand{\EQS}{eqs.\xspace}

\newcommand{\FIG}{Fig.\xspace}
\newcommand{\FIGS}{Figs.\xspace}
\newcommand{\EG}{\textit{e.g.}\xspace}
\newcommand{\IE}{\textit{i.e}.\xspace}
\newcommand{\REF}{Ref.\xspace}
\newcommand{\SEC}{sec.\xspace}
\def\gsim{\raise0.3ex\hbox{$\;>$\kern-0.75em\raise-1.1ex\hbox{$\sim\;$}}}
\newcommand{\Atlas}{ATLAS\xspace}
\newcommand{\CMS}{CMS\xspace}
\newcommand{\Xenon}[1]{\textsc{Xenon{#1}}\xspace}
\newcommand{\gBL}[1]{{\ensuremath{g_{BL}^{#1}}}\xspace}
\newcommand{\gmix}{{\ensuremath{\bar{g}}}\xspace}
\newcommand{\gsqsum}{{\ensuremath{g_{\Sigma}^{2}}}\xspace}
\newcommand{\rsnu}{R-sneutrino\xspace}
\newcommand{\rsnus}{R-sneutrinos\xspace}
\newcommand{\lsnu}{L-sneutrino\xspace}
\newcommand{\lsnus}{L-sneutrinos\xspace}
\newcommand{\mup}{{\ensuremath{{\mu}^{\prime}}}\xspace}
\newcommand{\mZp}{{\ensuremath{M_{Z^{\prime}}}}\xspace}
\newcommand{\mZpSq}{{\ensuremath{M_{Z^{\prime}}^{2}}}\xspace}
\newcommand{\mReSnuSq}[0]{{\ensuremath{m^{2}_{{\tilde{\nu}}^{S}}}}\xspace}
\newcommand{\mImSnuSq}[0]{{\ensuremath{m^{2}_{{\tilde{\nu}}^{P}}}}\xspace}
\newcommand{\ssnu}[1]{{\ensuremath{{\tilde{\nu}}_{#1}^{S}}}\xspace}
\newcommand{\psnu}[1]{{\ensuremath{{\tilde{\nu}}_{#1}^{P}}}\xspace}

\title{Dark matter scenarios in the minimal SUSY \BL model}

\preprint{FR-PHENO-2012-013, Bonn-TH-2012-16}

\author{
Lorenzo Basso$^{1,a}$, Ben O'Leary$^{2,b}$, Werner Porod$^{2,c}$,
Florian Staub$^{3,d}$ \\
$^1$Albert-Ludwigs-Universit\"at - Fakult\"at f\"ur Mathematik und Physik,\\
 D-79104 Freiburg i.\ Br., Germany\\
$^2$Institut f\"ur Theoretische Physik und Astrophysik, Universit\"at
W\"urzburg,\\
97074  W\"urzburg, Germany\\
$^3$Physikalisches Institut der Universit\"at Bonn, \\
53115 Bonn, Germany\\
$^a$Email: \email{lorenzo.basso@physik.uni-freiburg.de} \\
$^b$Email: \email{ben.oleary@physik.uni-wuerzburg.de} \\
$^c$Email: \email{porod@physik.uni-wuerzburg.de} \\
$^d$Email: \email{fnstaub@th.physik.uni-bonn.de} \\
}
\abstract{We perform a study of the dark matter 
 candidates of a
  constrained version of the minimal
  $R$-parity-conserving supersymmetric  model
  with a gauged \UBL. It
  turns out that there are four additional candidates for dark matter
  in comparison to the MSSM: two kinds of neutralino, which
  either correspond to
  the gaugino of the \UBL or to a fermionic bilepton, as well
  as ``right-handed'' CP-even and -odd
  sneutrinos. The correct dark matter
  relic density of the neutralinos can be obtained
   due to different
  mechanisms 
  including new co-annihilation regions and resonances. The large additional
  Yukawa couplings required to
  break the \UBL radiatively often lead to large annihilation cross sections
  for the sneutrinos. The correct treatment of gauge kinetic mixing is crucial to
  the success of some scenarios. All candidates are consistent with the
  exclusion limits of \Xenon{100}.} 

\begin{document}
\maketitle
\tableofcontents

\section{Introduction}
The LHC has been running now for more than 2 years
 and the \Atlas
\cite{Aad:2009wy} and \CMS \cite{Chatrchyan:2008aa}
experiments at the LHC have collected about 5 fb${}^{-1}$ of data.
While so far there has been
 no hint of the presence of supersymmetry (SUSY)
\cite{Aad:2011ib,ATLAS:2011ad,Aad:2011cwa,Chatrchyan:2011ek,Chatrchyan:2011qs,
 Chatrchyan:2011zy},
there is an indication for a Higgs boson in the mass range of
124--127~GeV \cite{ATLAS:2012ae,Chatrchyan:2012tx}. Both observations
can be explained within the minimal supersymmetric
 standard model (MSSM) and even within its constrained, $R$-parity-conserving
 version (CMSSM); however, masses in the
multi-TeV range are needed.  In addition, the CMSSM provides
 a candidate particle -- the lightest neutralino --
  to explain
the observation that roughly 23\% of the energy density of the
universe consists of non-baryonic matter
\cite{Ellis:1983ew,Ellis:2003cw}. However, the observation of massive
neutrinos \cite{Fukuda:1998mi,Ahmad:2002jz,Eguchi:2002dm,Abe:2008aa}
is not covered by the MSSM and 
 requires an extension, \EG a kind of seesaw mechansim 
\cite{Ma:1998dn,Minkowski:1977sc,Schechter:1980gr,Cheng:1980qt,Foot:1988aq}
or $R$-parity violation \cite{Romao:1999up,Hirsch:2000ef,Diaz:2003as}.
Furthermore, also the explanation of the origin of $R$-parity 
or the baryon asymmetry of the  universe might demand an extension of the 
MSSM, see \cite{Ibanez:2012wg,Blum:2008ym} and references therein.

All-in-all, there has been a growing interest in non-minimal
 SUSY scenarios. For instance, it has been shown that in the
 next-to-minimal supersymmetric standard model (NMSSM) and in 
the generalized
NMSSM (GNMSSM)  it is easier and more natural to
 obtain Higgs masses in the
preferred mass range without the need to make the
superpartners extremely
heavy \cite{Moroi:2011aa,Ellwanger:2011aa,Ross:2012nr}. Also in this context,
 there have
 been studies of extended gauge groups,
since they can offer heavier Higgs masses more easily
 \cite{Haber:1986gz,Drees:1987tp,Cvetic:1997ky,Zhang:2008jm,Ma:2011ea,
Hirsch:2011hg} as
well as new collider phenomenology
\cite{Gherghetta:1996yr,Erler:2002pr,Kang:2004bz,Langacker:2009su,Chang:2011be,
Corcella:2012dw,Athron:2011wu,Krauss:2012ku,Hirsch:2012kv}. 
One of the simplest
possibilities to extend the MSSM gauge sector
 is to add
an additional Abelian gauge group. We will focus here on the presence
of an $U(1)_{B-L}$ group which can be a result of an \(E_8 \times
E_8\) heterotic string theory (and hence M-theory)
\cite{Buchmuller:2006ik,Ambroso:2009sc,Ambroso:2010pe}. 
This model, the minimal $R$-parity-conserving \BL supersymmetric
 standard model (\BLSSM),
 was proposed in
\cite{Khalil:2007dr,FileviezPerez:2010ek} and neutrino masses are
obtained via a type I seesaw mechanism. Furthermore, it could
 help
 to understand the origin of $R$-parity and its
 possible spontaneous violation in supersymmetric models
\cite{Khalil:2007dr,Barger:2008wn,FileviezPerez:2010ek}
 as well as
the mechanism of leptogenesis \cite{Pelto:2010vq,Babu:2009pi}. While
the mass spectrum of the constrained version of this model
 has been
studied in detail in ref.~\cite{O'Leary:2011yq}, we will focus in this
work on the dark matter aspects of the model.

The model here considered contains an enhanced variety of candidates
for the particle responsible for the relic density compared to the
 MSSM: an extended neutralino
sector and an extended sneutrino sector as the model contains
right-handed neutrinos. If one introduces an additional $Z_2$ symmetry to
the model the right-handed neutrino can also be a
 valid dark matter
candidate \cite{Burell:2011wh}.
The ``right-handed'' sneutrinos (\rsnus for short) have been
 considered in various models
\cite{Gopalakrishna:2006kr,Arina:2007tm,Thomas:2007bu,Cerdeno:2008ep} 
including $U(1)$ extensions of the MSSM
 \cite{Bandyopadhyay:2011qm,Belanger:2011rs,Dumont:2012ee}.
However, in \UBL-extended supersymmetric models, sneutrinos have mostly been
 studied in the context of the inverse seesaw mechanism
\cite{Khalil:2011tb}. In this inverse-seesaw \BL model, the large Yukawa
 couplings
 $Y_{\nu}$ lead to an annihilation cross section that is large enough to get the
 correct relic density. It might even be possible that R-sneutrinos in \rsnus-extended 
models are connected to inflation  \cite{Allahverdi:2007wt}. 
  The \BL gaugino
 has also already been considered as
 dark matter \cite{Khalil:2008ps}.
However, we will restrict ourselves to the most predictive setup: a
 constrained version of the \BLSSM, the
 \CBLSSM.
Here a large Majorana mass term for the
right-handed neutrinos implies a splitting of the sneutrinos
into their scalar and pseudoscalar components with important
consequences for their properties as dark matter candidates.
As already shown in Ref.~\cite{O'Leary:2011yq}, there are new
possibilities for the lightest supersymmetric particle (LSP): the
lightest neutralino can be the gaugino of the \BL gauge group, the
\blino, or a fermionic partner of the bilepton scalars needed
to break the \UBL, \IE a bileptino,
 in addition to the normal MSSM neutralino LSP possibilities.

Much like a neutralino LSP in the CMSSM, the
 lightest neutralino in the \CBLSSM has
 in general so great an abundance
 that it would overclose the universe. This is solved
in the CMSSM in four distinct regions of the parameter space: the
 bulk region, the focus point region,
 the co-annihilation region and
the Higgs funnel. Similarly, we also find
  regions in the \BLSSM
with sufficient co-annihilation not only with stops 
 and staus but also with CP-even and -odd sneutrinos.
 There are also new resonances with scalar
Higgs fields, which can be either MSSM-like or
 correspond to the
extended Higgs sector, as well as with the $Z'$ boson. 
The low abundance of \rsnus is
related to the breaking of \UBL which requires certain Yukawa couplings
to be large. These also induce a Majorana mass term for the
 right-handed neutrinos and
 thus also a splitting of the sneutrinos
into  CP-even and CP-odd   mass eigenstates.

Furthermore, it has been pointed out that the presence of two Abelian
gauge groups in this model gives rise to kinetic mixing terms of the
form
\begin{equation}
\label{eq:offfieldstrength}
- \chi_{ab}  \hat{F}^{a, \mu \nu} \hat{F}^b_{\mu \nu}, \quad a \neq b
\end{equation}
that are allowed by gauge and Lorentz invariance \cite{Holdom:1985ag},
as $\hat{F}^{a, \mu \nu}$ and $\hat{F}^{b, \mu \nu}$ are
 gauge-invariant quantities by themselves,
 see \EG \cite{Babu:1997st}. Even
if these terms are absent at tree level at a particular scale, they
will in general be generated by RGE effects
\cite{delAguila:1988jz,delAguila:1987st}.  These terms can have a
sizable effect on the mass spectrum of this model
\cite{O'Leary:2011yq}. As we will see, gauge kinetic mixing in
 the context
of supersymmetric dark matter is even more important and several
scenarios do
not work if it is neglected.
 This agrees with previous
observations concerning kinetic mixing in the context of non-SUSY dark
matter \cite{Mambrini:2010dq,Chun:2010ve,Mambrini:2011dw}.

We start in section~\ref{sec:model} with an
 introduction to the
 \CBLSSM, including a discussion of the
 relevant masses and the
physics of gauge kinetic mixing. In section~\ref{sec:darkmatter}
 we present
in detail our results on the properties of the dark matter
 candidates.
Finally, we conclude in section~\ref{sec:conclusion}.
\section{The Model}
\label{sec:model}
In this section we discuss briefly the particle content and the
superpotential of the model under consideration. Furthermore, the
tree-level masses and mixings of the particles important for our dark
matter studies are given. For a detailed discussion of the masses of
all particles as well as of the corresponding one-loop corrections, we
refer to \cite{O'Leary:2011yq}. In addition, we show the main aspects
of $U(1)$ kinetic mixing since it can have important
 consequences for
the abundance of the dark matter candidate.

\subsection{Particle content and superpotential}
The model consists of three generations of matter particles including
right-handed neutrinos which can, for example, be embedded in $SO(10)$
16-plets. Moreover, below the GUT scale the usual MSSM Higgs doublets
are present as well as two fields $\eta$ and $\bar{\eta}$ responsible
for the breaking of the \UBL. Furthermore, $\eta$ is responsible for
generating a Majorana mass term for the right-handed neutrinos and
thus we interpret the \BL charge of this field as its lepton number,
and likewise for $\bar{\eta}$, and call these fields bileptons since
they carry twice the lepton number of (anti-)neutrinos.  We summarize
the quantum numbers of the chiral superfields with respect to $U(1)_Y
\times SU(2)_L \times SU(3)_C \times \UBL$ in Table~\ref{tab:cSF}.
\begin{table} 
\centering
\begin{tabular}{|c|c|c|c|c|c|} 
\hline \hline 
Superfield & Spin 0 & Spin \(\frac{1}{2}\) & Generations & \((U(1)_Y\otimes\,
SU(2)_L\otimes\, SU(3)_C\otimes\, \UBL)\) \\ 
\hline 
\(\hat{Q}\) & \(\tilde{Q}\) & \(Q\) & 3
 & \((\frac{1}{6},{\bf 2},{\bf 3},\frac{1}{6}) \) \\ 
\(\hat{d}^c\) & \(\tilde{d}^c\) & \(d^c\) & 3
 & \((\frac{1}{3},{\bf 1},{\bf \overline{3}},-\frac{1}{6}) \) \\ 
\(\hat{u}^c\) & \(\tilde{u}^c\) & \(u^c\) & 3
 & \((-\frac{2}{3},{\bf 1},{\bf \overline{3}},-\frac{1}{6}) \) \\ 
\(\hat{L}\) & \(\tilde{L}\) & \(L\) & 3
 & \((-\frac{1}{2},{\bf 2},{\bf 1},-\frac{1}{2}) \) \\ 
\(\hat{e}^c\) & \(\tilde{e}^c\) & \(e^c\) & 3
 & \((1,{\bf 1},{\bf 1},\frac{1}{2}) \) \\ 
\(\hat{\nu}^c\) & \(\tilde{\nu}^c\) & \(\nu^c\) & 3
 & \((0,{\bf 1},{\bf 1},\frac{1}{2}) \) \\ 
\(\hat{H}_d\) & \(H_d\) & \(\tilde{H}_d\) & 1
 & \((-\frac{1}{2},{\bf 2},{\bf 1},0) \) \\ 
\(\hat{H}_u\) & \(H_u\) & \(\tilde{H}_u\) & 1
 & \((\frac{1}{2},{\bf 2},{\bf 1},0) \) \\ 
\(\hat{\eta}\) & \(\eta\) & \(\tilde{\eta}\) & 1
 & \((0,{\bf 1},{\bf 1},-1) \) \\ 
\(\hat{\bar{\eta}}\) & \(\bar{\eta}\) & \(\tilde{\bar{\eta}}\) & 1
 & \((0,{\bf 1},{\bf 1},1) \) \\ 
\hline \hline
\end{tabular} 
\caption{Chiral superfields and their quantum numbers.}
\label{tab:cSF}
\end{table}

The superpotential is given by
\begin{align} 
\nonumber 
W = & \, Y^{ij}_u\,\hat{u}^c_i\,\hat{Q}_j\,\hat{H}_u\,
- Y_d^{ij} \,\hat{d}^c_i\,\hat{Q}_j\,\hat{H}_d\,
- Y^{ij}_e \,\hat{e}^c_i\,\hat{L}_j\,\hat{H}_d\,+\mu\,\hat{H}_u\,\hat{H}_d\, \\
 & \, \, 
+Y^{ij}_{\nu}\,\hat{\nu}^c_i\,\hat{L}_j\,\hat{H}_u\,
- \mu' \,\hat{\eta}\,\hat{\bar{\eta}}\,
+Y^{ij}_x\,\hat{\nu}^c_i\,\hat{\eta}\,\hat{\nu}^c_j\,
\label{eq:superpot}
\end{align} 
and we have the additional soft SUSY-breaking terms:
\begin{align}
\nonumber \mathscr{L}_{SB} = & \mathscr{L}_{MSSM}
 - \lambda_{\tilde{B}} \lambda_{\tilde{B}'} {M}_{B B'}
 - \frac{1}{2} \lambda_{\tilde{B}'} \lambda_{\tilde{B}'} {M}_{B'}
 - m_{\eta}^2 |\eta|^2 - m_{\bar{\eta}}^2 |\bar{\eta}|^2
 - {m_{\nu^c,ij}^{2}} (\tilde{\nu}_i^c)^* \tilde{\nu}_j^c \\
& - \eta \bar{\eta} B_{\mu'} + T^{ij}_{\nu}  H_u \tilde{\nu}_i^c \tilde{L}_j
 + T^{ij}_{x} \eta \tilde{\nu}_i^c \tilde{\nu}_j^c 
\end{align}
$i,j$ are generation indices. Without loss of generality one can take
$B_\mu$ and $B_{\mu'}$ to be real. The extended gauge group breaks to
$SU(3)_C \otimes U(1)_{em}$ as the Higgs fields and bileptons receive
vacuum expectation values (\vevs):
\begin{align} 
H_d^0 = & \, \frac{1}{\sqrt{2}} \left(\sigma_{d} + v_d  + i \phi_{d} \right),
\hspace{1cm}
H_u^0 = \, \frac{1}{\sqrt{2}} \left(\sigma_{u} + v_u  + i \phi_{u} \right)\\ 
\eta
= & \, \frac{1}{\sqrt{2}} \left(\sigma_\eta + v_{\eta} + i \phi_{\eta} \right),
\hspace{1cm}
\bar{\eta}
= \, \frac{1}{\sqrt{2}} \left(\sigma_{\bar{\eta}} + v_{\bar{\eta}}
 + i \phi_{\bar{\eta}} \right)
\end{align} 
We define $\tan\beta' = v_{\eta}/v_{\bar{\eta}}$
 in analogy to
the ratio of the MSSM \vevs ($\tan\beta = v_{u}/v_{d}$).

\subsection{The $Z'$ sector}
\label{subsec:kineticmixing}
As already mentioned in the introduction, the presence of two Abelian
gauge groups in combination with the given particle content gives 
rise to a new
effect absent in the MSSM or other SUSY models with just one Abelian
gauge group:  gauge kinetic mixing.

The details of this mechanism and how it affects the mass spectrum are
 elaborated in \REF~\cite{O'Leary:2011yq}. We merely note here that the
 off-diagonal coupling constant $\gmix$, which plays an important role in the
 mass matrices of the neutralinos and of the scalar bosons, is not negligible
 (typically about a third the size of the diagonal $U(1)$ couplings). 

 We also note that $\mZp \simeq \gBL{} x$  and thus we find an approximate
relation between $\mZp$ and $\mu'$
\begin{equation}
\label{eq:tadpole_MZp}
 \mZp^2 \simeq 
 - 2 |\mu'|^2 + \frac{4 (m_{\bar{\eta}}^2 - m_{\eta}^2 \tan^2 \beta')
- v^2 \gmix \gBL{} \cos\beta (1+\tan\beta') }{2 (\tan^2 \beta'
 - 1) }
\end{equation}

\subsection{The scalar Higgs sector}
\label{sec:ScalarsHiggsSector}
 In the scalar sector the gauge kinetic terms induce a
 mixing
between the $SU(2)$ doublet Higgs fields and the bileptons.
The mass matrix reads at tree level in the basis 
$(\sigma_d,\sigma_u,\sigma_\eta,\sigma_{\bar{\eta}})$:
\begin{align}
& m^2_{h,T} = \nonumber \\
&\left(\begin{array}{cccc}
m^2_{A^0} s^2_\beta + \gsqsum v^2_u & \,\,
-m^2_{A^0} c_\beta s_\beta - \gsqsum v_d v_u &
 \frac{\gmix \gBL{}}{2}   v_d v_{\eta} &
  -\frac{\gmix \gBL{}}{2} v_d v_{\bar{\eta}} \\
-m^2_{A^0} c_\beta s_\beta  - \gsqsum v_d v_u &
m^2_{A^0} c^2_\beta + \gsqsum v^2_d & \,\,
 - \frac{\gmix \gBL{}}{2}   v_u v_{\eta} &
  \frac{\gmix \gBL{}}{2} v_u v_{\bar{\eta}} \\
\frac{\gmix \gBL{}}{2}   v_d v_{\eta} &
 - \frac{\gmix \gBL{}}{2}   v_u v_{\eta}  &
  m^2_{A^0_\eta} c^2_{\beta'} + \gBL{2} v^2_\eta &
 \,\, -  m^2_{A^0_\eta} c_{\beta'} s_{\beta'} 
  - \gBL{2} v_\eta v_{\bar{\eta}} \\
-\frac{\gmix \gBL{}}{2} v_d v_{\bar{\eta}} &
 \frac{\gmix \gBL{}}{2} v_u v_{\bar{\eta}}  &
\,\, -  m^2_{A^0_\eta} c_{\beta'} s_{\beta'} 
  - \gBL{2} v_\eta v_{\bar{\eta}} &
 m^2_{A^0_\eta} s^2_{\beta'} + \gBL{2} v^2_{\bar{\eta}}
\end{array}
\right)
\end{align}
where we have defined
 $\gsqsum = \frac{1}{4} (g^2_1+g^2_2+{\gmix}^{2}),
 c_x = \cos(x)$,
 and $s_x = \sin(x)$ ($x=\beta,\beta')$, and used
the masses of the physical pseudoscalars
 $A^0$ and $A^0_\eta$ given
by
\begin{equation}
m^2_{A^0} = \frac{2 B_\mu}{\sin2\beta} \thickspace, \hspace{1cm} 
m^2_{A^0_\eta}  = \frac{2 B_{\mu'}}{\sin2\beta'} \thickspace.
\end{equation}
For completeness we note that the mass of charged Higgs
boson reads, as in the MSSM, as
\begin{equation}
m^2_{H^+} = B_\mu \left( \tan\beta+\cot\beta\right) + m^2_W
\end{equation}

\subsection{Neutralinos}
\label{sec:neutralino}
In the neutralino sector we find that the gauge kinetic effects lead
to a mixing between the usual MSSM neutralinos with the additional
states, similar to the mixing in the CP-even Higgs sector. In other
words, were these to be neglected, both sectors would decouple.  The
mass matrix reads in the basis \( \left(\lambda_{\tilde{B}},
  \tilde{W}^0, \tilde{H}_d^0, \tilde{H}_u^0, \lambda_{\tilde{B}{}'},
  \tilde{\eta}, \tilde{\bar{\eta}}\right) \)
\begin{equation} 
\label{eq:NeutralinoMM}
m_{\tilde{\chi}^0} = \left( 
\begin{array}{ccccccc}
M_1 & 0 & -\frac{1}{2} g_1 v_d & \frac{1}{2} g_1 v_u & \frac{1}{2} {M}_{B B'}
 & 0 & 0 \\ 
0 & M_2 & \frac{1}{2} g_2 v_d  & -\frac{1}{2} g_2 v_u  & 0 & 0 & 0 \\ 
-\frac{1}{2} g_1 v_d  & \frac{1}{2} g_2 v_d  & 0 & - \mu
 & -\frac{1}{2} \gmix v_d  & 0 & 0 \\ 
\frac{1}{2} g_1 v_u  & -\frac{1}{2} g_2 v_u & - \mu  & 0
 & \frac{1}{2} \gmix v_u  & 0 & 0 \\ 
\frac{1}{2} {M}_{B B'}  & 0 & -\frac{1}{2} \gmix v_d
 & \frac{1}{2} \gmix v_u  & {M}_{B} & - \gBL{} v_{\eta}
 & \gBL{} v_{\bar{\eta}} \\ 
0 & 0 & 0 & 0 & - \gBL{} v_{\eta}  & 0 & - {\mu'} \\ 
0 & 0 & 0 & 0 & \gBL{} v_{\bar{\eta}}  & - {\mu'} & 0\end{array} 
\right) 
\end{equation} 
It is well known that for real parameters such a matrix can be
diagonalized by an orthogonal mixing matrix $N$ such that $N^*
M^{\tilde\chi^0}_T N^\dagger$ is diagonal. For complex parameters one
has to diagonalize $M^{\tilde\chi^0}_T (M^{\tilde\chi^0}_T)^\dagger$.

In addition, we will refer to the
 bino- and wino-like states,
\IE the states built by the gauginos of the MSSM, often in the
following as `gaugino-like'. Note that this does not
 include the
\blino, the gaugino of the \BL sector.

In this model, for the chosen boundary conditions, the lightest
supersymmetric particle (LSP), \IE{} the dark matter candidate,
 is
always either the lightest neutralino or the lightest sneutrino. The
reason is that $m_0$ must be very  large in order to solve the tadpole
equations, and therefore all sfermions are heavier than the lightest
neutralino, with the possible exception of the sneutrinos.  A
neutralino LSP is in general a mixture of all seven gauge eigenstates.
However, its properties are typically dominated by only one or two
constituents. In this context, we can distinguish the following
extreme cases:
\begin{enumerate}
 \item $M_1 \ll M_2, \mu, M_{B'}, \mu'$: bino-like LSP
 \item $M_2 \ll M_1, \mu, M_{B'}, \mu'$: wino-like LSP
 \item $\mu \ll M_1, M_2, M_{B'}, \mu'$: Higgsino-like LSP
 \item $M_{B'} \ll M_1, M_2, \mu, \mu'$: \blino-like LSP
 \item $\mu' \ll M_1,M_2,\mu, M_{B'}$: bileptino-like LSP
\end{enumerate}

\begin{figure}[t]
 \begin{minipage}{0.99\linewidth}
\centering
\includegraphics[width=0.42\linewidth]{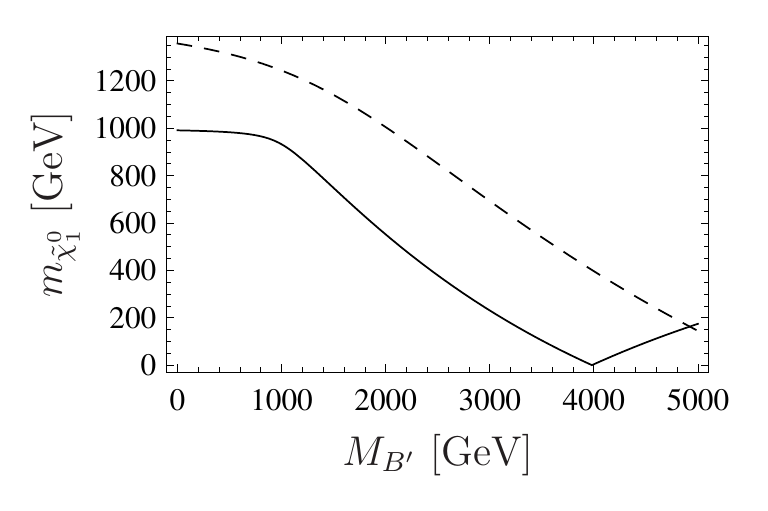}
\hfill
\includegraphics[width=0.42\linewidth]{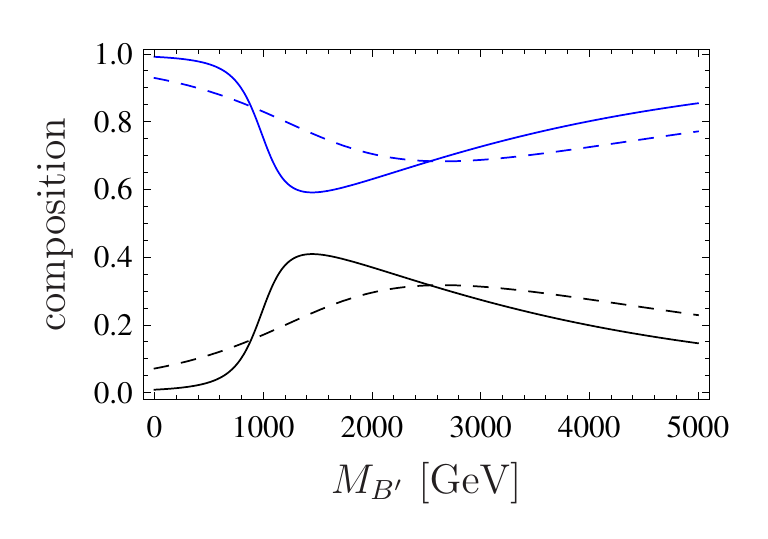} \\
\includegraphics[width=0.42\linewidth]{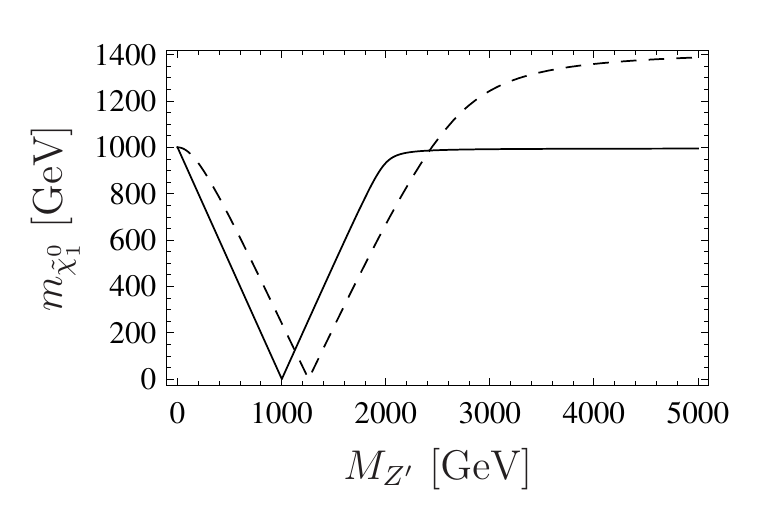}
\hfill
\includegraphics[width=0.42\linewidth]{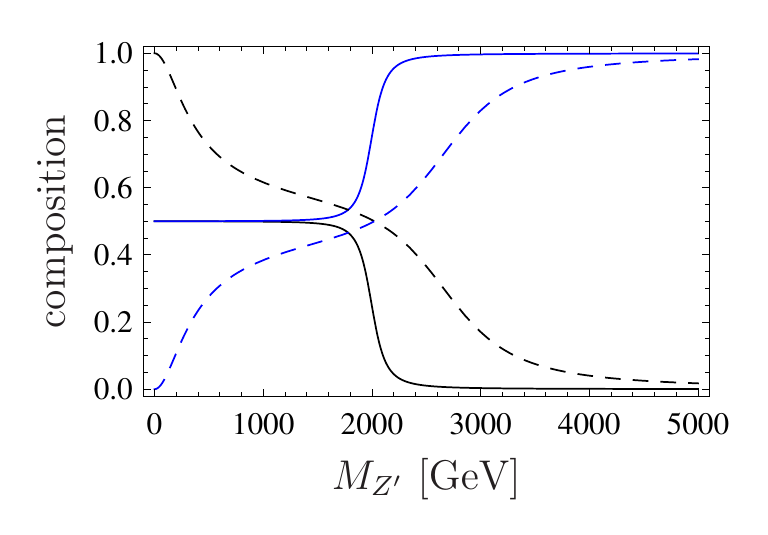} \\
\includegraphics[width=0.42\linewidth]{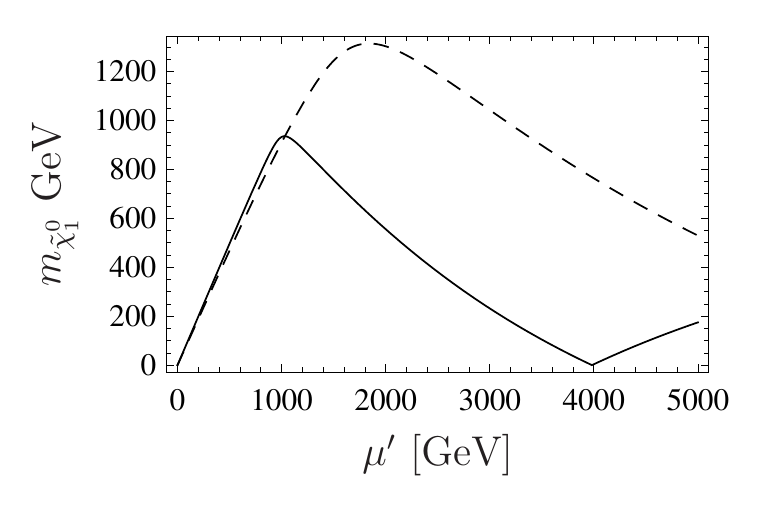}
\hfill
\includegraphics[width=0.42\linewidth]{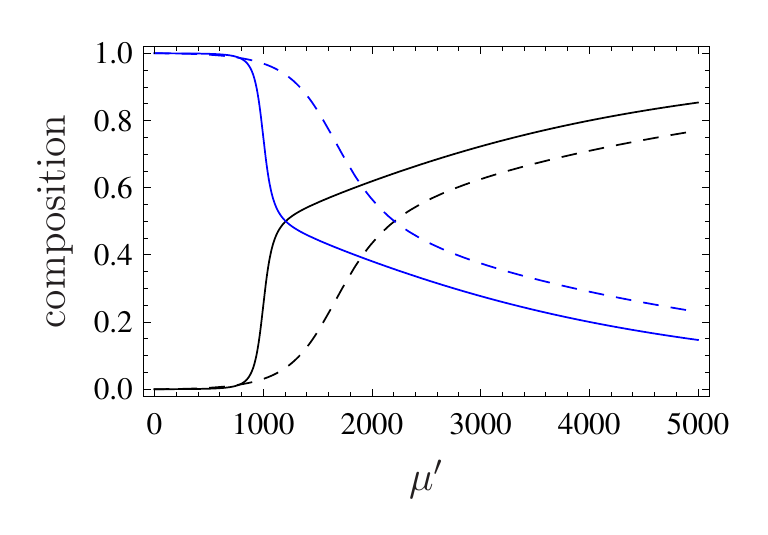} \\
\includegraphics[width=0.42\linewidth]{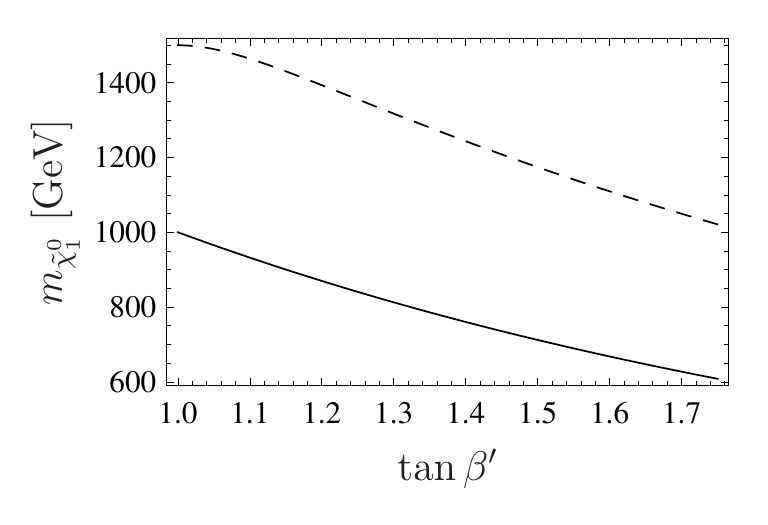}
\hfill
\includegraphics[width=0.42\linewidth]{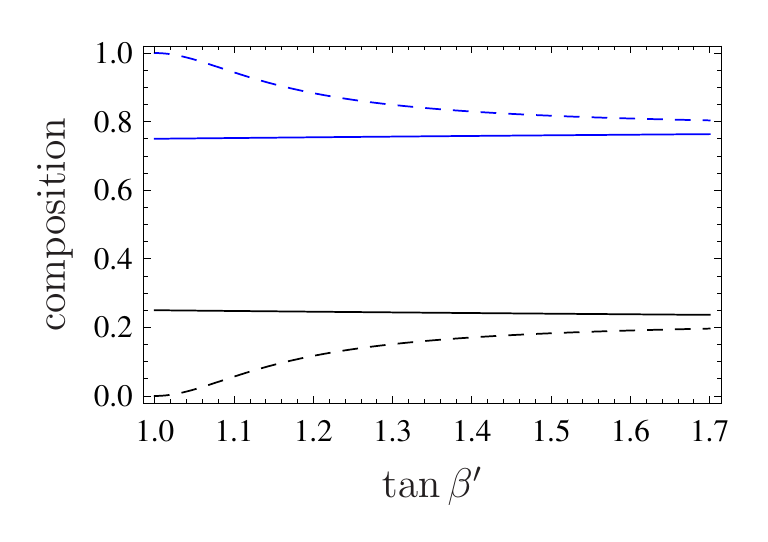} 
 \end{minipage}
\caption{Properties of the $3 \times 3$ $B-L$ neutralino sub-mass matrix.
 The left column shows the value of the lightest eigenvalues, the right column
 shows the \blino (black) and bileptino (blue) fraction of the lightest
 eigenstates for a variation of $M_{B'}$, $\mZp$, $\mu'$ and $\tan\beta'$.
 The solid lines correspond to a starting point with
 $M_{B'} = \mu' = \frac{1}{2} \mZp = 1$~TeV and $\tan\beta' = 1.1$, for the
 dashed lines $M_{B'} = 1$~TeV, $\mu' = 1.5$~TeV, $\mZp = 3$~TeV and
 $\tan\beta'$ = 1.4 has been chosen. }
\label{fig:neutralinoBLmatrix}
\end{figure}

If we neglect the kinetic mixing for a moment, the MSSM and $B-L$
sectors decouple and we can study the $3\times 3$
 sub-matrix to get
some feeling of the basic properties of the latter system. The matrix
can be re-written as
\begin{equation}
\left(\begin{array}{ccc}
 M_{B'} & - \mZp \sin\beta' & \mZp \cos\beta' \\
- \mZp \sin\beta' & 0 & - \mu' \\
\mZp \cos\beta' & -\mu' & 0
\end{array}\right).
\end{equation}
While $M_{B'}$, $\mZp$ and $\beta'$ are nearly independent, $\mu'$
is not a fundamental parameter of this model but it is connected
through the tadpole equations to the other three parameters.
Nevertheless, it is possible to change $\mu'$ by changing $m_0$ and
$A_0$ and therefore we will consider it here as independent. The
dependence of the smallest eigenvalue, as well as its \blino and
bileptino fractions, on these four parameters is depicted in
Figure~\ref{fig:neutralinoBLmatrix}. We can observe some interesting
features of that mass matrix, in particular in the
 phenomenologically 
interesting range of $\tan\beta'$ close to 1. 
Therefore we study the matrix in the limit $\tan\beta'=1$ where
one gets the following eigenvalues
\begin{eqnarray}
\label{eq:heavychi1}
m_1 &=&  - \mu' \\
m_{2,3} &=& \frac{1}{2} \left( M_{B'} + \mu' \mp 
\sqrt{(M_{B'} - \mu')^2 + 4 \mZp^2}\right)
\label{eq:heavychi2}
\end{eqnarray}
which is sufficient to understand the numerical results.
Note that the ordering of the eigenvalues at this stage
 is arbitrary.
From these equations one can easily derive two cases where one
eigenvalue is rather small

\begin{enumerate}
\item small $\mu'$
\item small $\mZp$ combined with small $M_{B'}$
\item $M_{B'}\mu' \simeq \mZp^2$
\end{enumerate}
This explains the features of the plots in
 \FIG~\ref{fig:neutralinoBLmatrix}: in the
 plot where $M_{B'}$ is varied, the lightest eigenvalue corresponds to $m_{2}$
 of \EQ~(\ref{eq:heavychi2}) for $M_{B'} \gsim 1$ TeV. This is also the case
 when \mZp is varied. The plots showing the \mup dependence show the lightest
 eigenvalue switching from being given by $m_{1}$ to $m_{2}$ at around
 $\mup = 1$ TeV. Increasing $\tan\beta'$ leads in 
general to a decrease of the lightest mass eigenvalue. 
For completeness we note that with these considerations
 one gets
a rough understanding of the extended neutralino sector. However,
for masses in the order of max($|\gmix v_u|,|M_{BB'}|$), the
 thus-far neglected mixing with the MSSM neutralinos becomes important.

\subsection{Sneutrinos}
\label{sec:model_sneutrino}
We focus here on the sneutrino sector as it shows two distinct features
compared to the MSSM. Firstly, it gets enlarged by the
 superpartners of the right-handed neutrinos.
 Secondly,
even more drastically, a splitting between the real and imaginary
parts of each sneutrino occurs resulting in twelve states: six scalar
sneutrinos and six pseudoscalar ones
\cite{Hirsch:1997vz,Grossman:1997is}. The origin of this splitting
is the $Y^{ij}_x\,\hat{\nu}^c_i\,\hat{\eta}\,\hat{\nu}^c_j$ term in the
superpotential, eq.~(\ref{eq:superpot}), which is a $\Delta L=2$ operator
after the breaking of $U(1)_{B-L}$.  
Therefore, we define
\begin{equation}
\tilde{\nu}^i_L = \frac{1}{\sqrt{2}}\left(\sigma^i_L
 + i \phi^i_L\right)\, \hspace{1cm} 
 \tilde{\nu}^i_R = \frac{1}{\sqrt{2}}\left(\sigma^i_R + i \phi^i_R\right)
\end{equation}
In the following we will denote the partners of the
left-handed and right-handed neutrinos by \lsnus and \rsnus, respectively.
The $6 \times 6$ mass matrices of the CP-even
(\mReSnuSq) and CP-odd (\mImSnuSq)
sneutrinos can be written in the basis
\(\left(\sigma_{L},\sigma_{R}\right)\) respectively
\(\left(\phi_{L},\phi_{R}\right)\) as
\begin{equation} 
\mReSnuSq = \left( 
\begin{array}{cc}
m_{LL} &m^{R,T}_{RL}\\ 
m^R_{RL} &m^R_{RR}\end{array} 
\right), \hspace{1cm} 
\mImSnuSq = \left( 
\begin{array}{cc}
m_{LL} &m^{I,T}_{RL}\\ 
m^I_{RL} &m^I_{RR}\end{array} 
\right) \hspace{0.5cm} .
\end{equation} 
While $m^I_{LL}=m^R_{LL}=m_{LL}$ holds\footnote{We have neglected the
splitting induced by the left-handed neutrinos as this is suppressed
by powers of the light neutrino
 mass over the sneutrino mass.}, 
the entries involving
 \rsnus differ by a few signs. It is possible to express them in a
compact form by
\begin{align} 
 m_{LL} &= \frac{1}{8} \Big({\bf 1} \Big( \Big(g_{1}^{2}+g_{2}^{2}
+{\gmix}^{2} \Big) \Big(- v_{u}^{2}  + v_{d}^{2}\Big)
 + \gmix \gBL{} \Big(-2 v_{\bar{\eta}}^{2}  + 2 v_{\eta}^{2}
  - v_{u}^{2}
  + v_{d}^{2}\Big)\nonumber \\ 
  & \, \, +2 \gBL{2} \Big(- v_{\bar{\eta}}^{2}  + v_{\eta}^{2}\Big)\Big)
+8 m_l^2 +4 v_{u}^{2} {Y^T_\nu  Y_{\nu}^{*}} \Big)\, ,\\ 
 m^{R,I}_{RL} &= \frac{1}{4} \Big(-2 \sqrt{2} v_d \mu Y_{\nu}^{*}
   + v_u \Big(2 \sqrt{2} T_{\nu}^{*} 
 \pm 4 v_{\eta} {Y_x  Y_{\nu}^{*}} \Big)\Big)\, , \\ 
 \label{eq:mRR}
m^{R,I}_{RR}
 & = \frac{1}{8} \Big({\bf 1}\Big( 2 \gBL{2} (v^2_{\bar{\eta}}
  - v^2_{\eta}) - \gmix \gBL{} \Big(- v_{u}^{2} 
 + v_{d}^{2}\Big)\Big)+8 {m_{\nu^c}^2}
 +2 v_{\bar{\eta}} \Big(\mp 4 \sqrt{2} Y_x {\mu'}^*  \Big) \nonumber \\ 
 & \, \, +4 v_{u}^{2} {Y_{\nu}  Y_\nu^\dagger} 
 +2 v_{\eta} \Big(\pm 4 \sqrt{2} T_x  + 8 v_{\eta}  {Y_x  Y_x^*}\Big)\Big)\, .
\end{align} 
The upper signs correspond to the scalar and the lower ones to the
pseudoscalar matrices and we have assumed
 CP conservation.  In the case
of complex trilinear couplings or $\mu$-terms, a mixing between the
scalar and pseudoscalar particles occurs, resulting in 12 mixed states
and consequently in a $12\times 12$ mass matrix.  It particular the
term $\sim v_{\bar{\eta}} Y_x {\mu'}^*$ is potentially large and
induces a large mass splitting between the scalar and pseudoscalar
states. Also the corresponding soft SUSY-breaking term
 $\sim v_{\eta} T_x$ can
lead to a sizable mass splitting in the case of large 
 $|T_x|$, \EG for large $|A_0|$ at the GUT-scale
where $T_x = A_0 Y_x$ holds.

\begin{figure}[t]
 \begin{minipage}{0.99\linewidth}
\centering
\includegraphics[width=0.44\linewidth]{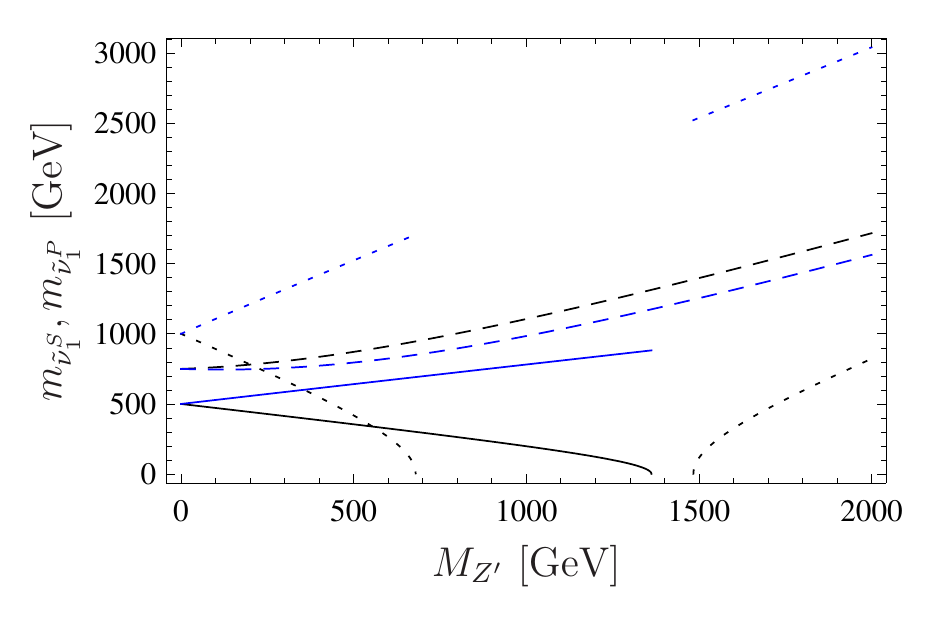}
\hfill
\includegraphics[width=0.44\linewidth]{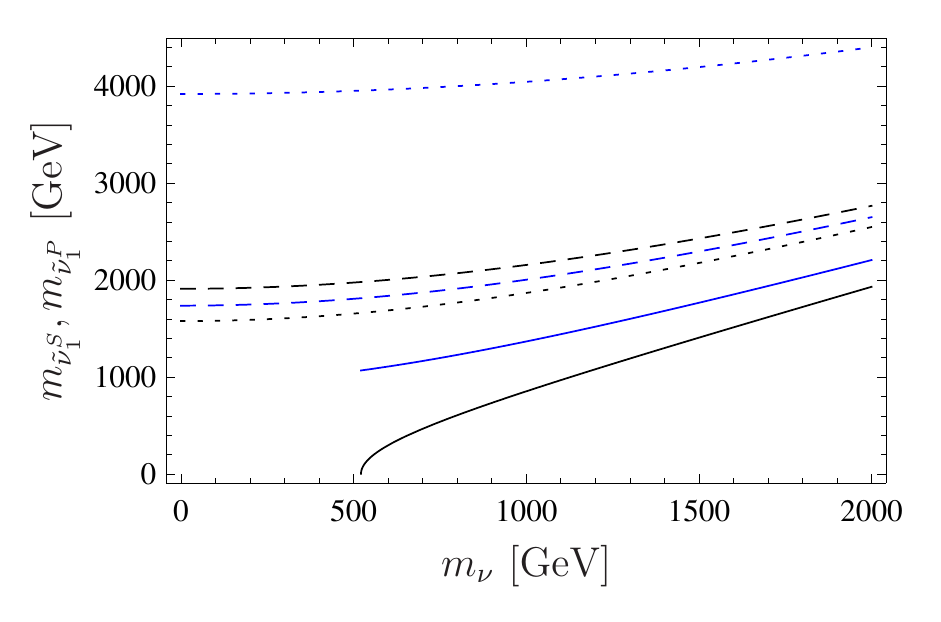} \\
\includegraphics[width=0.44\linewidth]{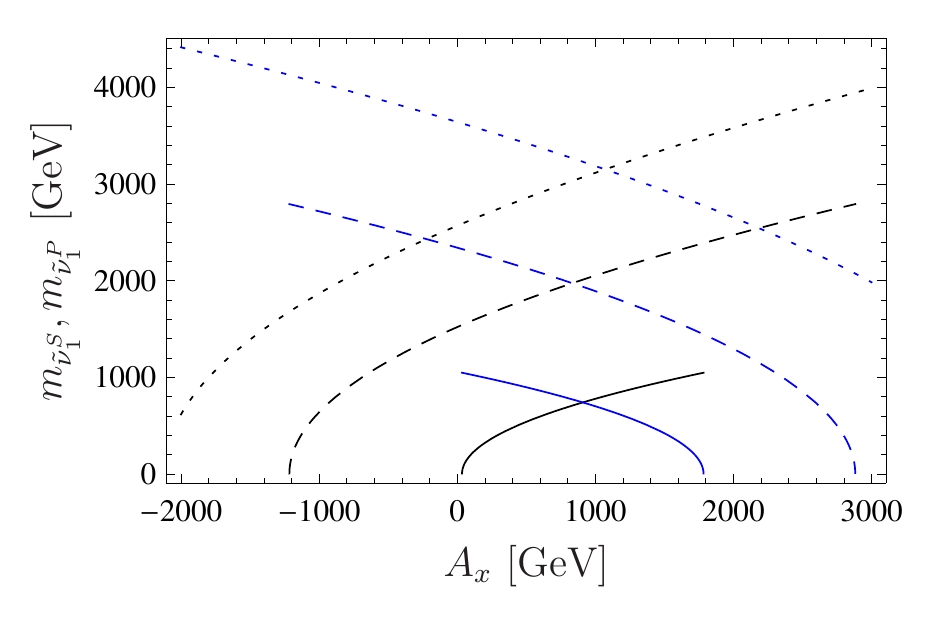}
\hfill
\includegraphics[width=0.44\linewidth]{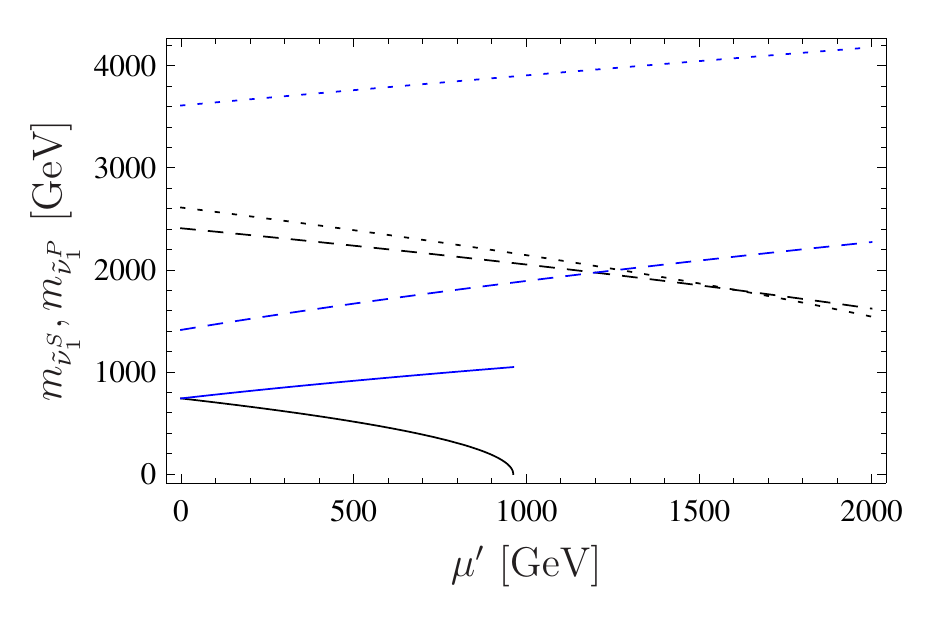} \\
\includegraphics[width=0.44\linewidth]{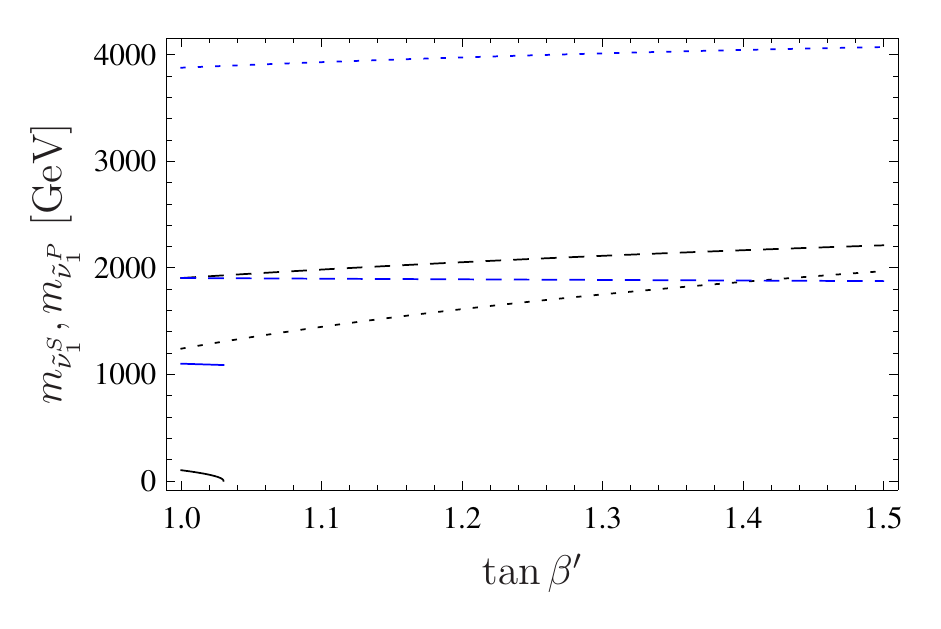}
\hfill
\includegraphics[width=0.44\linewidth]{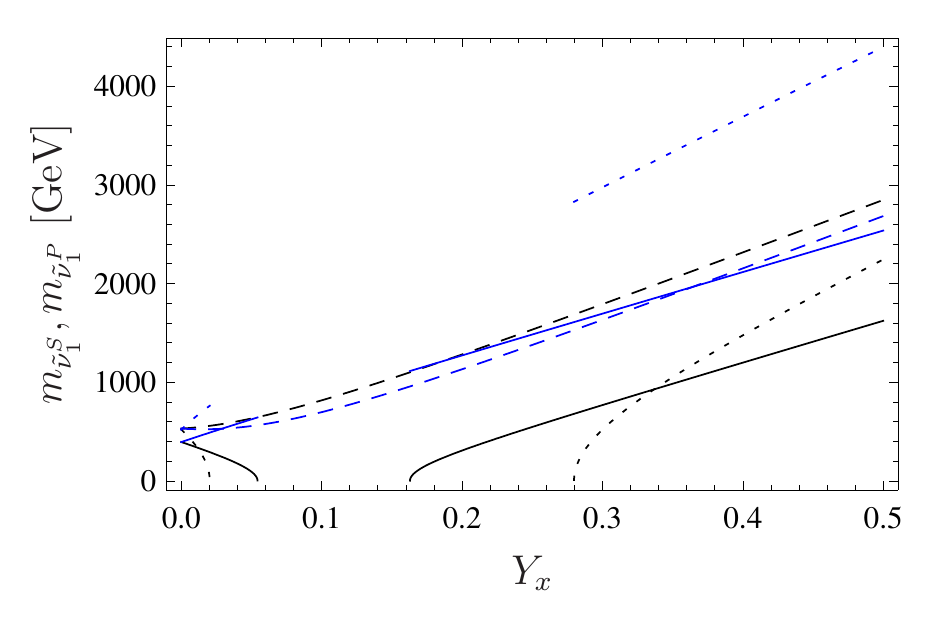} \\
 \end{minipage}
\caption{Mass dependence of the  CP-even (black) and CP-odd 
(blue) \rsnus. As the
 starting points,  we used
$\mZp = 2$~TeV, $A_x = 0$, $\mu' = 1$~TeV, $m_{\nu^c} = 0.5$~TeV, 
$\tan\beta' = 1.1$ and $Y_x = 0.15$ (solid lines);
$\mZp = 2.5$~TeV, $A_x = 1$~TeV, $\mu' = 1$~TeV, $m_{\nu^c} = 
0.75$~TeV, $\tan\beta' = 1.2$ and $Y_x = 0.35$ (dashed lines);
$\mZp = 3$~TeV, $A_x = -1$~TeV, $\mu' = 1.5$~TeV, $m_{\nu^c} = 1$~TeV, 
$\tan\beta' = 1.4$ and $Y_x = 0.45$ (dotted lines).
Note that the lines are only shown for regions in which
 neither the CP-even nor -odd sneutrino is tachyonic. }
\label{fig:sneutrinoMasses}
\end{figure}

To gain also some feeling for the behavior of the sneutrino masses we
can consider a simplified setup: neglecting kinetic mixing as well as
left-right mixing,  the
masses of the \rsnus can be expressed as
\begin{align}
\label{eq:mSnuA}
\mReSnuSq \simeq & \,\, m_{\nu^c}^2
 + \mZpSq \left( \frac{1}{4} \cos(2 \beta')
                 + \frac{2 Y_x^2}{\gBL{2}} \sin\beta'^2 \right)
 + \mZp \frac{\sqrt{2} Y_x}{\gBL{}}
    \left(A_\nu \sin\beta'-\mu' \cos\beta' \right)\, ,\\ 
\mImSnuSq \simeq & \,\, m_{\nu^c}^2
 + \mZpSq \left( \frac{1}{4} \cos(2 \beta')
                 + \frac{2 Y_x^2}{\gBL{2}} \sin\beta'^2 \right)
 - \mZp \frac{\sqrt{2} Y_x}{\gBL{}}
    \left(A_\nu \sin\beta'-\mu' \cos\beta' \right)\, .
\label{eq:mSnuB}
\end{align}
In addition, we treat the parameters $A_x$, $m_{\nu^c}^2$, $\mZp$, $\mu'$,
$Y_x$ and $\tan\beta'$ as independent. The different effects on
 the sneutrino masses are shown in Figure~\ref{fig:sneutrinoMasses}
and can easily be understood by inspecting \EQS~(\ref{eq:mSnuA})
and (\ref{eq:mSnuB}). The first two terms give always a positive
contribution whereas the third one gives either a positive
or a negative one depending on the sign of 
$A_x \sin\beta'-\mu' \cos\beta'$. For example choosing 
$Y_x$ and $\mu'$ positive, one finds that  the CP-odd
 (CP-even)
sneutrino is the lighter one for $A_x < 0$ ($A_x > 0$).
For completeness we note that for 
$A_x \sin\beta'\simeq \mu' \cos\beta'$, the mass splitting is 
rather small compared to the masses and thus one has effectively
a complex sneutrino.

\subsection{Constrained model}
We will consider in the following a scenario motivated by minimal
supergravity assuming a GUT unification
 of all soft SUSY-breaking scalar mass parameters
as well as a unification of all gaugino
mass parameters
\begin{align}
 m^2_0 = & m^2_{H_d} = m^2_{H_u} = m^2_{\eta} = m^2_{\bar{\eta}} \\
m^2_0 \mathbf{1} = & m_D^2  =  m_U^2  = m_Q^2 
= m_E^2  = m_L^2  = m_{\nu^c}^2  \\
 M_{1/2} = & M_1 = M_2 = M_3 = M_{\tilde{B}'}
\end{align}
Also, for the trilinear soft SUSY-breaking coupling, the ordinary
 \msugra-inspired conditions are assumed
\begin{align}
 T_i = A_0 Y_i, \hspace{1cm} i = e,d,u,x,\nu \thickspace . 
\end{align}

Furthermore, we assume that there are no off-diagonal gauge couplings
or gaugino mass parameters present at the GUT scale
\begin{align}
 g_{B Y} = & g_{Y B} = 0 \\
 M_{B B'} = & 0
\end{align}
This choice is motivated by the possibility that the two Abelian groups are a
remnant of a larger product group which gets broken at the GUT
scale as stated in the introduction.
In that case \(g_{Y Y}\) and \(g_{B B}\) correspond to the physical
couplings $g_1$ and $\gBL{}$,  which we assume
 to unify with $g_2$:
\begin{equation}
 g^{GUT}_1 = g^{GUT}_2 = \gBL{GUT} \thickspace .
\end{equation}
where we have already taken into account the correct GUT normalization
as discussed in section \ref{subsec:kineticmixing}.

In addition, we consider the mass of the $Z'$ and $\tan\beta'$ as 
inputs and use the following set of free parameters
\begin{eqnarray}
& m_0, \thickspace M_{1/2},\thickspace A_0,\thickspace \tan\beta,\thickspace
 \tan\beta',\thickspace \sign(\mu),\thickspace \sign(\mu'),\thickspace \mZp,
 \thickspace  Y_x \thickspace \mbox{and} \thickspace Y_{\nu} . &
\end{eqnarray}
\(Y_{\nu}\) is constrained by neutrino data and must therefore be
very small in comparison to the other couplings in this model, 
as required by the embedded TeV-scale type-I seesaw mechanism. 
Therefore we neglect it in the following. $Y_x$ can always
 be taken
diagonal and thus effectively we have 9 free parameters and 2 signs.

\section{$B-L$ Dark Matter}
\label{sec:darkmatter}
Astrophysical observations and the data from WMAP
\cite{Komatsu:2010fb} put the existence of
non-baryonic dark matter in the Universe on solid grounds. 
The best-fit value from a
combined analysis of the cosmic microwave background (CMB), supernovae
observations and baryonic acoustic oscillations (BOA) predicts a dark
matter density of \cite{Jarosik:2010iu}
\begin{equation}
\Omega h^2 = 0.1123\pm 0.0035
\end{equation}
at the 1$\sigma$ level.  As already mentioned, there are four distinct
regions in the parameter space of the CMSSM which lead to a neutralino
density consistent with this observation: (i) the bulk region with
light sfermions enabling a sufficient t-channel annihilation, (ii) the
co-annihilation region with a second, 
 particle close in mass to the LSP with stronger interactions
\cite{Ellis:1998kh,Boehm:1999bj,Ellis:2001nx,Edsjo:2003us}, (iii) the
Higgs funnel characterised by a resonance of the LSP with the
 pseudoscalar Higgs bosons \cite{Griest:1990kh} and
 (iv) the focus point
region where the large Higgsino fraction of the LSP
  increases the coupling to the SM gauge
 bosons and third generation quarks
\cite{Feng:1999zg,Feng:2000gh}.

As stated above, the extended neutralino and sneutrino
sectors yield additional possibilities
 for explaining the relic density.
Here we discuss the details of how
 the correct value can be
obtained. As in  the usual CMSSM, this requires some
 special mechanisms, either resonances or co-annihilation.
 The \CBLSSM requires
in general large values of $m_0$ to get a consistent solution for the
tadpole equations. Therefore it
 does not seem to be possible to find a
bulk-like region, despite the presence of new D-term contributions to
the masses of the sfermions. However, it turns out that there are
different manifestations of the other mechanisms to 
reduce the relic
density for a neutralino LSP which is either mainly \blino or
 bileptino.
 
Another dark matter candidate in the CMSSM is the lightest sneutrino.
However, due to its coupling to the $Z$ boson, a pure
 ``left-handed''
sneutrino LSP is already ruled out by direct dark matter searches
\cite{Falk:1994es}. In contrast, as shown in
\SEC~\ref{sec:model_sneutrino}, the LSP in the \BLSSM
 can be a CP-even or -odd \rsnu with a very
suppressed coupling to the $Z$ boson. We will start our discussion
with the sneutrino LSPs in section~\ref{sec:SneutrinoDM} and we will
 present the results for the \blino and
 bileptino LSPs in
\SEC~\ref{sec:fermionDM}. Finally, we will  discuss
the impact of direct
detection experiments on the different dark matter candidates in
\SEC~\ref{sec:DD} and comment briefly on the impact of
 the \BLSSM
on MSSM-like dark matter candidates in \SEC~\ref{sec:impactMSSM}.
Before we start, some remarks about the numerical calculation are
 in order.

\paragraph{Numerical setup} To check the properties of the new dark
matter candidates arising in the \BLSSM, we have used
 the implementation of the model in \SPheno
\cite{Porod:2003um,Porod:2011nf} based on the corresponding output of
\SARAH \cite{Staub:2008uz,Staub:2009bi,Staub:2010jh}. This
implementation provides a precise mass calculation using two-loop RGEs
and one-loop corrections to all masses. Also, all effects of kinetic
mixing are taken into account during the RGE running by using the
results presented in \REF~\cite{Fonseca:2011vn}. For more details
about the calculation of the mass spectrum, we refer the interested
reader to \REF~\cite{O'Leary:2011yq}.  The calculation of the relic
density of the LSP is done with \MO \cite{Belanger:2006is} version
\texttt{2.4.5} based on the \CalcHep output of \SARAH. The data transfer
between \SPheno and \MO happens via the \texttt{SLHA+} functionality of
\CalcHep \cite{Belanger:2010st} which enables \CalcHep to read the
SLHA spectrum file written by \SPheno. To perform the scans we have
used the Mathematica package {\tt SSP} \cite{Staub:2011dp}.
 In particular, all presented benchmark scenarios in the following
 include scalar Higgs bosons in 
the preferred mass range of 122-128~GeV. All scalar masses are calculated
using the full one-loop corrections as well as the dominant 
two-loop corrections known from the MSSM.

\subsection{Sneutrino dark matter}
\label{sec:SneutrinoDM}
Since there is  a large mass splitting only 
 for \rsnus, the LSP can  be either a CP-even or -odd
 \rsnu. Neglecting the tiny $Y_\nu$ neutrino Yukawa
couplings, the only tree-level interactions are with the
Higgs particles and the $Z'$ boson  and with the
corresponding superpartners. However, the $Z'$ boson cannot
contribute to the dark matter annihilation because it just couples to
one CP-even and one CP-odd sneutrino at a time, but it is only
 possible
to get one of them lighter than the neutralinos at the same time.
Furthermore, in
 the \CBLSSM, the typically large value of $m_{0}$ leads to very heavy
 Higgs pseudoscalars and
therefore the main annihilation properties are fixed by the
interaction with the scalar Higgs fields. We can write the
corresponding three- and four-point interactions in the limit of
vanishing $Y_\nu$ neutrino Yukawa couplings and diagonal $Y_x$ as
{\allowdisplaybreaks
\begin{align}
\Gamma_{\tilde{\nu}^{{S,P}}_i \tilde{\nu}^{{S,P}}_j h_k h_l}
 \simeq
 & \frac{i}{4} \Big(-16 \sum_{c=1}^{3}|Y_{x,{c c}}|^2 Z^{X}_{i 3 + c}
 Z^{X}_{j 3 + c}  Z_{{k 3}}^{H} Z_{{l 3}}^{H} \nonumber \\ 
 &+g_{B} \sum_{a=1}^{3}Z^{X}_{i 3 + a} Z^{X}_{j 3 + a}
  \Big(2 g_{B} (Z_{{k 3}}^{H} Z_{{l 3}}^{H}  - Z_{{k 4}}^{H} Z_{{l 4}}^{H})
  + \gmix (Z_{{k 1}}^{H} Z_{{l 1}}^{H}
  - Z_{{k 2}}^{H} Z_{{l 2}}^{H}) \Big)\nonumber \\ 
 &+\sum_{a=1}^{3}Z^{X}_{i a} Z^{X}_{j a}  \Big(2 g_{B} \Big(\gmix
 + \gBL{}\Big)\Big( Z_{{k 4}}^{H} Z_{{l 4}}^{H}
- Z_{{k 3}}^{H} Z_{{l 3}}^{H} \Big) \nonumber \\
 & \label{eq:vertexSvSvHH} \hspace{2cm}
 - \Big(\gmix \Big(\gmix
 + \gBL{}\Big) + g_{1}^{2} + g_{2}^{2}\Big)(Z_{{k 1}}^{H} Z_{{l 1}}^{H}
-Z_{{k 2}}^{H} Z_{{l 2}}^{H}) \\
\Gamma_{\tilde{\nu}^{{S,P}}_i \tilde{\nu}^{{S,P}}_j h_k} \simeq
 & -\frac{i}{4} \Big(\pm \gBL{}
 \sum_{a=1}^{3}Z^{X}_{i 3 + a} Z^{X}_{j 3 + a}\Big(2 \gBL{}
 (v_{\bar{\eta}} Z_{{k 4}}^{H}  - v_{\eta} Z_{{k 3}}^{H})
  - \gmix (v_d Z_{{k 1}}^{H}  - v_u Z_{{k 2}}^{H}) \Big)\nonumber \\ 
 &+2 \Big(\Big(\pm 8 v_{\eta} \sum_{c=1}^{3}|Y_{x,{c c}}|^2
 Z^{X}_{i 3 + c} Z^{X}_{j 3 + c}
   + 2 \sqrt{2} \sum_{b=1}^{3}Z^{X}_{i 3 + b} Z^{X}_{j 3 + b}
 \Re(T^*_{x,{b b}})  \Big)Z_{{k 3}}^{H} \nonumber \\ 
 &- \sqrt{2} \Big({\mu_{\eta}} \sum_{b=1}^{3}Y^*_{x,{b b}}
 Z^{X}_{i 3 + b} Z^{X}_{j 3 + b} 
  + {\mu_{\eta}}^* \sum_{b=1}^{3}Z^{X}_{i 3 + b}
 Z^{X}_{j 3 + b} Y_{x,{b b}}  \Big)Z_{{k 4}}^{H} \Big)\nonumber \\ 
 &+\sum_{a=1}^{3}Z^{X}_{i a} Z^{X}_{j a}  \Big(2 \gBL{}
 \Big(\gmix + \gBL{}\Big)\Big(v_{\eta} Z_{{k 3}}^{H}
- v_{\bar{\eta}} Z_{{k 4}}^{H} \Big) \nonumber \\
 & \label{eq:vertexSvSvH} \hspace{2cm} \pm \Big(\gmix
 \Big(\gmix
 + \gBL{}\Big) + g_{1}^{2} + g_{2}^{2}\Big)(v_d Z_{{k 1}}^{H}
-v_u Z_{{k 2}}^{H})  \Big)\Big)
\end{align}} with $X={S,P}$ for the rotation matrices in case of 
CP-even ($Z^S$) and CP-odd ($Z^P$)  sneutrinos,
 respectively. In \EQS~(\ref{eq:vertexSvSvHH} ) and (\ref{eq:vertexSvSvH})
the upper (lower) signs are for CP-even (CP-odd) sneutrinos.  The
 bilepton \vevs are usually larger than the light
 bilepton
or the sneutrino LSP masses because of the  large
$Z'$ mass. Thus we
can expect that diagrams involving two three-point interactions
dominate over those with one four-point interaction. Furthermore,
there is also the possibility of a resonance between the sneutrino and
a Higgs particle. Since there are qualitative differences between the
behaviour of CP-even and -odd sneutrinos, we discuss them
 separately in
the following, starting with the CP-even case.

\subsubsection{CP-even sneutrino LSP}
\label{sec:CPevenDM}
\begin{figure}[t]
 \begin{minipage}{0.99\linewidth}
\centering
\includegraphics[width=0.48\linewidth]{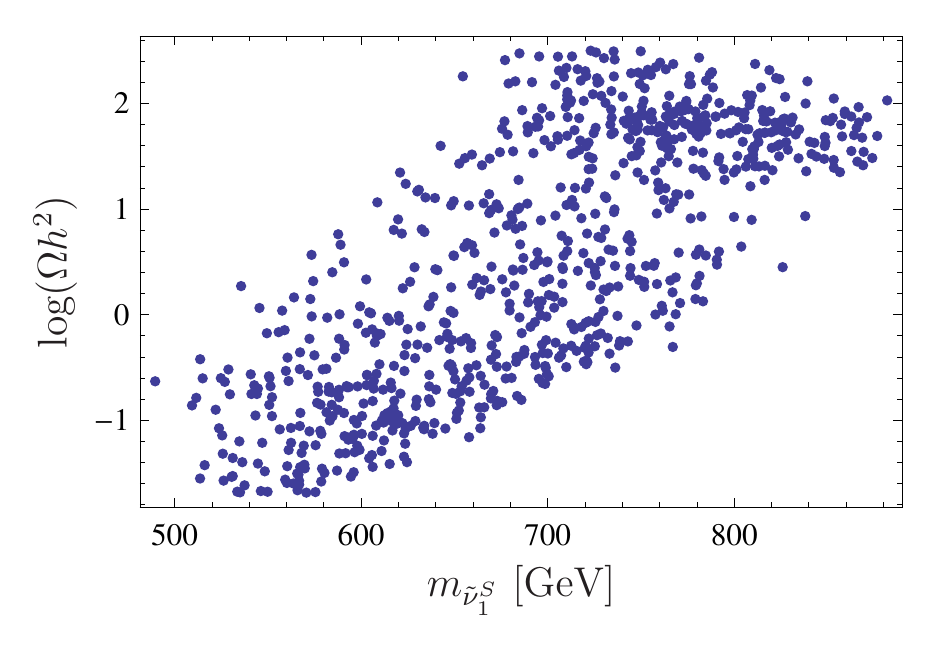}
\hfill
\includegraphics[width=0.48\linewidth]{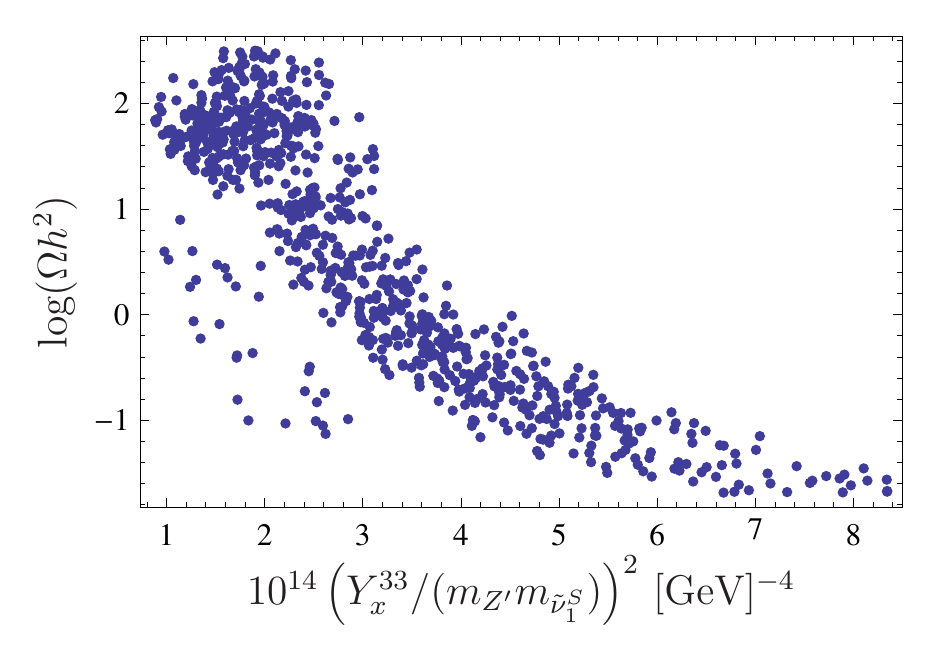} 
 \end{minipage}
 \caption{CP-even sneutrino dark matter. The left plot shows the
   dependence of the relic density $\log(\Omega h^2)$ on the mass of
   the LSP $m_{\ssnu{1}}$. The right hand side gives
   $\log(\Omega h^2)$ as function of $10^{14}\left(Y_x^{33}/(\mZp
     m_{\ssnu{1}})\right)^2$.  The chosen parameter ranges are
   $m_0 = [1.7,1.9]$~TeV, $M_{1/2} = [1.5,1.8]$~TeV, $\tan\beta =
   [6,11]$, $A_0 = -1.4$~TeV, $\tan\beta' = [1.16,1.20]$, $\mZp =
   [2.5,3.0]$~TeV, $Y_x^{33} = [0.10,0.42]$, $Y_x^{11} = Y_x^{22} =
   0.42$ }
\label{fig:CPeven}
\end{figure}
In \FIG~\ref{fig:CPeven} (left-hand side) the relic density
 is shown as a function of the CP-even  sneutrino mass.
 As seen in
\SEC~\ref{sec:model_sneutrino}, scalar sneutrinos can be the LSP for
large negative values of  the combination
 $T_x \sin\beta' - Y_x \mu' \cos\beta'$.
This can be  obtained for large positive values of $Y_x$ and
large negative values of $A_0$. Therefore, the sneutrino interactions
coming from F-terms dominate and for a fixed value of the $Z'$ mass
their annihilation cross section is determined mostly by $Y_x$,
increasing with larger $Y_x$. However, the sneutrino  mass depends
strongly on $Y_x$ due to the RGE evolution and it gets smaller with
increasing $Y_x$. All-in-all, the relic density drops  for smaller
sneutrino masses. To make the dependence of the annihilation cross
section on $Y_x$ more visible, on the  right-hand side of
\FIG~\ref{fig:CPeven} we plot the relic density as a function of
$\left(Y_x^{33} /(\mZp m_{\ssnu{1}})\right)^2$, \IE we have
divided out the dependence on the sneutrino mass and on the $Z'$ mass,
finding a clear correlation.

\begin{figure}[t]
 \begin{minipage}{0.99\linewidth}
\centering
\includegraphics[width=0.48\linewidth]{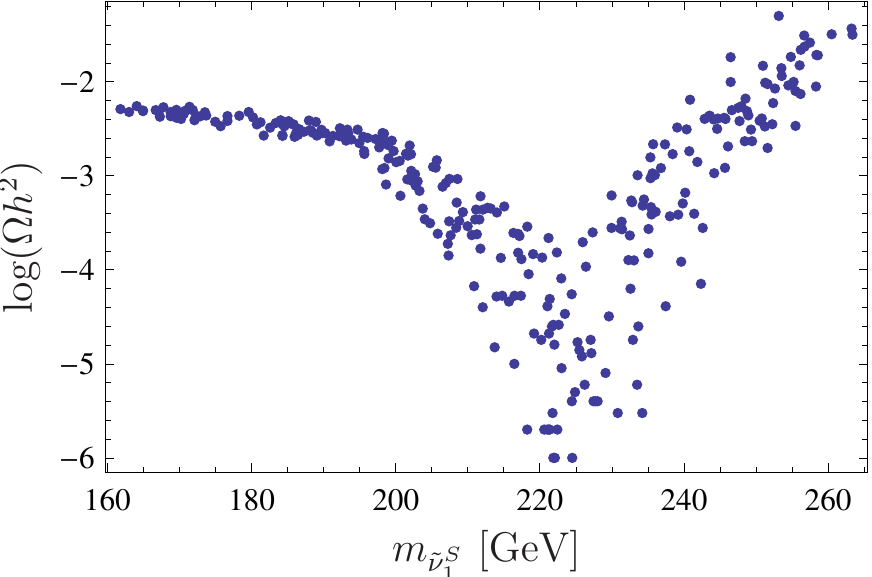}
\hfill
\includegraphics[width=0.48\linewidth]{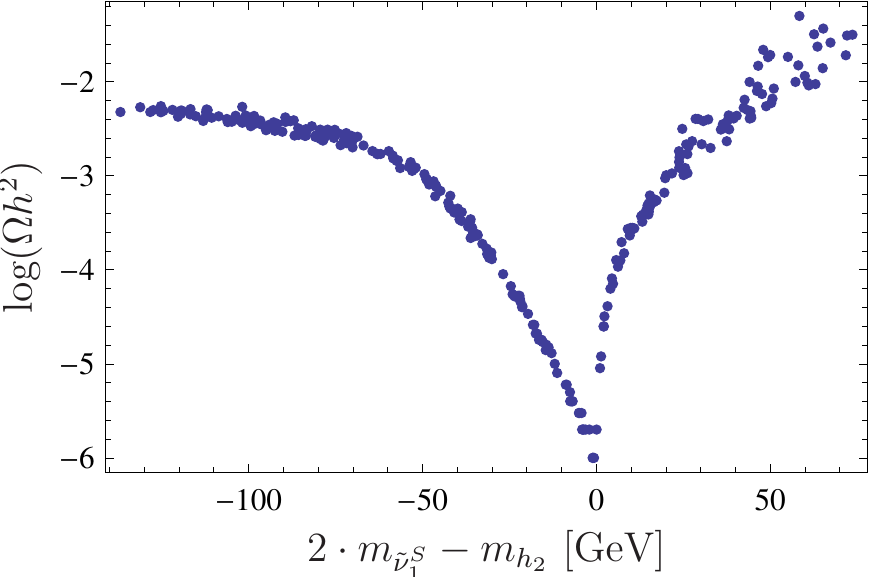} 
 \end{minipage}
 \caption{Rather light CP-even sneutrinos: $\log(\Omega h^2)$ vs.
   $m_{\ssnu{1}}$ (left) and $\log(\Omega h^2)$ vs. $2\cdot
   {m_{\ssnu{1}}} - m_{h_2} $ (right). The chosen parameter
   ranges were $m_0 = [1.55,1.65]$~TeV, $M_{1/2} =
   [500,550]$~GeV, $\tan\beta = [13,15]$, $A_0 = [-2.65,-2.5]$~TeV,
   $\tan\beta' = [1.33,1.36]$, $\mZp = [2.0,2.4]$~TeV, $Y_x^{33} =
   [0.33,0.35]$, $Y_x^{11} = Y_x^{22} = 0.42$ }
\label{fig:lightCPeven}
\end{figure}
The most important annihilation channel is the one with a bilepton
pair ($h_2 h_2$) in the final state which can reach up to 98 per-cent.
As can be seen, the sneutrino usually has a mass of several hundred
GeV. The reason is that there is an upper bound on the entries
of $Y_x$ from the requirement that there should be no
Landau pole up to the GUT scale. It is still
 possible to get lower masses by tuning $|A_0|$ and/or $\tan\beta'$ and one
 can even find sneutrino masses below 200~GeV. However, in this region of
 parameter
space the relic density is usually too small,
as shown in \FIG~\ref{fig:lightCPeven}, because of the large
annihilation cross section in two bileptons.  Even if this final
 state is
kinematically forbidden and the mass of the sneutrino is well below
the resonance point, the annihilation cross section for final
states containing SM vector bosons and Higgs bosons
 is still too large. Typically we find the following ratios 
 for the three dominant final states, 
 provided there is no
kinematical suppression: $W^+ W^- : Z Z : h_1 h_1 \simeq 2:1:1$.
We want to stress that this is a consequence of gauge
 kinetic mixing, as otherwise these final states would be strongly suppressed
 and the calculated relic density would be several
 orders of magnitude too large, and much larger than the measured value.
For completeness we note that these parameter points with a very small
sneutrino abundance are not ruled out, because the dark
matter can still be formed by another particle like the axino or the
axion \cite{Nomura:2008ru,Covi:2001nw}, or even by primordial black
holes \cite{Ivanov:1994pa}.

\subsubsection{CP-odd sneutrino LSP}
\label{sec:CPoddDM}
\begin{figure}[!ht]
 \begin{minipage}{0.99\linewidth}
\centering
\includegraphics[width=0.48\linewidth]{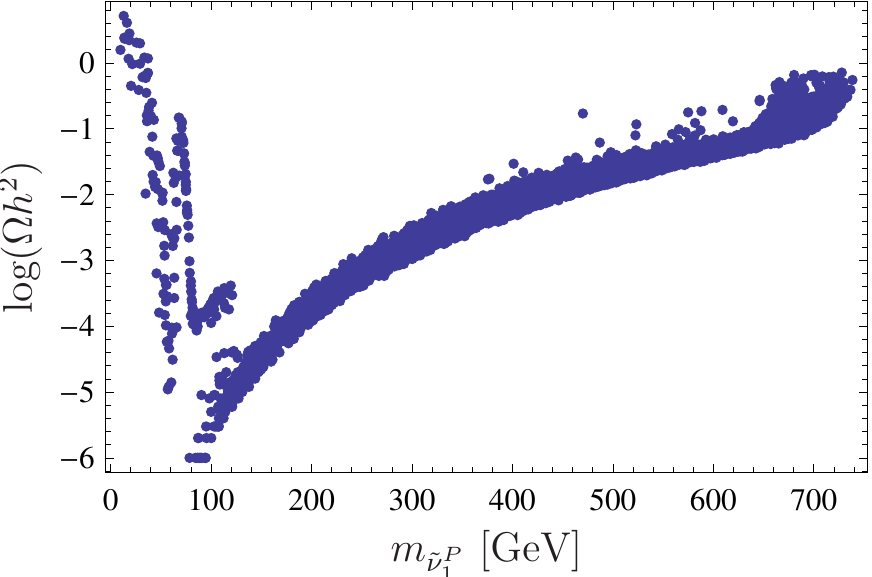}
\hfill
\includegraphics[width=0.48\linewidth]{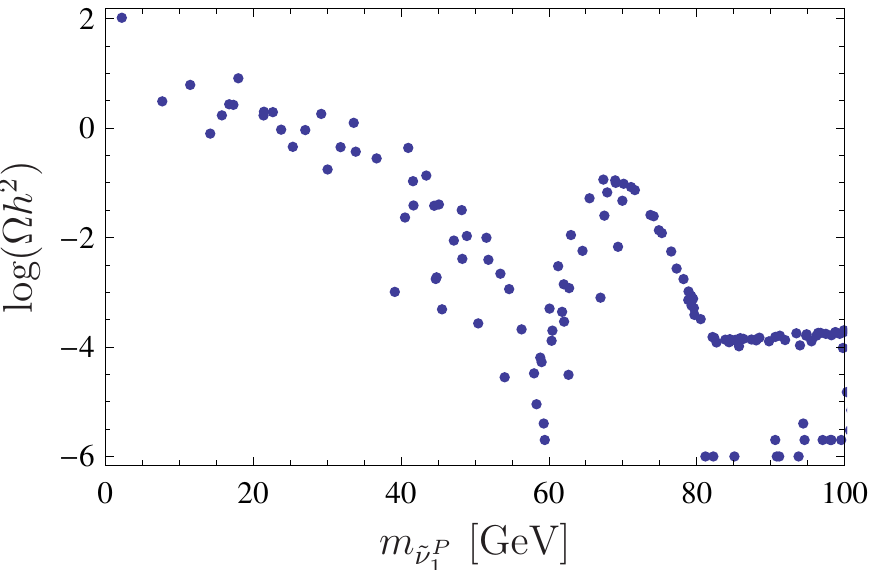} 
 \end{minipage}
 \caption{CP-odd sneutrino dark matter. Left: $\log(\Omega h^2)$ vs.
   $m_{\psnu{}}$. Right: zoom on low masses. Parameter
   ranges: $m_0 = [780,820]$~GeV, $M_{1/2} = [1.30,1.45]$~TeV,
   $\tan\beta = [8,12]$, $A_0 = [2.7,3.0]$~TeV, $\tan\beta' =
   [1.11,1.145]$, $\mZp = [2.5,3.0]$~TeV, $Y_x^{33} = [0.13,0.20]$,
   $Y_x^{11} = Y_x^{22} = 0.42$. }
\label{fig:pseudoscalar}
\end{figure}
For CP-odd sneutrinos to be the LSP, a necessary
 requirement  is
$T_x \sin\beta' - Y_x \mu' \cos\beta' > 0$. It turns out that
this can be achieved for $A_0>0$ and if one of the entries in
$Y_x$ is smaller compared to the others which avoids also the
problem of a Landau pole. We easily find sneutrino LSPs with a mass
of 50 GeV and below\footnote{We note that such light sneutrinos
are not constrained by LEP data as they are mainly SM singlets.} 
as can be seen from \FIG~\ref{fig:pseudoscalar}.
  Note the clear correlation between the relic density and the mass of the
 sneutrino
between 100 and 700~GeV. The reason is that the mass of the sneutrino
decreases with decreasing $Y_x$ while the annihilation cross section
increases: the D-terms in \EQ~(\ref{eq:vertexSvSvH}) are positive
while the F-terms and the trilinear soft SUSY-breaking terms are
 negative.
The smaller $Y_x$ is the more
the gauge interactions dominate increasing the cross
section. In general the final states from the annihilation of 
 CP-odd sneutrinos
  are similar to those from  CP-even sneutrinos: if kinematically
allowed, the final state with two bilepton dominates followed by
 those with SM vector bosons and MSSM Higgs
 bosons. However, for
sneutrino masses below about $m_Z/2$, only SM fermions show up
 as final states with the following branching ratios:
 $\bar{b} b$ ($\simeq 78$\%), 
$\bar{\tau} \tau$ ($\simeq 16$\%) and $\bar{c} c$ ($\simeq 5$\%)
for the dominant channels. However, there is one important difference:
around 60~GeV there is a pronounced dip
because of the resonance with the light MSSM-like Higgs.
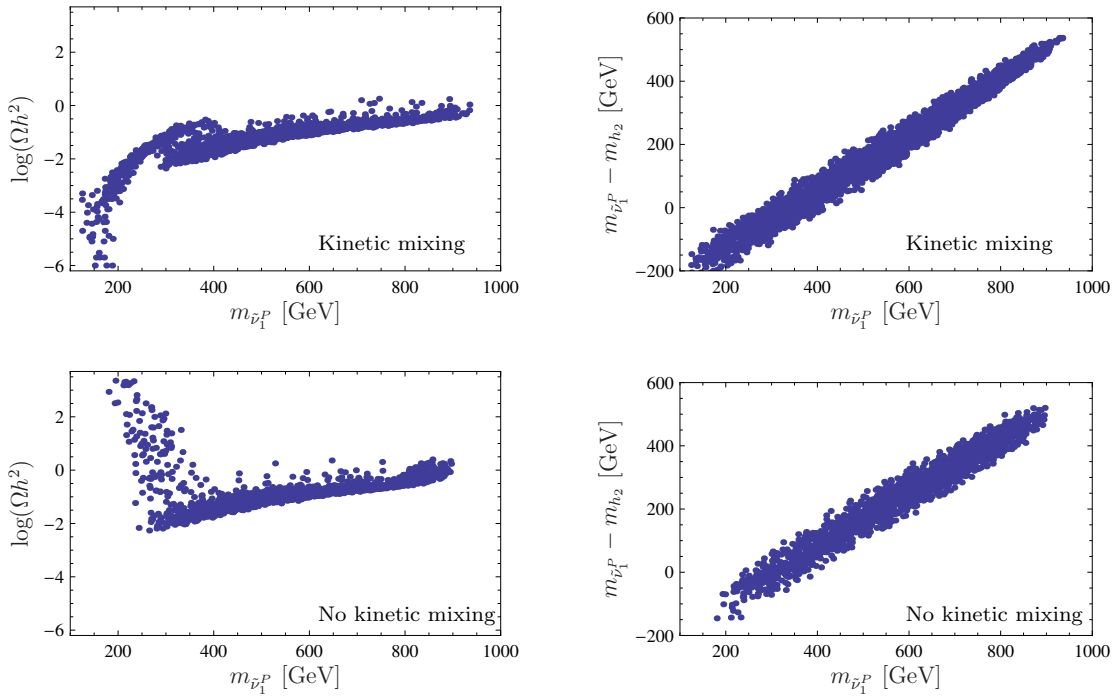
\begin{figure}[t]
\begin{picture}(0,0)
\put(125,40){\scriptsize Kinetic mixing}
\put(345,40){\scriptsize Kinetic mixing}
\put(125,-100){\scriptsize No kinetic mixing}
\put(345,-100){\scriptsize No kinetic mixing}
\end{picture}
 \begin{minipage}{0.99\linewidth}
\includegraphics[width=0.48\linewidth]{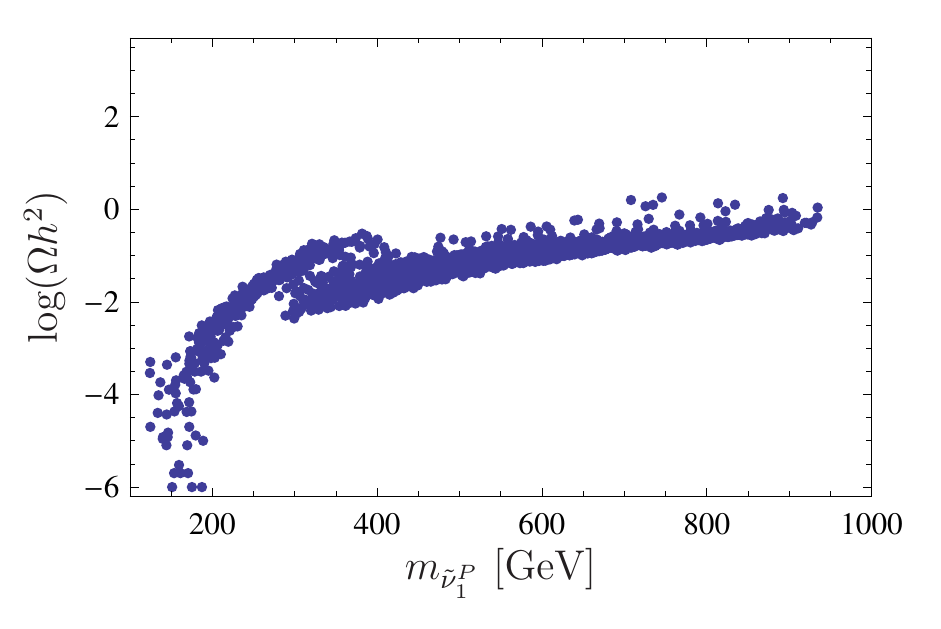}
\hfill
\includegraphics[width=0.48\linewidth]{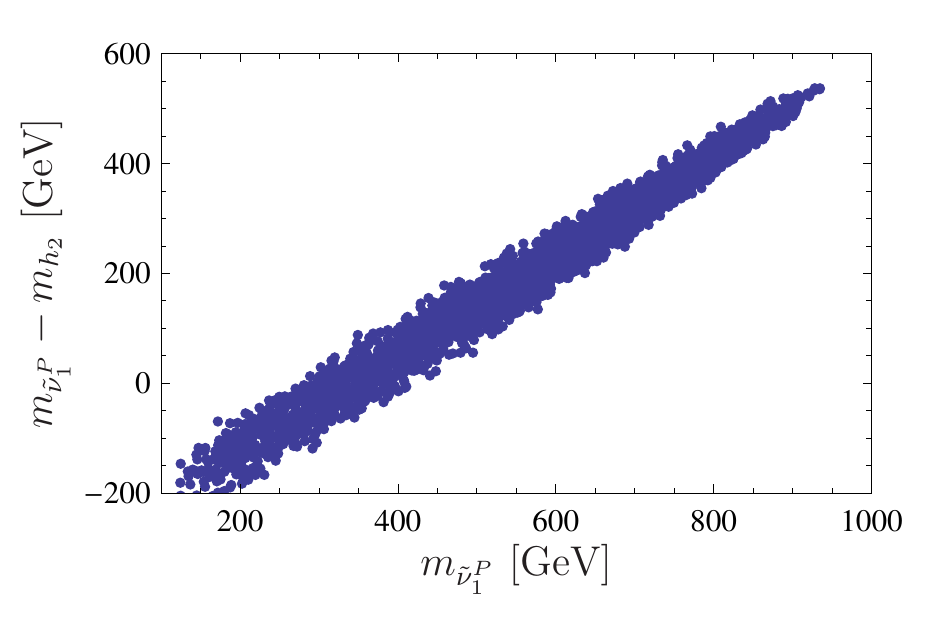} \\
\centering
\includegraphics[width=0.48\linewidth]{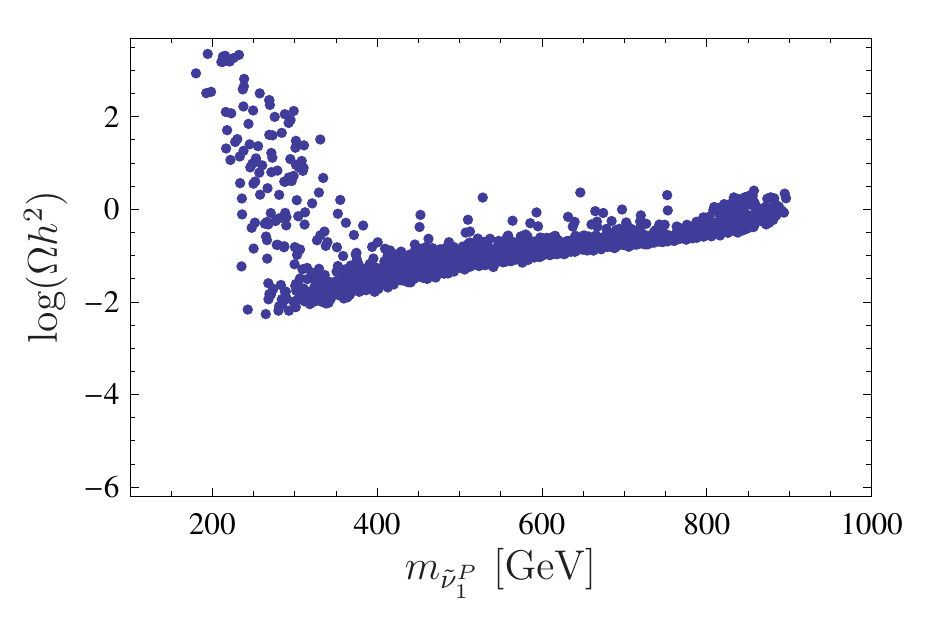}
\hfill
\includegraphics[width=0.48\linewidth]{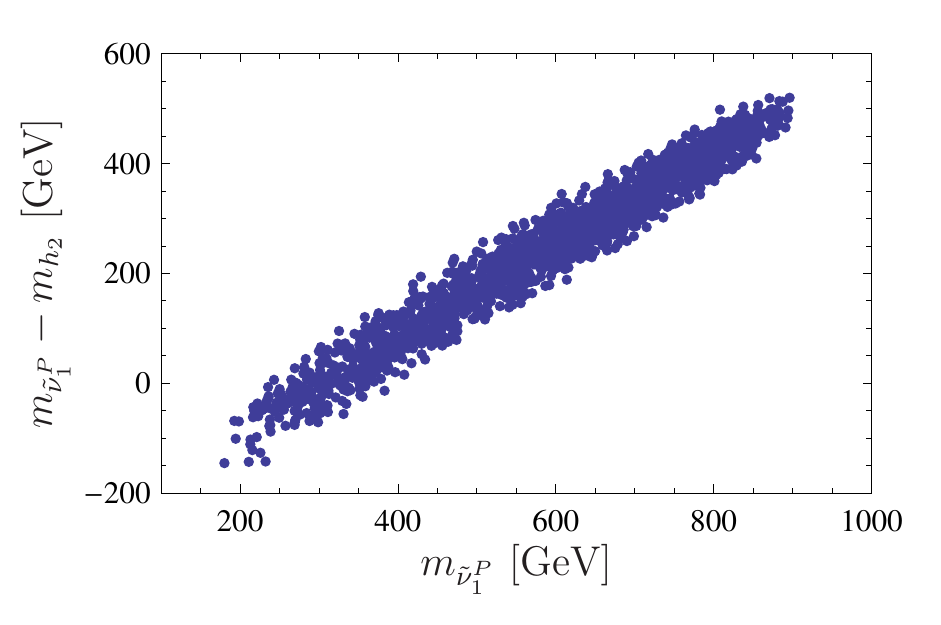} 
 \end{minipage}
 \caption{First row (left): $\log(\Omega h^2)$ vs. the mass of the
   lightest CP-odd sneutrino. The jump around 300~GeV is an caused by
   the mass of the bilepton as shown on the right side. The second row
   contains the same information, but kinetic mixing has been
   neglected. Parameters: $m_0 = [0.8,1.2]$~TeV, $M_{1/2} =
   [1.7,1.9]$~TeV, $\tan\beta = [10,15]$, $A_0 = [4.5,5]$~TeV,
   $\tan\beta' = [1.3,1.35]$, $\mZp = [2.7,3.0]$~TeV, $Y_x^{33} =
   [0.36,0.40]$, $Y_x^{11} = 0.42$, $Y_x^{22} = 0.33$.}
\label{fig:PSsneutrino_KM_NKM}
\end{figure}
For completeness, we mention that  a resonance
 is possible with not only the
MSSM-like Higgs, as shown in
 \FIG~\ref{fig:pseudoscalar},
but also with the bileptons. However, since the preferred final
 states are either SM gauge bosons or the light MSSM Higgs, the
 bilepton has to have a non-vanishing doublet fraction. Therefore,
  kinetic mixing is crucial for this case as well. A bilepton resonance
  in general suppresses the relic density even
 more than a doublet resonance and the
abundance of sneutrinos would be even smaller.

\subsection{\blino and bileptino dark matter}
\label{sec:fermionDM}
We turn now to the new fermionic dark matter candidates
 arising in the  \BLSSM: the \blino and bileptino.
 Much like a neutralino LSP in the
CMSSM, the relic density of  a neutralino LSP in the \BLSSM is in general
 too large.
Therefore, special mass configurations are needed to reduce the
abundance to the correct amount. We found as possible
 scenarios:
(i) Higgs resonances, (ii) $Z'$ resonance, (iii) sneutrino
co-annihilation (iv) stop and stau co-annihilation and (v)
 annihilation by a t-channel neutralino similar to the focus point in 
the CMSSM. As we will see during our discussion of the different
mechanisms, kinetic mixing turns out to be of
 large importance.

\subsubsection{Higgs resonances}
\label{sec:HiggsResonance}
A resonant annihilation of neutralino LSP with a Higgs particle is
already well-known from the Higgs funnel region in the CMSSM, which is
characterised by $2\cdot m_{\tilde{\chi}^0_1} = m_A$ ($m_A$ the
mass of the  pseudoscalar)\footnote{In some parts of the parameter
space  resonances via the scalar Higgs bosons $h^0$ and $H^0$ 
are also open in models with non-universal Higgs masses, see \EG
\cite{Das:2010kb}.}. However, the  extended Higgs sector
offers the possibility  of additional
 resonances. We find that in the bilepton sector the resonance
is usually via the  lightest scalar which is relatively
light as $\tan\beta'$ is close to 1 \cite{O'Leary:2011yq,Hirsch:2011hg}.
However, the pseudoscalar as well as the heavier scalar bilepton
are usually heavier than the $Z'$ and thus can only be effective
for very heavy LSPs. Therefore, we concentrate here on the resonances
with scalars. The interaction between a neutralino and the scalar
Higgs can be parametrised by
\begin{equation}
\label{eq:FFS}
 \Gamma^L \frac{1-\gamma_5}{2} + \Gamma^R \frac{1+\gamma_5}{2}
\end{equation}
with the coefficients
\begin{align}
 \Gamma^R_{\tilde{\chi}^0_{{i}}\tilde{\chi}^0_{{j}}h_{{k}}}  =
  & \,\frac{i}{2} \Big(Z_{{k 1}}^{H} \Big(\Big(g_1 N_{{i 1}}  - g_2 N_{{i 2}}
  + \gmix N_{{i 5}} \Big)N_{{j 3}}
  + N_{{i 3}} \Big(g_1 N_{{j 1}}  - g_2 N_{{j 2}}
  + \gmix N_{{j 5}} \Big)\Big)\nonumber \\ 
 &- Z_{{k 2}}^{H} \Big(\Big(g_1 N_{{i 1}}  - g_2 N_{{i 2}}
  + \gmix N_{{i 5}} \Big)N_{{j 4}}  + N_{{i 4}} \Big(g_1 N_{{j 1}}
  - g_2 N_{{j 2}}  + \gmix N_{{j 5}} \Big)\Big)\nonumber \\ 
 &+2 \Big(Z_{{k 3}}^{H} \Big(\Big(\gBL{} N_{{i 5}}
  + N_{{i 6}} \gBL{} N_{{j 5}} \Big)
 - Z_{{k 4}}^{H} \Big(\gBL{} N_{{i 5}} N_{{j 7}}
  + N_{{i 7}} \gBL{} N_{{j 5}} \Big)\Big)\Big)  \label{eq:vertexNNHR} \\
\Gamma^L_{\tilde{\chi}^0_{{i}}\tilde{\chi}^0_{{j}}h_{{k}}}  =  &
 - (\Gamma^R_{\tilde{\chi}^0_{{i}}\tilde{\chi}^0_{{i}}h_{{k}}})^* 
  \label{eq:vertexNNHL} 
\end{align}
Here, $Z^N$ is the $7\times 7$ neutralino mixing matrix and $Z^H$ the
$4\times 4$ rotation matrix of the scalars. To have a non-vanishing
coupling between a \BL neutralino and the bilepton component of
 the
Higgs, the neutralino must always be an admixture of \blino and
bileptino.
  As can be seen from \FIG~\ref{fig:neutralinoBLmatrix},
a \blino LSP has a sizeable bileptino fraction whereas the
\blino contribution to a bileptino LSP is rather small.
 Therefore, one expects the Higgs resonance to be more effective in 
 the case of a \blino LSP.

\begin{figure}[t]
\begin{picture}(0,0)
\put(40,-30){\scriptsize Kinetic mixing}
\put(350,-30){\scriptsize No kinetic mixing}
\end{picture}
 \begin{minipage}{0.99\linewidth}
\centering
\includegraphics[width=0.48\linewidth]{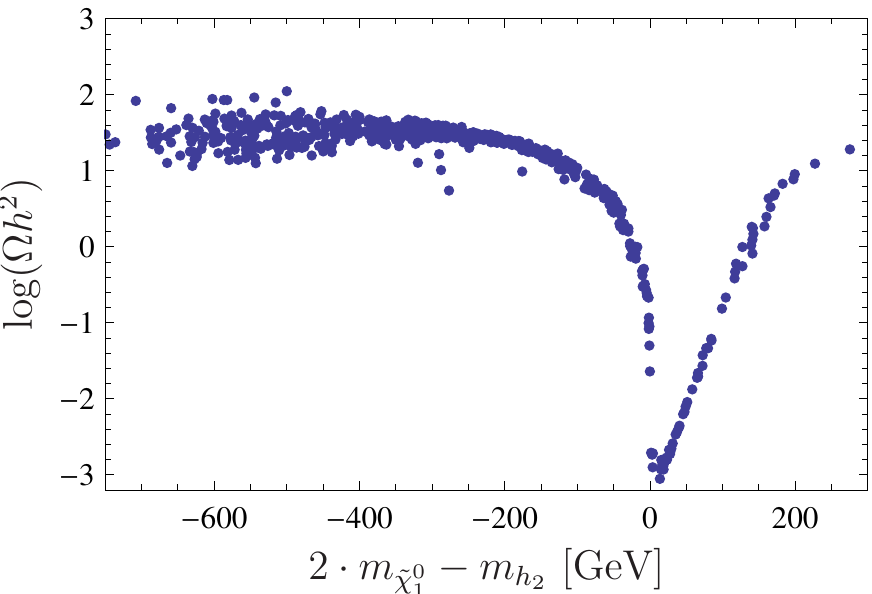} 
\hfill
\includegraphics[width=0.48\linewidth]{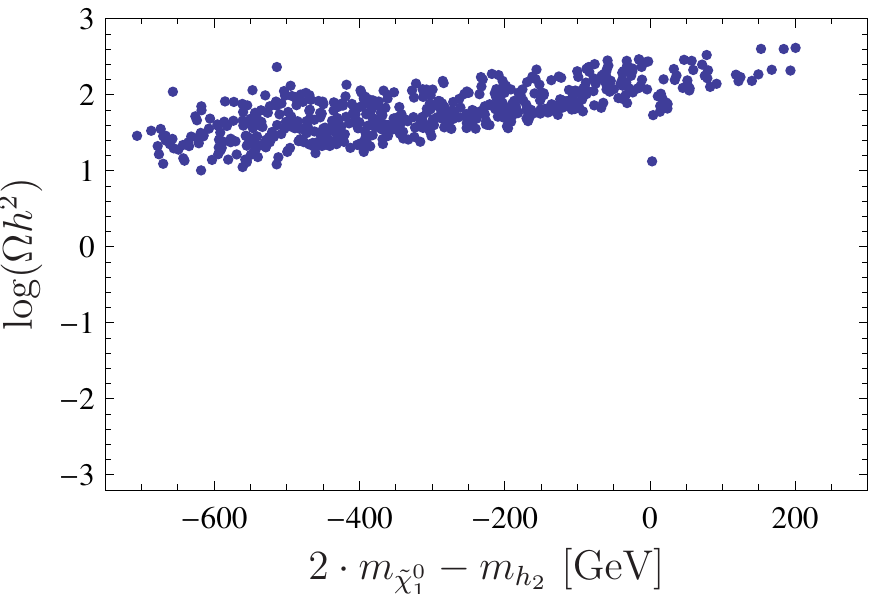}
 \end{minipage}
 \caption{\blino resonance with bilepton. The plots show $\log(\Omega
   h^2)$ as function of $2\cdot m_{\tilde{\chi}^0_1} - m_{h_2}$ in
   case of kinetic mixing (left) and without kinetic mixing (right).
   The parameter ranges were chosen to be $m_0 = [3.5,4.0]$~TeV,
   $M_{1/2} = [1.5,2.0]$~TeV, $\tan\beta = [30,40]$, $A_0 =
   [-5.5,-4]$~TeV, $\tan\beta' = [1.5,1.7]$, $\mZp = [2.0,2.5]$~TeV,
   $Y_x^{33} = [0.40,0.45]$, $Y_x^{11} = 0.42$, $Y_x^{22} = 0.377$.  }
\label{fig:blino_bileptonResonance}
\end{figure}
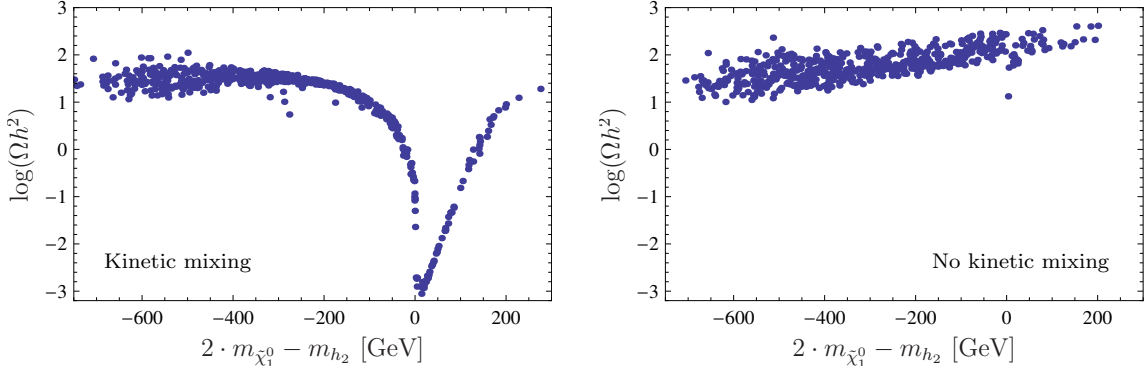
\paragraph{\blino} We first take a look  at the more obvious
 resonance
of a \blino with a bilepton Higgs. Because of the sizeable
 bileptino
fraction, the coupling between the LSP and the bilepton is large
enough to cause a pronounced resonance. This is shown in the left-hand
plot of \FIG~\ref{fig:blino_bileptonResonance}. If we pick a point with
a relic density of $\Omega h^2 = 0.116$, with  LSP mass
  of $m_{\tilde{\chi}^0_1} = 377.5$~GeV
 and bilepton mass $m_{h_2} =
844.5$~GeV, the main annihilation channels are
\begin{center}
\begin{tabular}{lc}
 $\tilde{\chi}^0_1 \tilde{\chi}^0_1 \, \to \, W^+ W^-$ & (48.5\%) \\ 
 $\tilde{\chi}^0_1 \tilde{\chi}^0_1 \, \to \, h_1 h_1$ & (24.7\%) \\
 $\tilde{\chi}^0_1 \tilde{\chi}^0_1 \, \to \, Z Z$ & (24.0\%) \\
 $\tilde{\chi}^0_1 \tilde{\chi}^0_1 \, \to \, t \bar{t}$ & (2.7\%) \\
\end{tabular}
\end{center}
For this point the MSSM-like Higgs mass is also
 in the preferred
range, as $m_{h_1} = 126.8$. Obviously, only SM/MSSM
 final states
appear and the \BL-specific states like right-handed neutrinos are not
important because their masses are too large. However, the particles
in the final states couple to the bilepton at tree level only through
 kinetic mixing. Therefore, if we switch off the
 kinetic mixing, the resonance also disappears, as shown on the right side of
\FIG~\ref{fig:blino_bileptonResonance}.

\begin{figure}[t]
 \begin{minipage}{0.99\linewidth}
\centering
\includegraphics[width=0.48\linewidth]{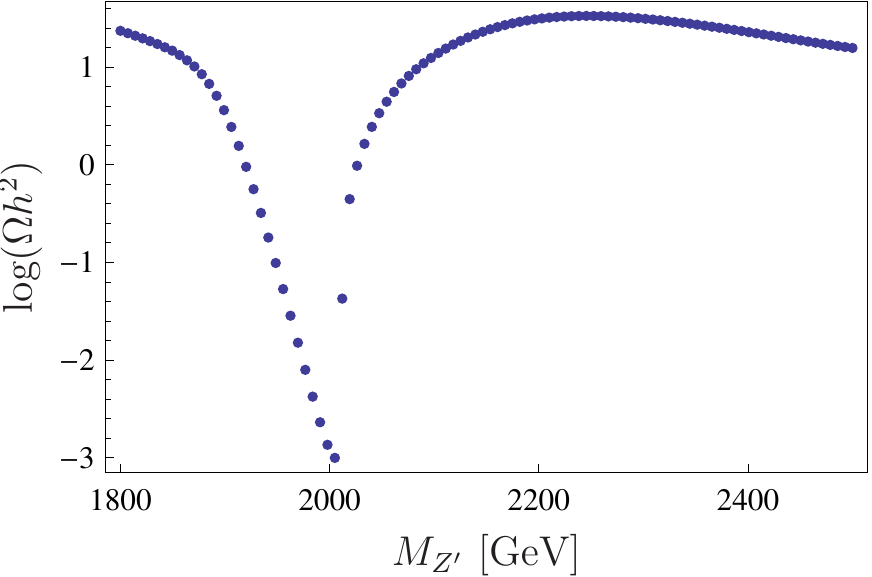}
\hfill
\includegraphics[width=0.48\linewidth]{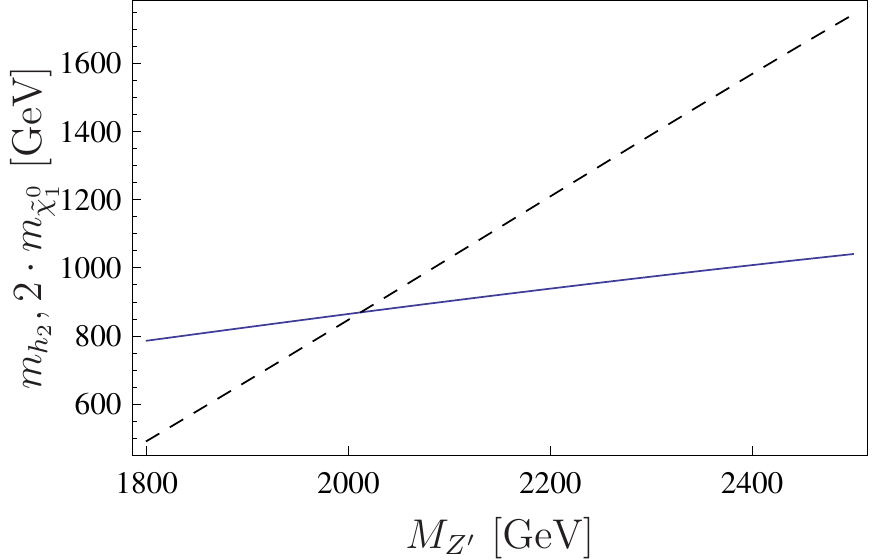} 
 \end{minipage}
 \caption{\blino resonance with bilepton due to a variation of
   $\mZp$. The left figures shows log($\Omega h^2$) as a function of
   $\mZp$, the right figure show the dependence of the mass of the
   light bilepton Higgs ($m_{h_2}$) (solid, blue line) and
   twice the mass of the LSP (dashed, black line).  }
\label{fig:blino_MZp}
\end{figure}

In general, it is often rather easy to find a resonance with a
bilepton Higgs and the \blino. The two masses scale differently with a
variation of $\mZp$ or $\tan(\beta')$ and they can be easily tuned.
This is shown in \FIG~\ref{fig:blino_MZp}. While the bilepton mass has
only a mild dependence on $\mZp$, the \blino mass scales much
 more with a variation of the $Z'$ mass. This strong
 dependence is
mostly given by the change of the bilepton fraction
because $\mu'$ is very sensitive to $\mZp$, see \EQ~(\ref{eq:tadpole_MZp}).

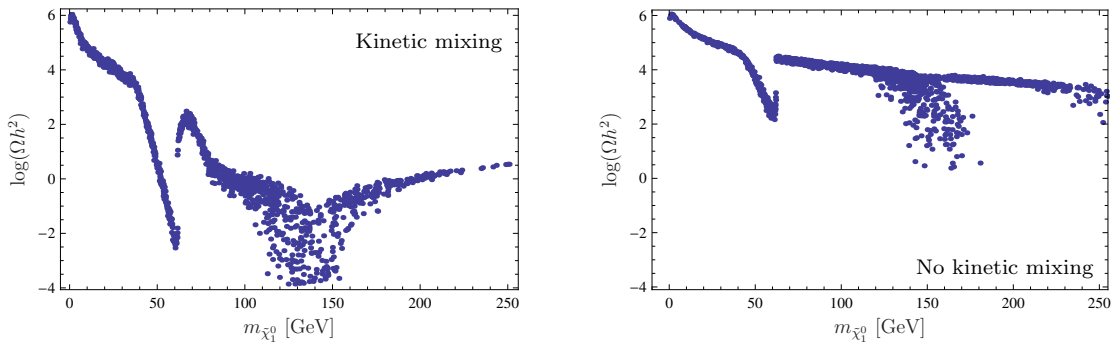
\begin{figure}[t]
\begin{picture}(0,0)
\put(140,50){\scriptsize Kinetic mixing}
\put(350,-35){\scriptsize No kinetic mixing}
\end{picture}
 \begin{minipage}{0.99\linewidth}
\centering
\includegraphics[width=0.48\linewidth]{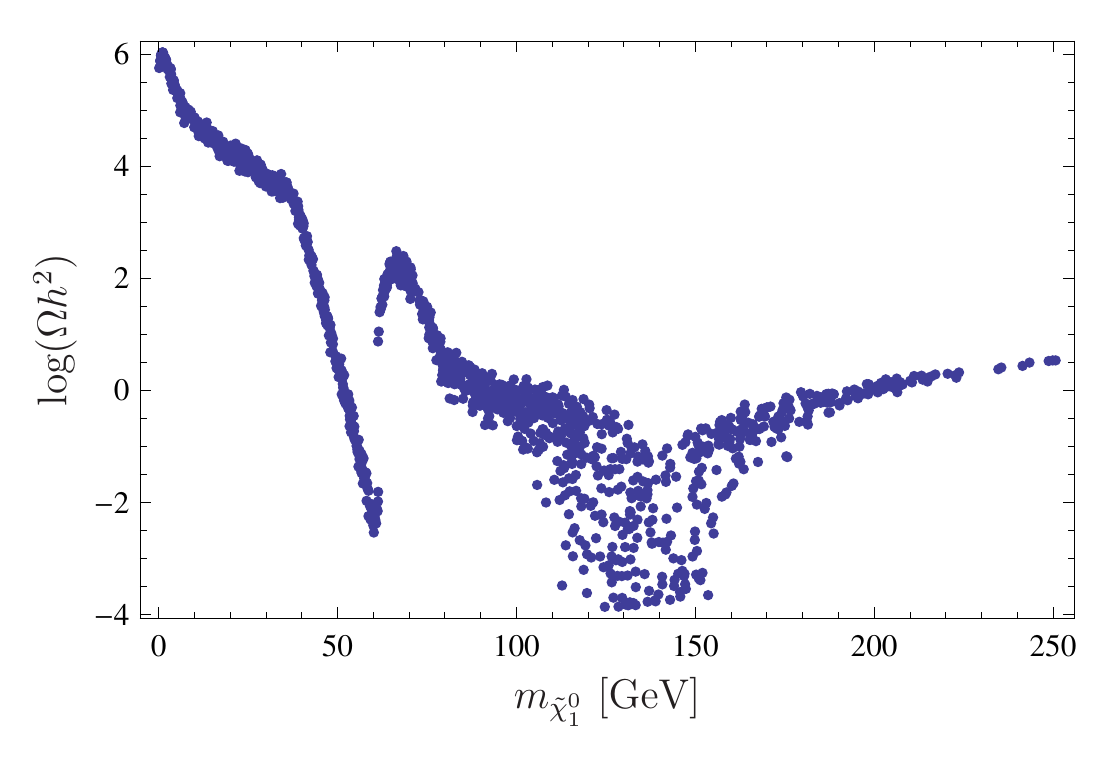} \hfill
\includegraphics[width=0.48\linewidth]{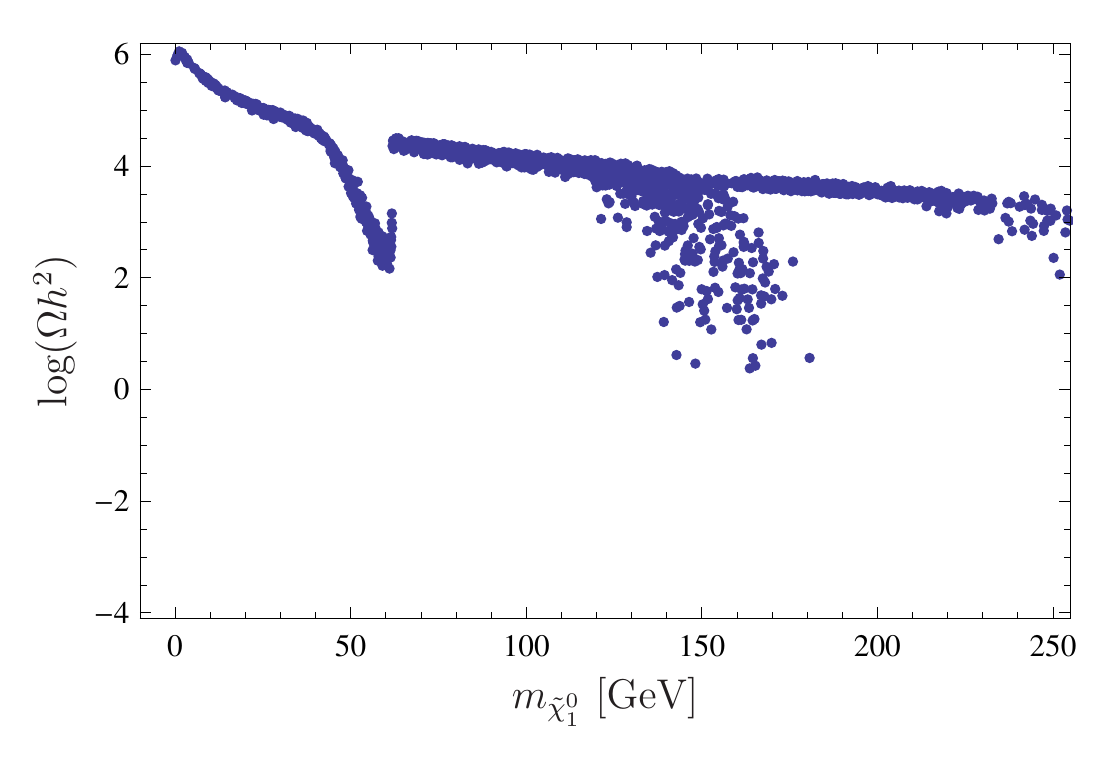}
 \end{minipage}
 \caption{\blino relic density log($\Omega h^2)$ vs. the mass of the
   LSP. The narrow dip around 60~GeV corresponds to resonance with the
   MSSM-like Higgs, the broader dip between 120 and 150~GeV is due to
   resonance with the light bilepton Higgs. The chosen parameter
   ranges were $m_0 = [4.0,5.0]$~TeV,
   $M_{1/2} = [1.8,2.0]$~TeV,
   $\tan\beta = [35,40]$, $A_0 = [4,5]$~TeV, $\tan\beta' =
   [1.19,1.21]$, $\mZp = [2.2,2.4]$~TeV, $Y_x^{33} = [0.40,0.45]$,
   $Y_x^{11} = 0.42$, $Y_x^{22} = 0.377$.  Left: with kinetic mixing,
   right without kinetic mixing.  }
\label{fig:blino_HiggsResonance}
\end{figure}

However, resonances of a \blino with a bilepton Higgs are not the only
possibility. Resonances with the light MSSM Higgs can appear
 too, especially for a light LSP, with a mass of $50 - 70$~GeV. As
already pointed out in \cite{O'Leary:2011yq}, one feature of this \BL
model is that we can have a rather light gaugino LSP with gaugino
unification at the GUT scale without being in conflict with the mass
limits of the chargino. The reason is that the mass of a \blino LSP is
not as strongly correlated to the mass of the wino-like chargino as
the bino, because of the rather strong \blino-bilepton mixing. This
mixing can cause a smaller mass than na\"{\i}vely expected from the
 running
value of the gaugino mass term $M_{B'}$.   Therefore, scanning over a 
broader parameter range one finds both Higgs
resonances in the dark matter abundance, as depicted in
\FIG~\ref{fig:blino_HiggsResonance}. Here the resonance with the
bilepton Higgs is broader than the one with the MSSM-like Higgs
because the bilepton mass has, of course, a larger dependence on
 the
variation of the $B-L$-specific parameters $\tan(\beta')$ and
 $\mZp$, and varies in the shown parameter range between 240 and 330~GeV, while
the MSSM-like Higgs mass lies between 122 and 125~GeV. The importance
of the final states is similar to the branching ratios of the SM-like
Higgs: $b \bar{b}$ (75.7\%), $\tau \bar{\tau}$ (18.3\%) and $c
\bar{c}$ (5.9\%), but \MO doesn't take loop-induced decays into two
gluons or two photons into account, leading to some theoretical
uncertainty.

The right plot in \FIG~\ref{fig:blino_HiggsResonance} shows that both
Higgs resonances can be visible even  without kinetic mixing, but the
reduction of the neutralino abundance is not large enough to explain
dark matter.

\begin{figure}[t]
 \begin{minipage}{0.99\linewidth}
\centering
\includegraphics[width=0.48\linewidth]{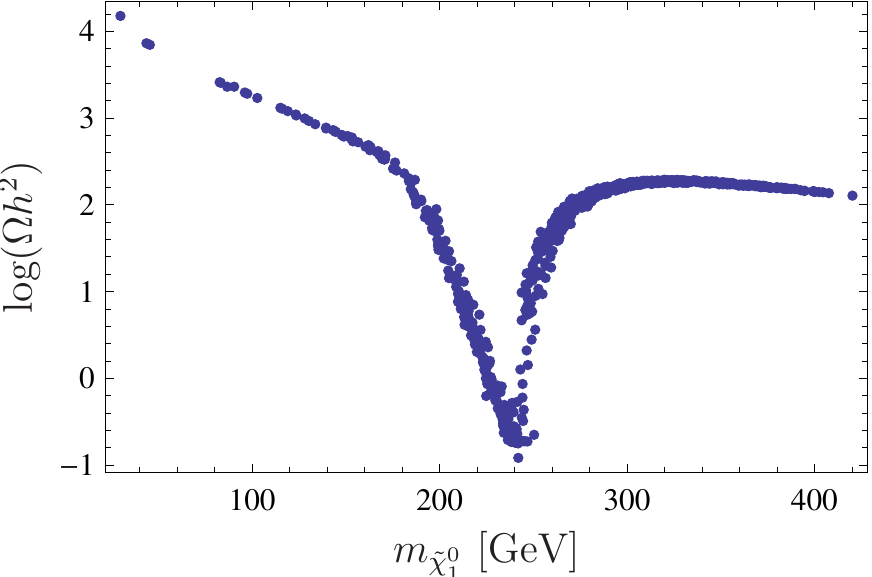} 
 \end{minipage}
 \caption{Bileptino resonance with a bilepton. The chosen
   parameter ranges were $m_0 = [1.45,1.5]$~TeV, $M_{1/2} =
   [1.08,1.1]$~TeV, $\tan\beta = [18,20]$, $A_0 = [-3.52,-3.5]$~TeV,
   $\tan\beta' = [1.52,1.53]$, $\mZp = [2.48,2.52]$~TeV, $Y_x^{33} =
   [0.41,0.42]$, $Y_x^{11} = Y_x^{22} = 0.42$.}
\label{fig:bileptinoBilepton}
\end{figure}

\paragraph{Bileptino} A bileptino LSP can also have a
resonance with a bilepton Higgs. However, there are some qualitative
and quantitative differences: the main difference is that often
  the relic density is not as  strongly  reduced as
for a \blino LSP. One has to be very close to the resonance point to
get $\Omega h^2$ below 0.12, see \FIG~\ref{fig:bileptinoBilepton}. The
reason is that the coupling of the bileptino to the bilepton is
 actually through its \blino component, which is typically below $1\%$.
 Another difficulty is that the masses of the bileptino and the
 bilepton are sensitive to
the same parameters. Therefore, it is much more difficult to tune the
parameters to get a resonance than for the \blino case: the parameter
ranges chosen for \FIG~\ref{fig:bileptinoBilepton} had to be much
smaller than those for instance used in
\FIG~\ref{fig:blino_bileptonResonance}. On the other hand, 
 the final states are very similar to those of a \blino-bileptino
 resonance and include SM gauge bosons and the light doublet Higgs.

\begin{figure}[t]
 \begin{minipage}{0.99\linewidth}
\centering
\includegraphics[width=0.48\linewidth]{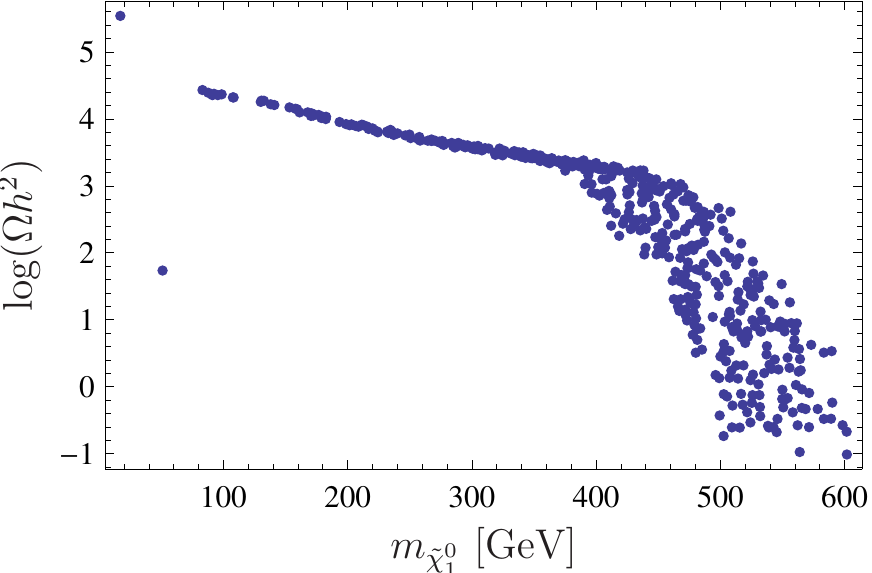}
\hfill
\includegraphics[width=0.48\linewidth]{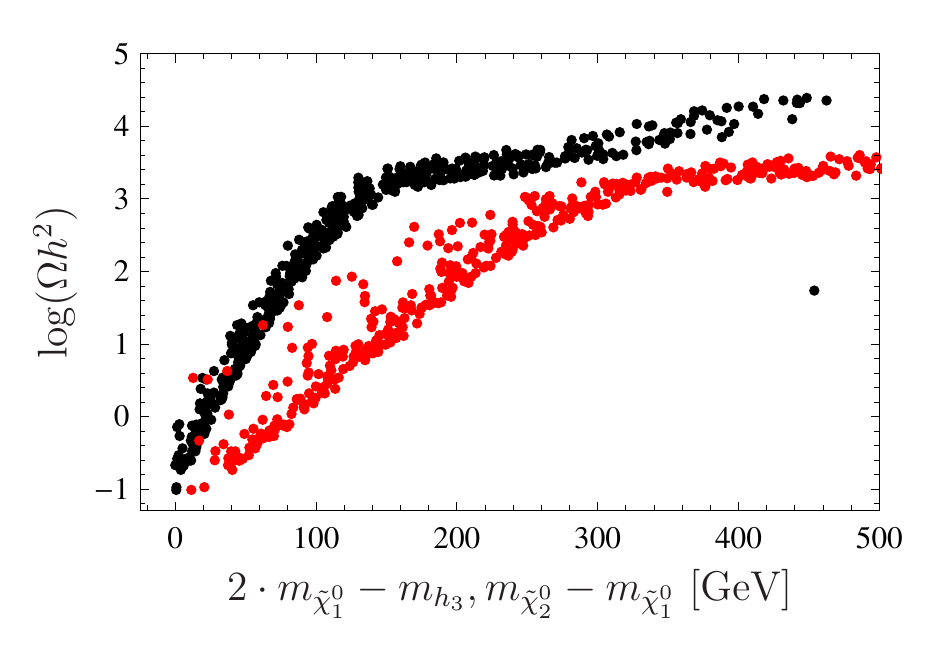} 
 \end{minipage}
 \caption{Indirect bileptino resonance with third Higgs which is
   MSSM-like: the $\tilde{\chi}^0_2$ annihilates resonantly with the
   doublet Higgs and due to the small mass splitting between
   $\tilde{\chi}^0_1$ and $\tilde{\chi}^0_2$ there is a
   co-annihilation of the LSP with the second neutralino. Left:
   $\log(\Omega h^2)$ vs. $m_{\tilde{\chi}^0_1}$, right: $\log(\Omega
   h^2)$ as function of $m_{h_3} - 2 m_{\tilde{\chi}^0_1}$ (black) and
   $m_{\tilde{\chi}^0_2} - m_{\tilde{\chi}^0_1}$ (red). The parameter
   ranges were
   $m_0 = [850,900]$~GeV, $M_{1/2} = [1.0,1.2]$~TeV,
   $\tan\beta = [38,41]$, $A_0 = [1.5,1.6]$~TeV, $\tan\beta' =
   [1.10,1.11]$, $\mZp = [2.7,2.8]$~TeV, $Y_x^{33} = [0.42,0.43]$,
   $Y_x^{11} = 0.38$, $Y_x^{22} = 0.42$. }
\label{fig:bileptino_resonance}
\end{figure}

Given that the bilepton
 resonance is generally unable to allow sufficient bileptino annihilation, one
 would expect that there would be even less scope for a sufficiently strong
 doublet Higgs resonance. The coupling between a
doublet Higgs and a bileptino is highly
 suppressed: firstly, by the
small \blino fraction of the bileptino;
 secondly, by the small bilepton
fraction of the doublet Higgs.  Nevertheless, an
 `indirect' doublet resonance can be important. If the
second-lightest neutralino is MSSM-like and very close in mass to the
LSP, and the mass of the Higgs fulfils $2 m_{\tilde{\chi}^0_1} \simeq
m_h \simeq 2 m_{{\tilde{\chi}}^{0}_{2}}$ there is, of course, also a
 resonance between $\tilde{\chi}^0_2$ and the doublet Higgs. Because of this
 resonance, the second neutralino annihilates very
 efficiently and the small mass difference
with the lightest neutralino leads to a co-annihilation between
 the neutralinos.  In \FIGS~\ref{fig:bileptino_resonance} we show an
example where this mass configuration appears and
 leads to the correct
relic density. As can be seen, the mass difference between the first
and second neutralinos as well as the distance to the resonance point
have to be very small to obtain a relic density of $\Omega h^2 = 0.1$.
Because of the large value of $\tan\beta$ for the shown parameter
range, the dominant final channels are (for $m_{\tilde{\chi}^0_1} =
563.9$~GeV, $m_{\tilde{\chi}^0_2} = 564.8$~GeV, $m_{h_3} =
1148.5$~GeV)
\begin{center}
\begin{tabular}{lc}
 $\tilde{\chi}^0_2 \tilde{\chi}^0_2 \, \to \, b \bar{b}$ & (78.8\%) \\ 
 $\tilde{\chi}^0_2 \tilde{\chi}^0_2 \, \to \, \tau \bar{\tau}$ & (20.8\%) \\
\end{tabular}
\end{center}
and involve only the second neutralino. The first channel involving
the LSP is $\tilde{\chi}^0_1 \tilde{\chi}^0_2 \to \tau \bar{\tau}$ and
contributes only 0.008\%.

\subsubsection{$Z'$ resonance}
\label{sec:ZpResonance}
The other possibility for a resonance is a neutral, massive gauge
boson, \IE either the $Z$ or $Z'$. However, it turns out
 that a resonance with the $Z$ boson is usually not important because the
coupling of the $Z$ boson to the neutralino is proportional to the
Higgsino fraction, which is usually negligible for a \BL neutralino.
Also the $B'$ contribution to the $Z$ boson due to kinetic mixing is too
constrained to cause a significant resonance effect.
 Therefore we can
concentrate here on the heavy $Z'$ boson. One general drawback of this
mechanism is that the LSP has to be very heavy because of bounds on the
$Z'$ mass. There is a lower limit of roughly 1.6~TeV for
 the mass of the $Z'$ in this model and our chosen setup \cite{Krauss:2012ku}.

\begin{figure}[t]
 \begin{minipage}{0.99\linewidth}
\centering
\includegraphics[width=0.48\linewidth]{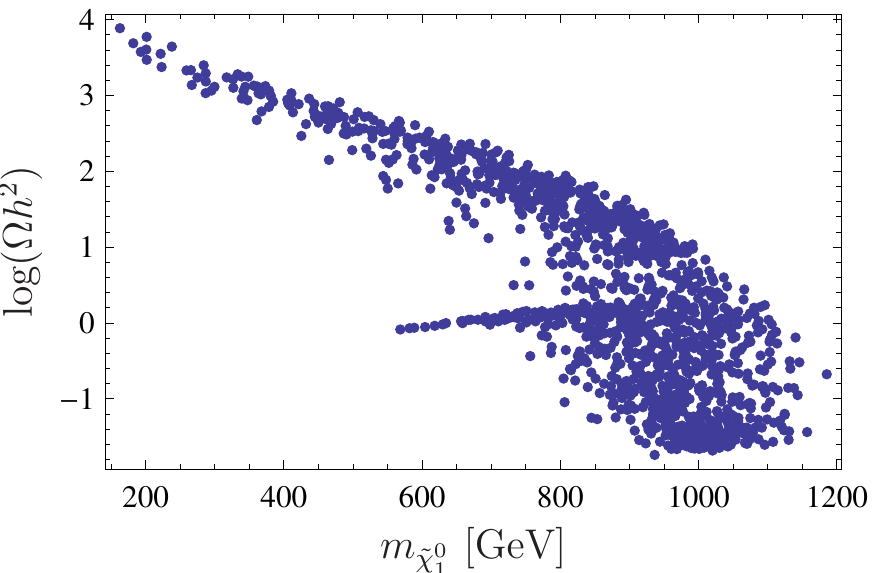}
\hfill
\includegraphics[width=0.48\linewidth]{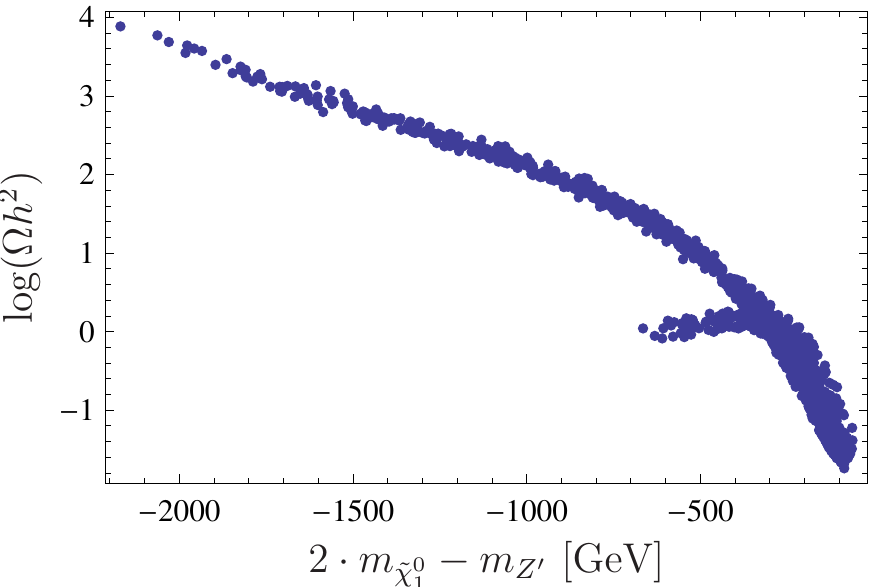} \\
\includegraphics[width=0.48\linewidth]{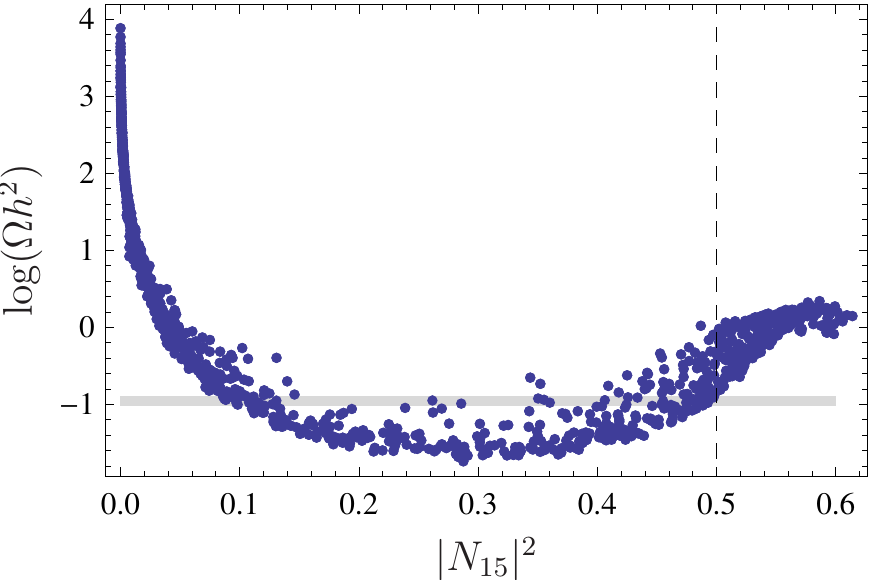}
\hfill
\includegraphics[width=0.48\linewidth]{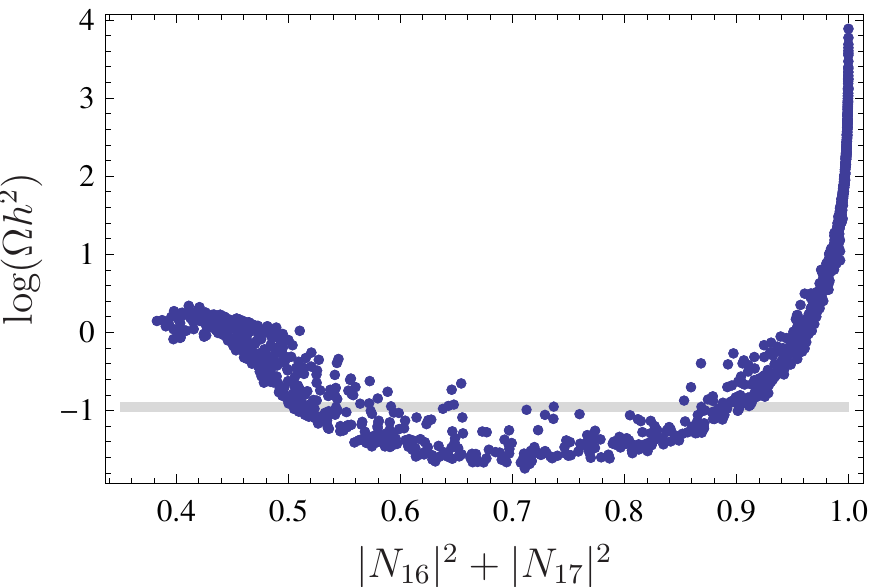} 
 \end{minipage}
 \caption{Bileptino resonance with $Z'$. Top Left: $\log(\Omega h^2)$
   vs. $m_{\tilde{\chi}^0_1}$. Top right: $\log(\Omega h^2)$ vs.
   $2\cdot m_{\tilde{\chi}^0_1} - \mZp$. The second row shows the
   dependence of the relic density on the \blino fraction (left) and
   the bileptino fraction (right). The parameter ranges
   were $m_0
   = [0.8,1.3]$~TeV, $M_{1/2} = [1.9,2.4]$~TeV, $\tan\beta = [10,25]$,
   $A_0 = [1.3,1.7]$~TeV, $\tan\beta' = [1.06,1.13]$, $\mZp =
   [1.7,2.5]$~TeV, $Y_x^{33} = [0.2,0.3]$, $Y_x^{11} = Y_x^{22} =
   0.42$.}
\label{fig:bileptino_Zp}
\end{figure}

The coupling between the $Z'$ boson and two neutralinos is given by
\begin{equation}
\gamma_\mu \left(\Gamma^L \frac{1-\gamma_5}{2}
 + \Gamma^R \frac{1+\gamma_5}{2}\right)
\end{equation}
with
\begin{align}
\Gamma^L_{\tilde{\chi}^0_{{i}}\tilde{\chi}^0_{{j}}{Z'}_{{\mu}}}  =  & \,
\frac{i}{2} \Big(N^*_{j 3} \Big(\Big(g_1 \sin\Theta_W
   + g_2 \cos\Theta_W  \Big)\sin{\Theta'}_W
   + \gmix \cos{\Theta'}_W  \Big)N_{{i 3}} \nonumber \\ 
 &- N^*_{j 4} \Big(g_1 \sin\Theta_W  \sin{\Theta'}_W
   + g_2 \cos\Theta_W  \sin{\Theta'}_W
   + \gmix \cos{\Theta'}_W  \Big)N_{{i 4}} \nonumber \\ 
 &+2 \gBL{} \cos{\Theta'}_W  \Big(N^*_{j 6} N_{{i 6}}
  - N^*_{j 7} N_{{i 7}} \Big)\Big) \label{eq:vertexNNZpL}\\ 
 \Gamma^R_{\tilde{\chi}^0_{{i}}\tilde{\chi}^0_{{j}}{Z'}_{{\mu}}}  =&\,
\left((\Gamma^L_{\tilde{\chi}^0_{{j}}\tilde{\chi}^0_{{i}}{Z'}_{{\mu}}} 
\right)^*
 \label{eq:vertexNNZpR}
\end{align}
Analogously to the MSSM where a non-vanishing Higgsino fraction is
needed for a coupling to the $Z$ boson, we need here a non-zero
bileptino fraction. Since the gauge couplings are fixed by the
 GUT condition, the only free parameters for the $Z'$ resonance are
$\mZp$ and the bileptino fraction, \IE a decreasing
 bileptino
fraction has to be compensated by a lower $Z'$ mass. Our result is
that to satisfy the bounds on the $\mZp$, the LSP has to be at least
45\% bileptino-like. Thus, a \blino LSP has to be strongly
 mixed with
the bileptino, and if we demand a $\mZp$ of at least
 2 TeV it is not
possible to find a LSP with the correct relic density which is more
\blino than bileptino, see \FIG~\ref{fig:bileptino_Zp}.

We see that over a large range of values of the
 bileptino fraction, the relic density is consistent with the limits on dark
 matter. Only if the
bileptino fraction drops below 50\% or gets larger than 95\%
 does the relic
density grow rapidly. The reason for the high neutralino abundance
 in the case of a very small \blino admixture is that the mass of the LSP
drops quickly with the increasing bileptino fraction for the
 given
parameter range: the upper mass limit for the LSP is roughly 1.0~TeV
(98\% bileptino), 0.8~TeV (99\% bileptino) and
 0.5~TeV (100\%
bileptino). Hence, the distance to the resonant point gets too
 large.

In case of a dominant $Z'$ resonance, the annihilation channels are
just the same as the branching ratios of the $Z'$ boson and given by
\begin{center}
\begin{tabular}{lcclc}
 $\tilde{\chi}^0_1 \tilde{\chi}^0_1 \, \to \, \sum_i \ell_i \bar{\ell}_i $
 & (36.1\%) & \hspace{1cm} &
 $\tilde{\chi}^0_1 \tilde{\chi}^0_1 \, \to \,  \sum_i d_i \bar{d}_i $
 & (29.6\%) \\
 $\tilde{\chi}^0_1 \tilde{\chi}^0_1 \, \to \, \sum_i \nu_i \nu_i $
 & (23.0\%) & \hspace{1cm} &
$\tilde{\chi}^0_1 \tilde{\chi}^0_1 \, \to \, \sum_i u_i \bar{u}_i $
 & (8.4\%) \\
$\tilde{\chi}^0_1 \tilde{\chi}^0_1 \, \to \, h_1 h_1  $
 & (1.5\%) & \hspace{1cm} & \\
\end{tabular}
\end{center}
where we have summed over the different generations of charged leptons
($\ell_i$), down-type quarks ($d_i$), up-type quarks ($u_i$) and light
neutrinos ($\nu_i$), $i=1,2,3$. This scenario is very appealing
because it is very predictive due to the small number of important
parameters. However, due to the sizable couplings to quarks, this
mechanism is also under some pressure from direct detection
experiments, as discussed in \SEC~\ref{sec:DD}.

\subsubsection{Sneutrino co-annihilation}
As we have seen in \SEC~\ref{sec:CPevenDM} - \ref{sec:CPoddDM}, the
 \rsnus annihilate very  efficiently due to the large
F- and D-term couplings with the bileptinos.
 Not only does  this allow a low abundance for a sneutrino LSP, but
 it also allows a sneutrino NLSP to have an important effect on the
 dark matter
relic density: if the neutralino LSP is close in mass to the sneutrino
NLSP, a new co-annihilation comes into play.

The important ingredients to understand the behaviour of 
 sneutrino
co-annihilation are the couplings between the sneutrinos and the Higgs
fields given in \EQS~(\ref{eq:vertexSvSvHH}) - (\ref{eq:vertexSvSvH})
as well as the interaction of the neutralino with a neutrino/sneutrino
pair. Ignoring the contributions of the tiny neutrino Yukawa
couplings, we can write this chiral coupling using \EQ~(\ref{eq:FFS})
as
\begin{align}
\Gamma^L_{\tilde{\chi}^0_i \nu_j \tilde{\nu}^{{S,P}}_k} =
 & \frac{i}{2} \Big(\Big(g_1 N^*_{i 1}  - g_2 N^*_{i 2}
  + \Big(\gmix
 + \gBL{}\Big)N^*_{i 5} \Big)\sum_{a=1}^{3}U^{V,*}_{j a} Z^{X}_{k a}
 \nonumber \\
 & \hspace{1cm}   \mp \gBL{} N^*_{i 5}
 \sum_{a=1}^{3}U^{V,*}_{j 3 + a} Z^{X}_{k 3 + a}
  -2 \sqrt{2} N^*_{i 6} \sum_{b=1}^{3}U^{V,*}_{j 3 + b} Z^{X}_{k 3 + b}
 Y_{x,{b b}}  \Big) \\
\Gamma^R_{\tilde{\chi}^0_i \nu_j \tilde{\nu}^{{S,P}}_k} =
 & \frac{i}{2} \Big(- \gBL{}
 \sum_{a=1}^{3}Z^{X}_{k 3 + a} U_{{j 3 + a}}^{V}  N_{{i 5}}
  \pm \sum_{a=1}^{3}Z^{X}_{k a} U_{{j a}}^{V}  \Big(g_1 N_{{i 1}}
  - g_2 N_{{i 2}}  + \Big(\gmix
 + \gBL{}\Big)N_{{i 5}} \Big)\nonumber \\ 
 & \hspace{1cm} \mp 2 \sqrt{2}
 \sum_{b=1}^{3}Y^*_{x,{b b}} Z^{X}_{k 3 + b} U_{{j 3 + b}}^{V}
  N_{{i 6}} \Big)
\end{align}
with $X = {S,P}$ for the rotation matrices of the CP-even
 or -odd sneutrinos and the upper signs correspond to the CP-even and the lower
ones to the CP-odd interactions.

\begin{figure}[t]
 \begin{minipage}{0.99\linewidth}
\centering
\includegraphics[width=0.48\linewidth]{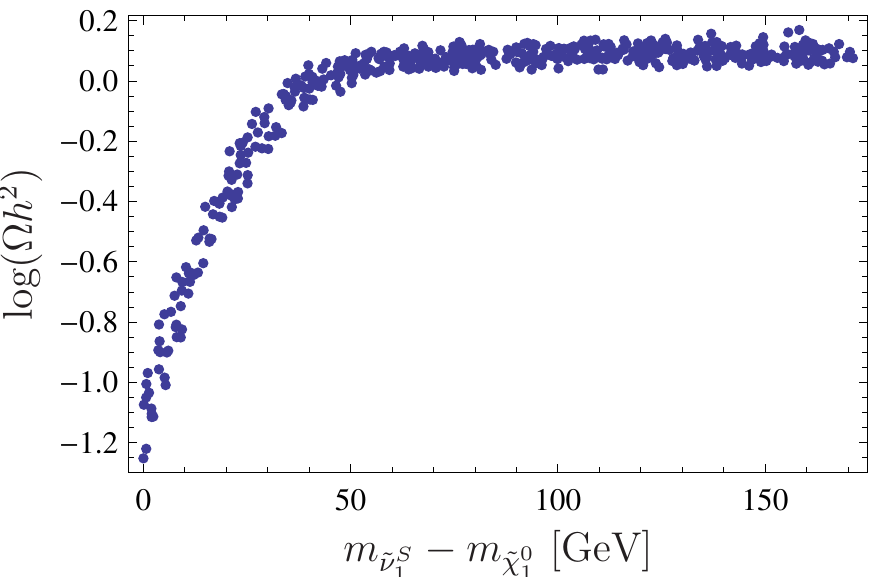}
\hfill
\includegraphics[width=0.48\linewidth]{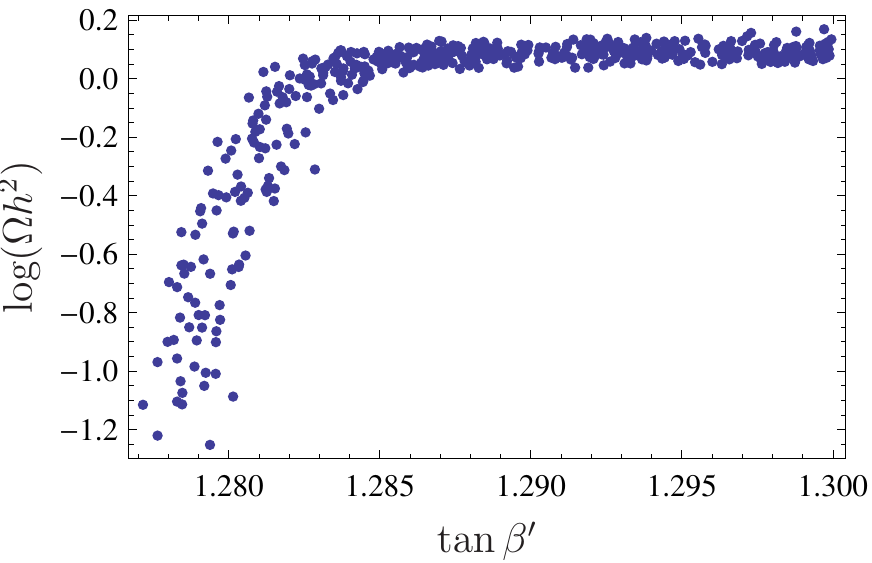} 
 \end{minipage}
 \caption{\blino LSP and co-annihilation with a scalar sneutrino. The
   left plot shows the dependence of the relic density $\log(\Omega
   h^2)$ on the mass splitting between the lightest neutralino and the
   lightest scalar sneutrino, the right plot gives $\log(\Omega h^2)$
   as function $\tan\beta'$. The mass of the LSP is between 670 and
   780~GeV, while the co-annihilation can happen in the range
   between 670 and 720~GeV. The other parameters were
   $m_0 =
   [2.75,2.8]$~TeV, $M_{1/2} = [1.7,1.75]$~TeV, $\tan\beta = 29$, $A_0
   = -1.45$~TeV, $\tan\beta' = [1.26,1.30]$, $\mZp =
   [2.25,2.3]$~TeV, $Y_x^{33} = [0.37,0.38]$, $Y_x^{11} = 0.42$,
   $Y_x^{22} = 0.45$ }
\label{fig:blino_CPevenCoann}
\end{figure}

\paragraph{\blino} Co-annihilation of a \blino LSP with both CP-even
 and CP-odd sneutrinos is possible. However,
 the co-annihilation
with a CP-even sneutrino is more likely because
 the larger values of $Y_x$
 usually required to obtain a \blino LSP  also
 prefer a light CP-even sneutrino. An example of
 co-annihilation between the lightest
scalar sneutrino and a \blino LSP is given in
\FIG~\ref{fig:blino_CPevenCoann}. Since both the \blino
 and the sneutrino masses are very sensitive to
 $\tan\beta'$, small changes in that parameter can be sufficient to obtain 
 an efficient co-annihilation:
for the shown range of $\tan\beta' = [1.285,1.300]$ in the right-hand
plot of \FIG~\ref{fig:blino_CPevenCoann}, the \blino mass covers the
range $[670,780]$~GeV and the sneutrino mass lies between
$[680,950]$~GeV.  In general, the mass difference between LSP and NLSP
has to be around 3~GeV or smaller. This difference is smaller than the
one known from the MSSM where  stau or stop co-annihilation
 works
 even with mass differences of  more than 15~GeV.  The reason is that the
 equivalent channels to the main co-annihilation channels in the MSSM, like
$\tilde{\chi}^0_1 \tilde{e}_1 \to Z \tau$, are $\tilde{\chi}^0_1
\ssnu{1} \to Z' \nu_4$. 
However, these are all kinematically
forbidden, and this has to be compensated by a smaller mass
difference. For a chosen point with $m_{\tilde{\chi}^0_1} =
726.9$~GeV, $m_{\ssnu{1}} = 727.9$~GeV and $\Omega h^2 =
0.107$, the most important (co-)annihilation channels are
\begin{center}
\begin{tabular}{lc}
 $\ssnu{1} \ssnu{1} \, \to \, h_2 h_2$ & (96.7\%) \\ 
 $\tilde{\chi}^0_1 \tilde{\chi}^0_1 \, \to \, h_2 h_2$ & (3.0\%) \\
\end{tabular}
\end{center}

To get the CP-odd sneutrino sufficiently light, at least one diagonal
entry of $Y_x$ has to be small. This has to be compensated
 by large values of $A_0$ to keep the \blino character of the LSP. For instance,
one point for which the CP-odd co-annihilation is 
 important is given by
\begin{center}
$m_0 = 1890$~GeV , \,
$M_{1/2} = 1736$~GeV , \,
$\tan\beta = 25$ , \,
$\mu > 0$, \,
$A_0 = 6345$~GeV , \,
$\tan\beta' = 1.277$ , \,
$\mu' > 0$, \,
$\mZp = 1695$~GeV  , \\
$Y_x^{11} = 0.42$ , \,
$Y_x^{22} = 0.40$ , \,
$Y_x^{33} = 0.02$ \, . 
\end{center}
The mass of the \blino LSP is $m_{\tilde{\chi}^0_1} = 239.5$~GeV, the
light CP-even sneutrino has a mass of $m_{\psnu{1}} =
247.5$~GeV and the relic density of this point is calculated to be
$\Omega h^2 = 0.108$. In contrast to the examples of scalar sneutrino
co-annihilation, the neutralino itself already has a
 sizable cross
section into two bileptons due to its light mass. This effect is
discussed in more detail in \SEC~\ref{sec:BLino_t_channel}. Therefore,
the co-annihilation channels contribute only a bit more than one third
 of the total annihilation:
\begin{center}
\begin{tabular}{lc}
 $\tilde{\chi}^0_1 \tilde{\chi}^0_1 \, \to \, h_2 h_2$ & (58.0\%) \\
 $\psnu{1} \psnu{1} \, \to \, h_2 h_2$ & (35.1\%) \\ 
 $\tilde{\chi}^0_1 \tilde{\chi}^0_1 \, \to \, h_1 h_2$ & (3.5\%) \\
$\psnu{1} \psnu{1} \, \to \, h_1 h_2$ & (1.4\%) \\ 
\end{tabular}
\end{center} 
In addition, the lightest right-handed neutrino is also lighter than
the LSP because of the small $Y_x^{33}$. Therefore, the mass splitting
between the LSP and NLSP can be larger than in the case of a
 CP-even sneutrino NLSP.

\begin{figure}[t]
 \begin{minipage}{0.99\linewidth}
\centering
\includegraphics[width=0.48\linewidth]{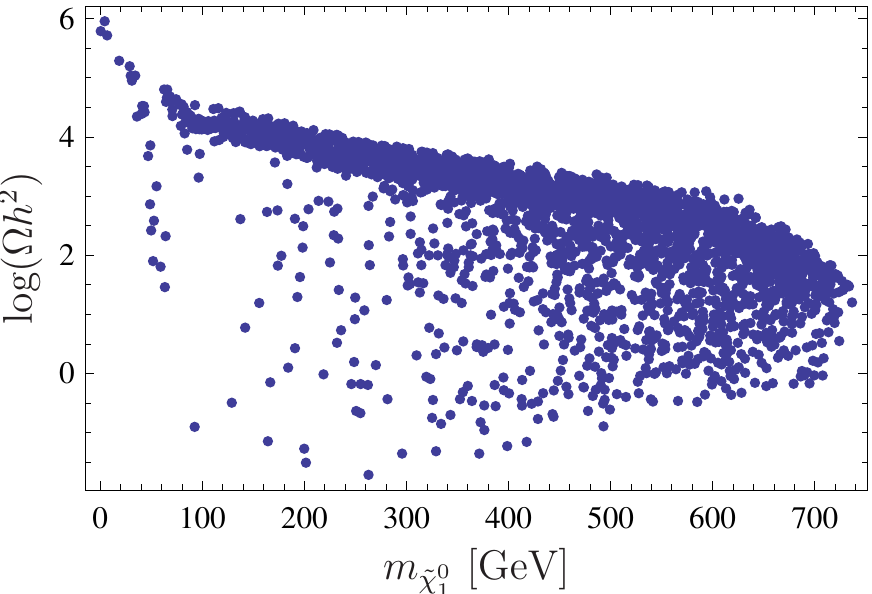}
\hfill
\includegraphics[width=0.48\linewidth]{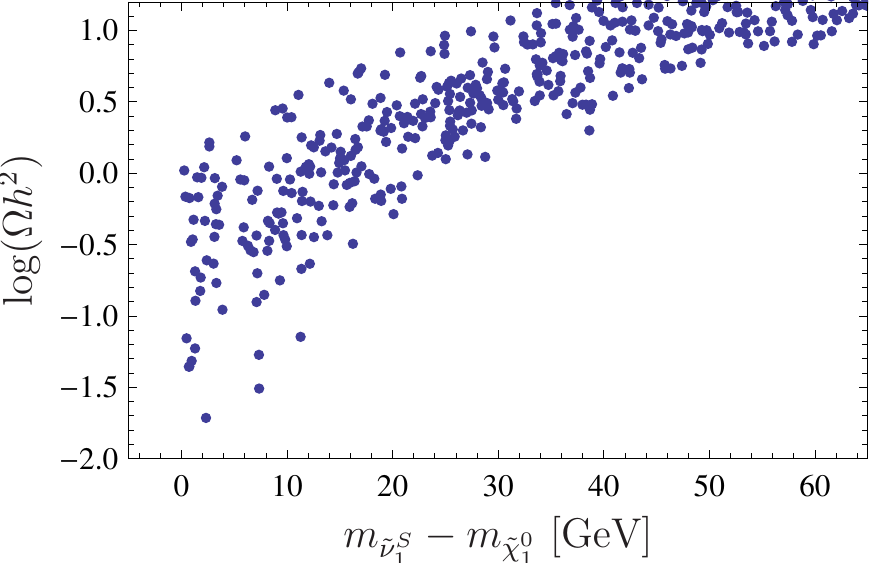}
 \end{minipage}
 \caption{Bileptino co-annihilation with
 pseudoscalar sneutrino.
   Left: $\log(\Omega h^2)$ vs. $m_{\tilde{\chi}^0_1}$. Right:
   $\log(\Omega h^2)$ vs.  $m_{\psnu{1}} -
   m_{\tilde{\chi}^0_1}$. Parameter ranges: $m_0 = [780,900]$~GeV,
   $M_{1/2} = [1.30,1.45]$~TeV, $\tan\beta = [6,12]$, $A_0 =
   [2.2,2.8]$~TeV, $\tan\beta' = [1.10,1.14]$, $\mZp =
   [2.5,3.2]$~TeV, $Y_x^{33} = [0.13,0.22]$, $Y_x^{11} = 0.42$.  }
\label{fig:bileptino_CPodd_coann}
\end{figure}

\paragraph{Bileptino} In contrast to the \blino, the bileptino
 prefers a co-annihilation with the CP-odd sneutrino to 
 with a CP-even
sneutrino: both states are lighter if the entries of $Y_x$ are not too
large and the mass degeneracy is easier to obtain.  In addition, the
example given in \FIG~\ref{fig:bileptino_CPodd_coann} shows that the
co-annihilation is possible over a wide range of the
 bileptino mass.

For the chosen parameters, the lightest scalar is a
 bilepton and the
most important annihilation channel is $\psnu{1}
\psnu{1} \, \to \, h_1 h_1 $. In addition, the small value of
$Y_x^{33}$ leads also to a right-handed neutrino which is lighter than
the LSP. Therefore, also co-annihilations of the form $\psnu{1}
\tilde{\chi}^0_1 \, \to \, h_i \nu_4 $ are sizable and the mass
splitting between LSP and NLSP can be of order
$\mathscr{O}(10~\text{GeV})$. For a chosen point with the correct
$\Omega h^2$ of 0.115, the channels which contribute more than 1\% to
the total annihilation read
\begin{center}
\begin{tabular}{lc}
$\psnu{1} \psnu{1} \, \to \, h_1 h_1 $ & (57.0\%) \\
$\psnu{1} \tilde{\chi}^0_1 \, \to \, h_1 \nu_4 $ & (22.1\%) \\
$\psnu{1} \psnu{1} \, \to \, h_1 h_2 $ & (14.1\%) \\
$\psnu{1} \tilde{\chi}^0_1 \, \to \, h_2 \nu_4 $ & (3.3\%) \\
$\psnu{1} \psnu{1} \, \to \, h_2 h_2 $ & (2.5\%) \\
\end{tabular}
\end{center}

\begin{figure}[!ht]
 \begin{minipage}{0.99\linewidth}
\centering
\includegraphics[width=0.48\linewidth
]{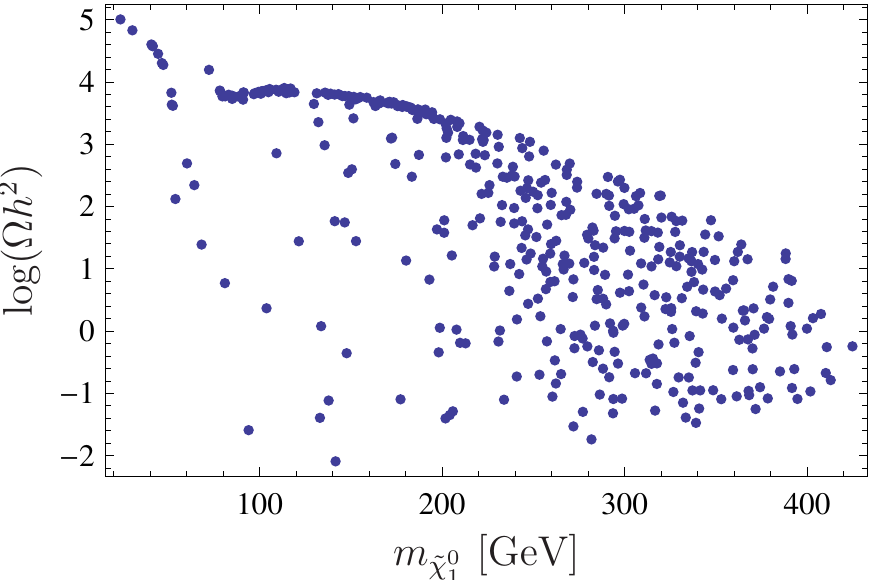}
\hfill
\includegraphics[width=0.48\linewidth
]{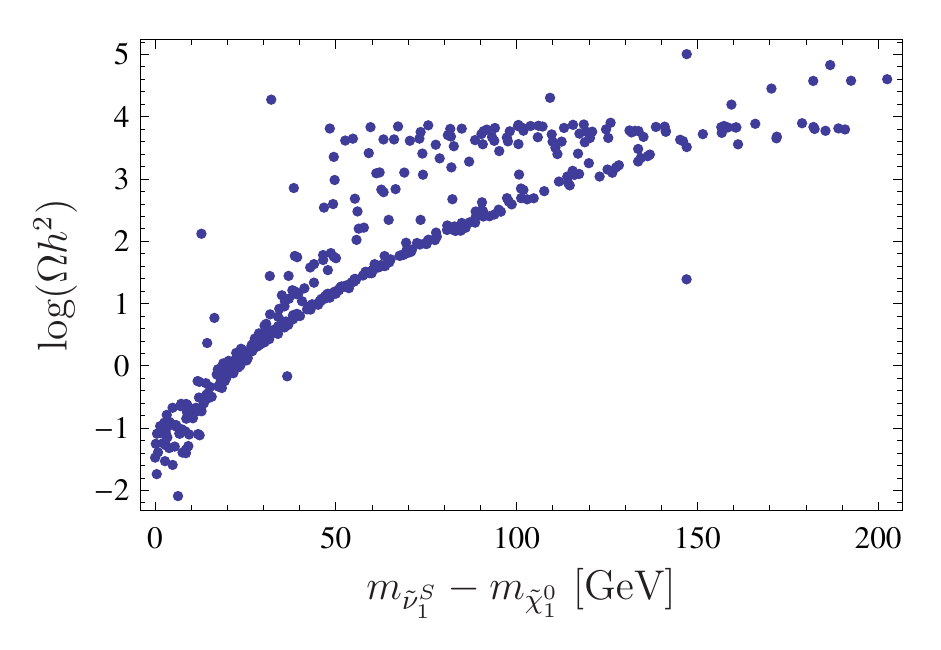} 
 \end{minipage}
 \caption{Bileptino co-annihilation with scalar sneutrino. Left: $\log(\Omega
   h^2)$ vs. $m_{\tilde{\chi}^0_1}$. Right: $\log(\Omega h^2)$ vs.
   $m_{\ssnu{1}} - m_{\tilde{\chi}^0_1}$. Parameter ranges:
   $m_0 = [790,800]$~GeV, $M_{1/2} = [1.50,1.55]$~TeV, $\tan\beta
   =17$, $A_0 = [2.70,2.75]$~TeV, $\tan\beta' = [1.14,1.15]$, $\mZp
   = [2.40,2.45]$~TeV, $Y_x^{33} = [-0.10,-0.08]$, $Y_x^{11} =
   Y_x^{22} = 0.42$.  }
\label{fig:bileptino_CPeven_coann}
\end{figure}

To find parameter points where a co-annihilation between the
 bileptino
and the CP-even sneutrino is realised it was necessary to choose
small, negative values for $Y_x^{33}$ and a positive $A_0$ of a few
TeV. This scenario is depicted in
\FIG~\ref{fig:bileptino_CPeven_coann}. We can see again that in
 a
small parameter region the LSP mass can vary between 100 and 450~GeV,
but it is always possible to get the correct relic density due to a
 CP-even sneutrino which is nearly degenerate in mass.

Despite the fact that the co-annihilation with scalar sneutrinos seems
 to be less generic than  with pseudoscalar
 sneutrinos, the efficiency is even better than for the example shown in
\FIG~\ref{fig:blino_CPevenCoann}:  a mass splitting between the
LSP and NLSP of only 10~GeV is needed to get the correct abundance. The
final states are comparable with the ones for the
 pseudoscalar
sneutrino since a light right-handed neutrino is present. For a chosen
point, the light bilepton is the second scalar Higgs and the
 channels read:
\begin{center}
\begin{tabular}{lc}
$\ssnu{1} \ssnu{1} \, \to \, h_2 h_2 $ & (66.4\%) \\
$\ssnu{1} \ssnu{1} \, \to \, h_1 h_2 $ & (20.1\%) \\
$\ssnu{1} \tilde{\chi}^0_1 \, \to \, h_2 \nu_4 $ & (7.1\%) \\
$\ssnu{1} \ssnu{1} \, \to \, h_1 h_1 $ & (3.8\%) \\
$\ssnu{1} \tilde{\chi}^0_1 \, \to \, h_1 \nu_4 $ & (1.2\%) \\
\end{tabular}
\end{center}

\subsubsection{Stop and stau co-annihilation}
Co-annihilation scenarios for \BL dark matter
 is not restricted only to other \BL states like the sneutrinos but may also occur
 with MSSM states.
For the \blino and for the bileptino the relic density can
 be strongly reduced if a charged or colored sfermion like a light stau
 or stop
is close enough in mass. The interaction of a neutralino LSP with
 a
light stau is given by
\begin{align}
\Gamma^L_{\tilde{\chi}^0_1 \tau \tilde{e}^*_1} =
 &  \frac{i}{2} \Big(-2 N^*_{1 3} Y_{e,{3 3}} Z_{{1 6}}^{E}
  + \sqrt{2} \Big(g_1 N^*_{1 1}  + g_2 N^*_{1 2}
  + \Big(\gmix
 + \gBL{}\Big)N^*_{1 5} \Big)Z_{{1 3}}^{E} \Big) \\
\Gamma^R_{\tilde{\chi}^0_1 \tau \tilde{e}^*_1} =
 &  -\frac{i}{2} \Big(2 Y^*_{e,{3 3}} Z_{{1 3}}^{E} N_{{1 3}}
  + \sqrt{2} Z_{{1 6}}^{E} \Big(2 g_1 N_{{1 1}}  + \Big(2 \gmix
  + \gBL{}\Big)N_{{1 5}} \Big)\Big)
\end{align}
where $Z^E$ is the $6 \times 6$ mixing matrix of the charged sleptons,
while the coupling to a squark/quark pair reads
\begin{align}
\label{eq:neutqsqL}
\Gamma^L_{\tilde{\chi}^0_1 t \tilde{u}^*_{1}} =
 & -\frac{i}{6} \Big(6 N^*_{1 4}
 \sum_{b=1}^{3}U^{u,*}_{L,{3 b}} \sum_{a=1}^{3}Y_{u,{a b}} Z_{{1 3 + a}}^{U}
 \nonumber \\
 &   + \sqrt{2} \Big(3 g_2 N^*_{1 2}  + g_1 N^*_{1 1}  + \Big(\gmix
 + \gBL{}\Big)N^*_{1 5} \Big)
 \sum_{a=1}^{3}U^{u,*}_{L,{3 a}} Z_{{1 a}}^{U}  \Big) \\
\Gamma^R_{\tilde{\chi}^0_1 t \tilde{u}^*_{1}} =
 &  \frac{i}{6}  \Big(-6
 \sum_{b=1}^{3}\sum_{a=1}^{3}Y^*_{u,{a b}} U_{R,{3 a}}^{u}  Z_{{1 b}}^{U}
  N_{{1 4}} \nonumber \\ & + \sqrt{2} \sum_{a=1}^{3}Z_{{1 3 + a}}^{U}
 U_{R,{3 a}}^{u}  \Big(4 g_1 N_{{1 1}}  + \Big(4 \gmix
  + \gBL{}\Big)N_{{1 5}} \Big)\Big)
\label{eq:neutqsqR}  
\end{align}
In both cases there is no tree-level coupling between a pure
 bileptino
and the MSSM particle, while the \blino couples proportionally to the
\BL charge. This may lead one to 
 think that co-annihilation scenarios with a MSSM particle work better for a
\blino than for a bileptino. However, this is not necessarily
 the case,
as we will discuss now. A general drawback of 
 stop co-annihilation
is that at least one stop has to be rather light and the loop
correction to the light MSSM Higgs gets reduced.
 Since we did not find a parameter point
 with stop co-annihilation and a Higgs mass above 115~GeV
 for either a \blino or bileptino LSP, we concentrate in the
following on  stau co-annihilation. For the shown
 examples, the mass
of the light doublet Higgs lies in the preferred range of 123-127~GeV.

\begin{figure}[hbt]
 \begin{minipage}{0.99\linewidth}
\centering
\includegraphics[width=0.48\linewidth]{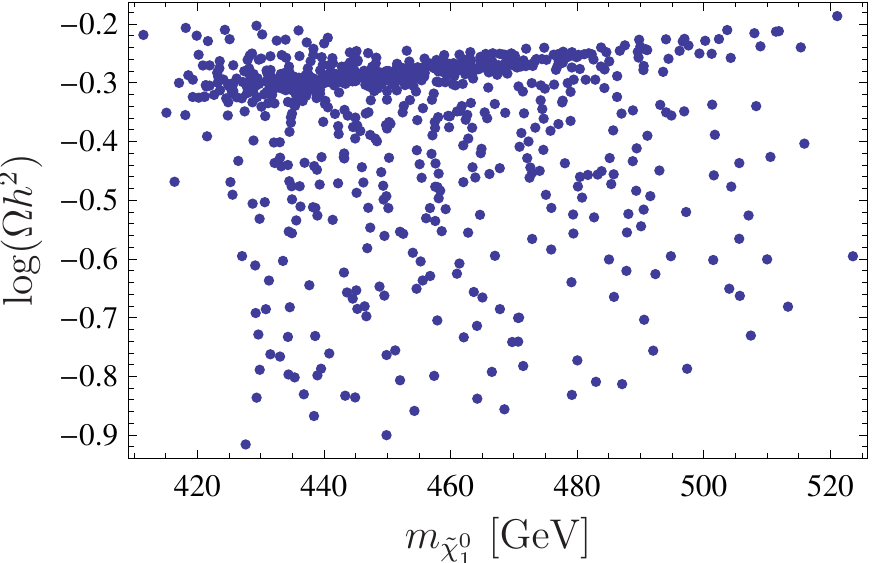}
\hfill
\includegraphics[width=0.48\linewidth]{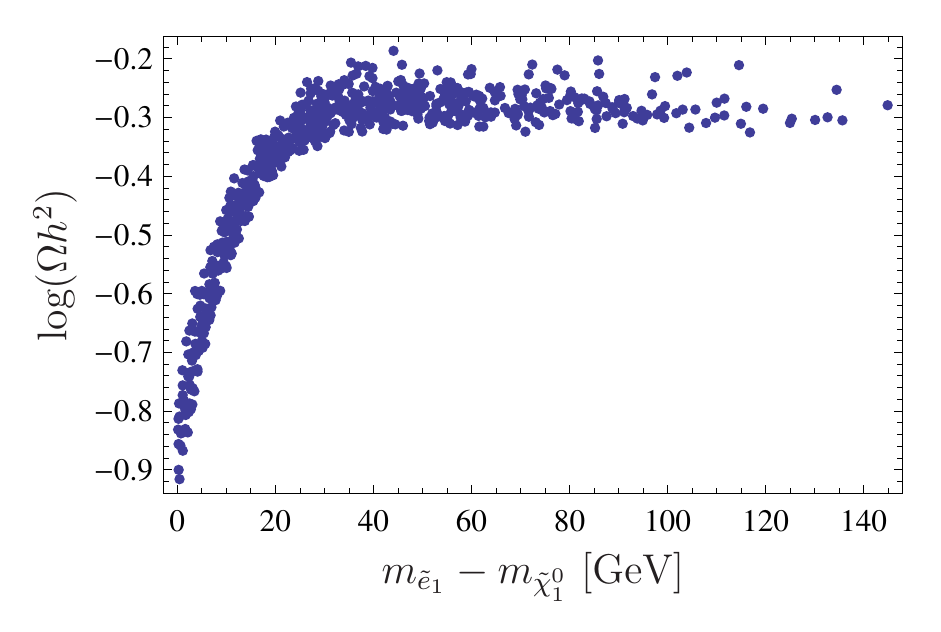} 
 \end{minipage}
 \caption{\blino co-annihilation with stau. Left: $\log(\Omega h^2)$ vs.
   $m_{\tilde{\chi}^0_1}$. Right: $\log(\Omega h^2)$ vs.
   $m_{\tilde{e}_1} - m_{\tilde{\chi}^0_1}$.
   Parameter ranges: $m_0 = [2150,2200]$~GeV,
   $M_{1/2} = [1700,1750]$~GeV, $\tan\beta = [50,52]$, $A_0 =
   [5550,5650]$~GeV, $\tan\beta' = [1.31,1.33]$, $\mZp =
   [2.0,2.1]$~TeV, $Y_x^{33} = [0.42,0.43]$, $Y_x^{11} =  Y_x^{22} =  0.42$.  }
\label{fig:BlinoStauCoann}
\end{figure}

\paragraph{\blino} To get a very light stau for values of $m_0$ in the
TeV range, large values of $A_0$ and of $\tan\beta$ are needed as in
the MSSM. An example for  stau co-annihilation with a \blino LSP is
shown in \FIG~\ref{fig:BlinoStauCoann}

Obviously, the mass difference between the LSP and NLSP has to be very
small and much smaller than normally necessary in the CMSSM. This is
even more surprising when we take into account that there are new
D-term contributions to the stau vertices for the interaction with
scalar Higgs particles of the form
\begin{align}
\Gamma^D_{\tilde{e}_1 \tilde{e}^*_1 h_k} =
 & \frac{i}{4} \Big(\sum_{a=1}^{3}Z^{E,*}_{1 3 + a} Z_{{1 3 + a}}^{E}
  \big((2 g_{1}^{2}  + \gmix (2 \gmix
  + \gBL{}))(v_d Z_{{k 1}}^{H}- v_u Z_{{k 2}}^{H}) \nonumber  \\ 
&  \hspace{2cm} + 2 \gBL{} (2 \gmix
  + \gBL{})( v_{\eta} Z_{{k 3}}^{H}
- v_{\bar{\eta}} Z_{{k 4}}^{H}  )\big) + \nonumber \\ 
 &+\sum_{a=1}^{3}Z^{E,*}_{1 a} Z_{{1 a}}^{E}  \big(2 \gBL{}
 (\gmix + \gBL{})(v_{\bar{\eta}} Z_{{k 4}}^{H}
  - v_{\eta} Z_{{k 3}}^{H} ) \nonumber \\
 &  \hspace{2cm} - (\gmix (\gmix + \gBL{})
 + g_{1}^{2}- g_{2}^{2})(v_d Z_{{k 1}}^{H}  -v_u Z_{{k 2}}^{H}) \big)\Big)
\end{align}
Similar contributions also exist 
 for the four point interactions
$\tilde{e}^*_1 \tilde{e}_1 h_k h_l$. Because of these new
contributions, the most important final state for  stau
co-annihilation can be the one with two bilepton fields. For instance,
the annihilation channels for one point with $\Omega h^2 = 0.116$,
$m_{\tilde{\chi}^0_1} = 426.0 $~GeV, $m_{\tilde{e}_1} = 426.4 $~GeV
are given by
\begin{center}
\begin{tabular}{lcclc}
 $\tilde{e}_1 \tilde{e}^*_1 \, \to \, h_2 h_2 $ & (17.7\%) & \hspace{1cm} &
$\tilde{e}_1 \tilde{e}^*_1 \, \to \, W^+ W^- $ & (12.0\%) \\
$\tilde{e}_1 \tilde{e}^*_1 \, \to \, \gamma \gamma $ & (9.9\%) & &
$\tilde{e}_1 \tilde{e}^*_1 \, \to \, \tau \bar{\tau}$ & (8.6\%) \\
$\tilde{e}_1 \tilde{e}^*_1 \, \to \, h_1 h_1 $ & (8.1\%) & &
$\tilde{e}_1 \tilde{\chi}^0_1 \, \to \, h_2 \tau $ & (7.7\%) \\
$\tilde{e}_1 \tilde{e}^*_1 \, \to \, Z Z $ & (6.8\%) & &
$\tilde{e}_1 \tilde{e}^*_1 \, \to \, \gamma Z $ & (5.2\%) \\
$\tilde{\chi}^0_1 \tilde{\chi}^0_1 \, \to \, h_2 h_2 $ & (4.6\%) & &
$\tilde{e}_1 \tilde{\chi}^0_1 \, \to \, \gamma \tau $ & (3.7\%) \\
$\tilde{e}_1 \tilde{e}^*_1 \, \to \, h_1 h_2 $ & (3.5\%) & &
$\tilde{e}_1 \tilde{e}^*_1 \, \to \, t \bar{t} $ & (3.5\%) \\
$\tilde{e}_1 \tilde{\chi}^0_1 \, \to \, Z \tau $ & (1.9\%) & &
$\tilde{e}_1 \tilde{\chi}^0_1 \, \to \, W^- \nu_2 $ & (1.9\%) \\
$\tilde{e}_1 \tilde{\chi}^0_1 \, \to \, h_1 \tau $ & (1.7\%) & &
$\tilde{e}_1 \tilde{e}^*_1 \, \to \, b \bar{b} $ & (1.1\%) \\
\end{tabular}
\end{center}

The reason that the mass difference between LSP and NLSP has to be so
small is that the mass spectrum  is in general very heavy: the second stau as
well as the lightest \lsnu
 have  masses of
1.95~TeV for the shown point, as consequence of the huge values of
$|A_0|$ and $m_0$. These heavy masses suppress the t-channel diagrams
$ \tilde{e}_1 \tilde{e}^*_1 \, \to \, h_i h_i $ and $ \tilde{e}_1
\tilde{e}^*_1 \, \to \, W^+ W^-$. Were the mass of the second stau of
order 1~TeV, the relic density of exactly the same point would be
$\Omega h^2 = 0.04$. However, those values are necessary to obtain a
sufficiently large $\mu'$ and the \blino character of the LSP.

\begin{figure}[!ht]
 \begin{minipage}{0.99\linewidth}
\centering
\includegraphics[width=0.48\linewidth]{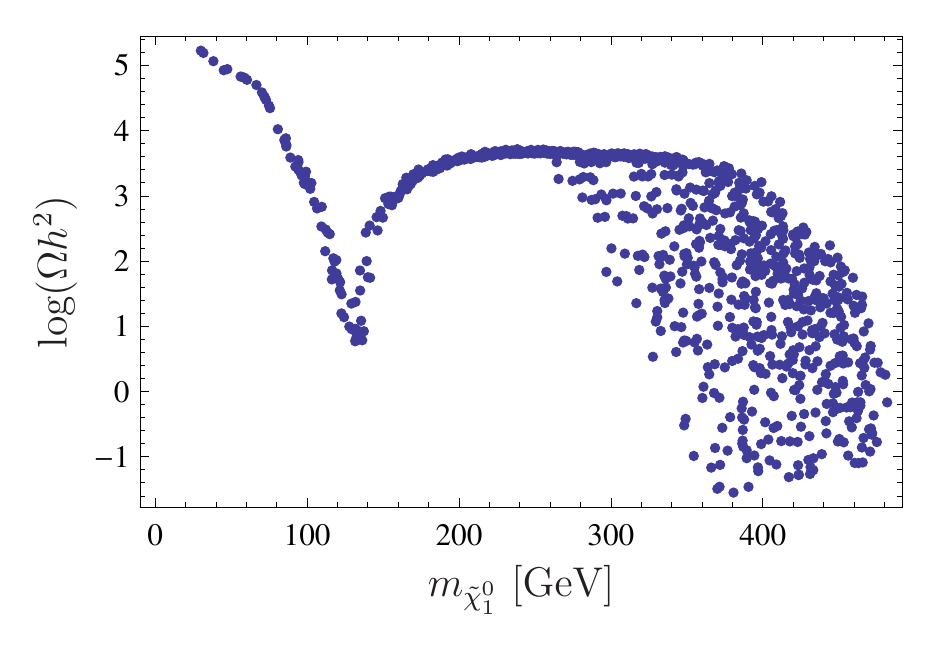}
\hfill
\includegraphics[width=0.48\linewidth]{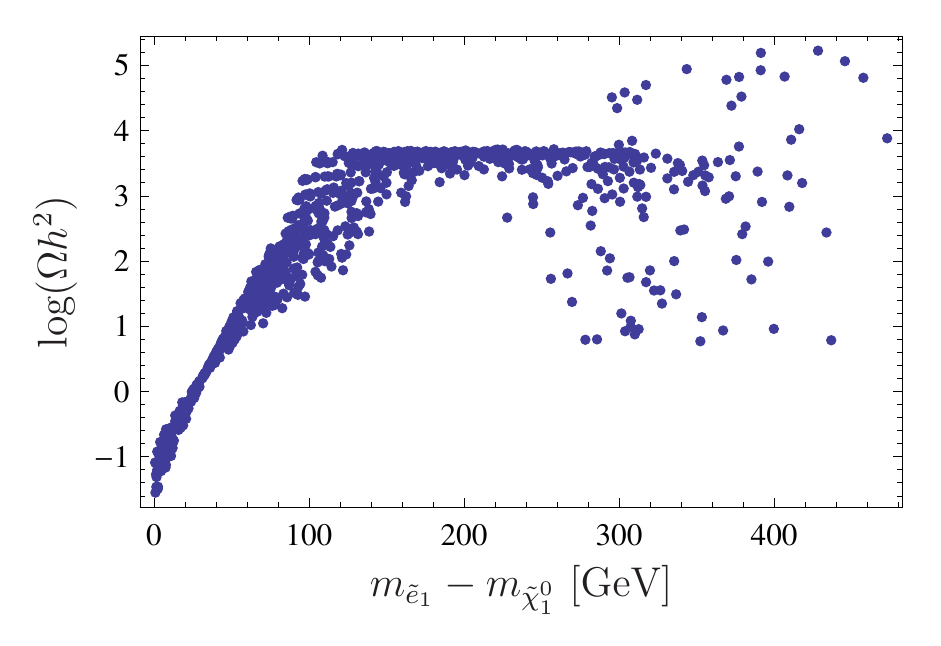}
 \\
\includegraphics[width=0.48\linewidth]{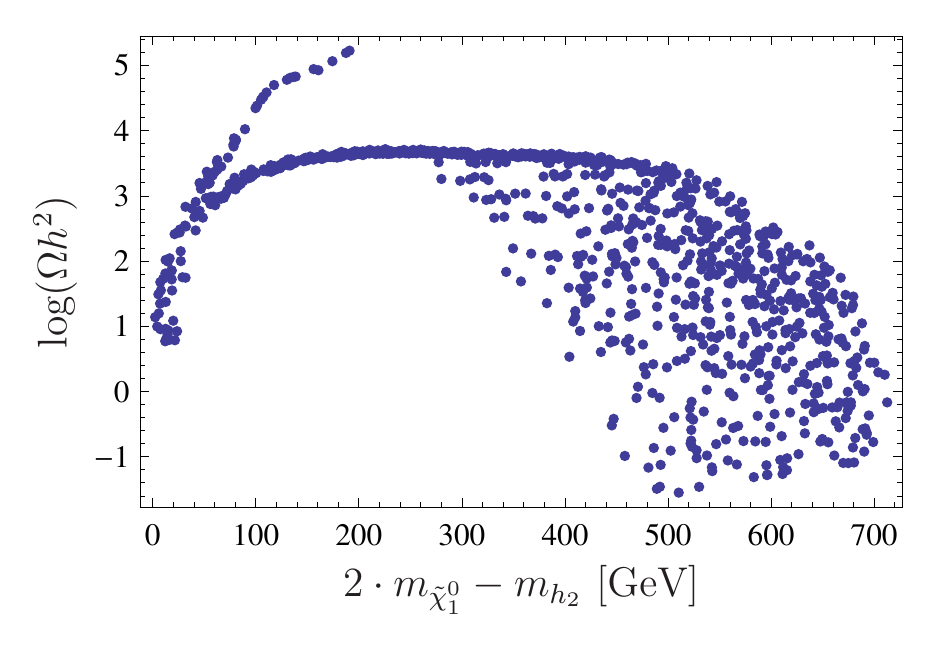} 
\hfill
\includegraphics[width=0.48\linewidth]{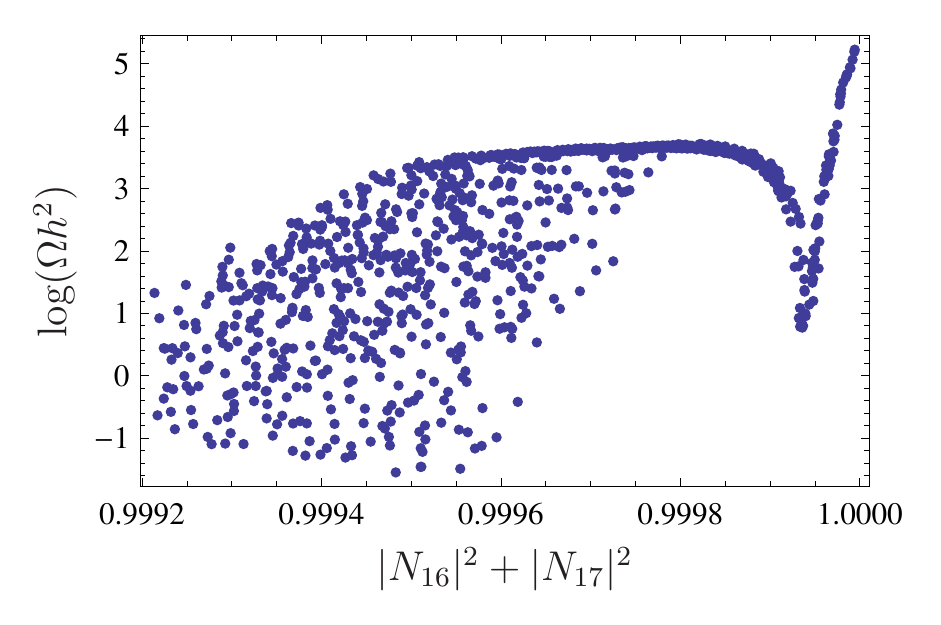}  
 \end{minipage}
 \caption{Bileptino co-annihilation with stau. Top left: $\log(\Omega h^2)$ vs.
   $m_{\tilde{\chi}^0_1}$. Top right: $\log(\Omega h^2)$ vs.
   $m_{\tilde{e}_1} - m_{\tilde{\chi}^0_1}$. Bottom left: $\log(\Omega
   h^2)$ vs. $m_{h_1} - 2\cdot m_{\tilde{\chi}^0_1}$. Bottom right:
   bileptino fraction.
   Parameter ranges: $m_0 = [1500,1550]$~GeV,
   $M_{1/2} = [960,990]$~GeV, $\tan\beta = [39,41]$, $A_0 =
   [-3650,-3600]$~GeV, $\tan\beta' = [1.21,1.23]$, $\mZp =
   [3.8,3.9]$~TeV, $Y_x^{33} = [0.40,0.41]$, $Y_x^{11} = 0.42$,
   $Y_x^{22} = 0.373$.  }
\label{fig:bileptino_stau_coann}
\end{figure}

\paragraph{Bileptino}
\label{sec:bileptino_stau}
In case of a bileptino LSP, 
 stau co-annihilation works as
efficiently as in  the MSSM, as can be seen in
\FIGS~\ref{fig:bileptino_stau_coann}: a mass splitting of 10~GeV is
still sufficient to reduce $\Omega h^2$ to roughly $0.1$. The main
difference compared to  a \blino LSP is that the second stau as
well as the \lsnus
 have a mass below 1.5~TeV. The dip
around 120~GeV is due to a resonance with the second scalar Higgs
which is a bilepton. However, since the LSP has only a very small
\blino fraction, the coupling to the bilepton is not large
 enough for
a sufficient annihilation even at the resonance, see also
\SEC~\ref{sec:HiggsResonance}. Hence, for the shown parameter range
only  stau co-annihilation leads to the correct relic density.

For the shown example, the pure bilepton final state is even
 more
important than in the example with 
a \blino LSP since the
t-channel diagrams are less suppressed. We get, for instance, for a
$m_{\tilde{e}_1} = 438.5$~GeV and $m_{\tilde{\chi}^0_1} = 433.3$~GeV,
a relic density of $\Omega h^2 = 0.095$ due to
\begin{center}
\begin{tabular}{lcclc}
 $\tilde{e}_1 \tilde{e}^*_1 \, \to \, h_2 h_2 $ & (52.5\%) & \hspace{1cm} &
$\tilde{e}_1 \tilde{e}^*_1 \, \to \, \tau \bar{\tau}$ & (11.1\%) \\
$\tilde{e}_1 \tilde{e}^*_1 \, \to \, W^+ W^- $ & (9.5\%) & &
$\tilde{e}_1 \tilde{e}^*_1 \, \to \, h_1 h_1 $ & (5.8\%) \\
$\tilde{e}_1 \tilde{e}^*_1 \, \to \, Z Z $ & (5.0\%) & &
$\tilde{e}_1 \tilde{e}^*_1 \, \to \, \gamma \gamma $ & (3.4\%) \\
$\tilde{e}_1 \tilde{e}^*_1 \, \to \, \gamma Z $ & (1.8\%) & &
$\tilde{e}_1 \tilde{e}^*_1 \, \to \, t \bar{t} $ & (1.6\%) \\
$\tilde{e}_1 \tilde{e}^*_1 \, \to \, h_2 \bar{\tau} $ & (1.3\%) & &
$\tilde{e}_1 \tilde{e}^*_1 \, \to \, h_2 \gamma $ & (1.2\%) \\
\end{tabular}
\end{center}

All-in-all, stau co-annihilation seems to work better
 with a \blino\ LSP 
than with a bileptino LSP despite the fact that there is no
 tree-level
coupling between  bileptino and the stau. However, this is
compensated by the lighter sfermion spectrum.
The
same statement also holds  in the case of stop
 co-annihilation.  However, in that
case the mass difference between the LSP and NLSP can be in general
larger than for the stau since  the gluons also
 contribute to the annihilation.

\subsubsection{Neutralino t-channel annihilation}
\label{sec:BLino_t_channel}
\begin{figure}[t]
 \begin{minipage}{0.99\linewidth}
\centering
\includegraphics[width=0.48\linewidth]{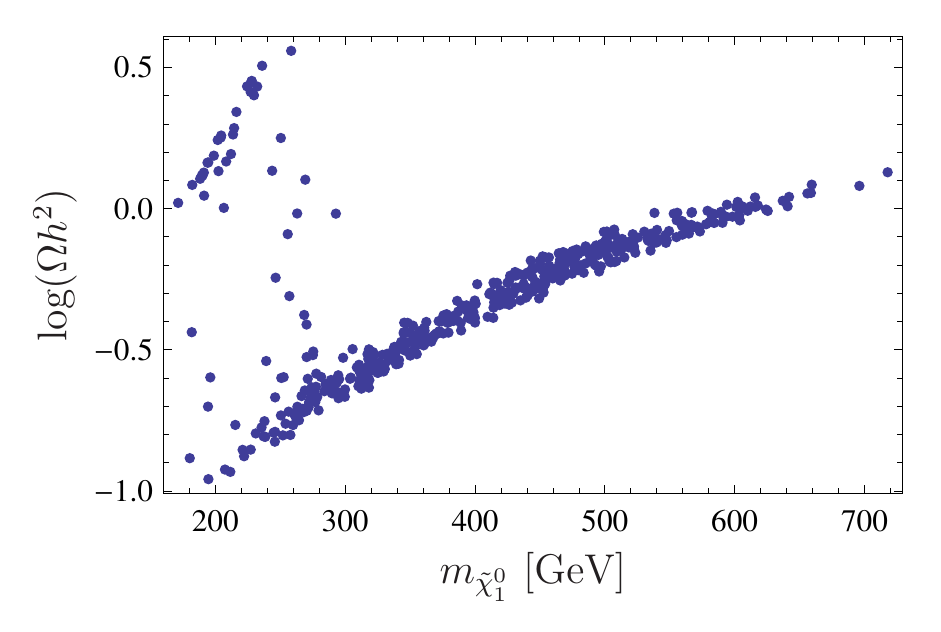}
\hfill
\includegraphics[width=0.48\linewidth]{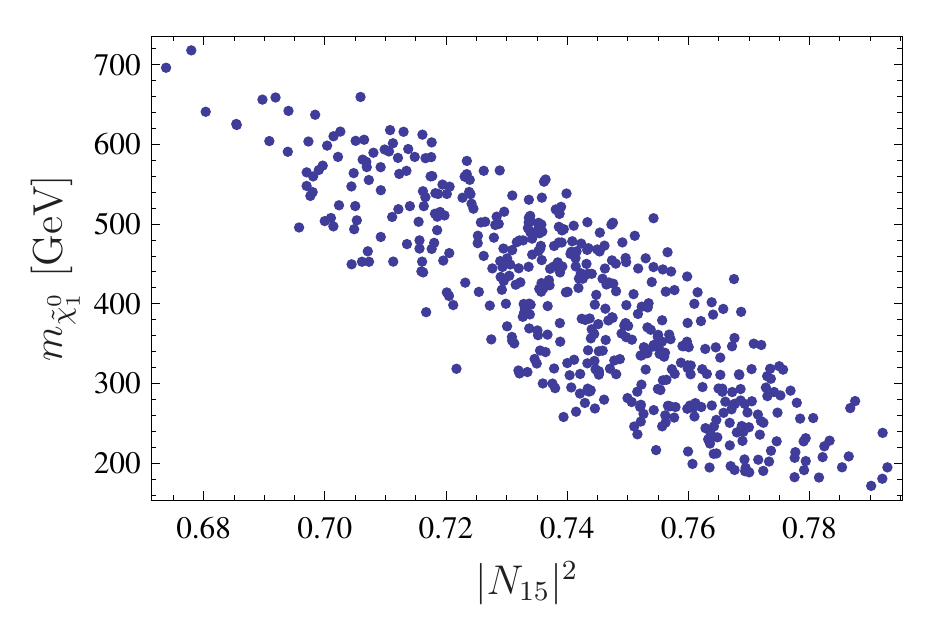} \\
\includegraphics[width=0.48\linewidth]{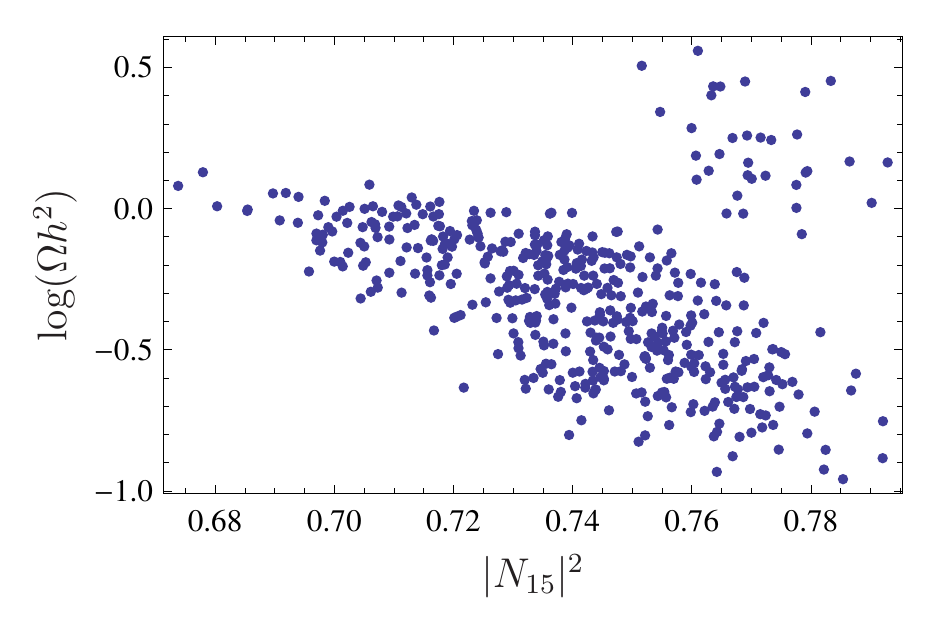}
\hfill
\includegraphics[width=0.48\linewidth]{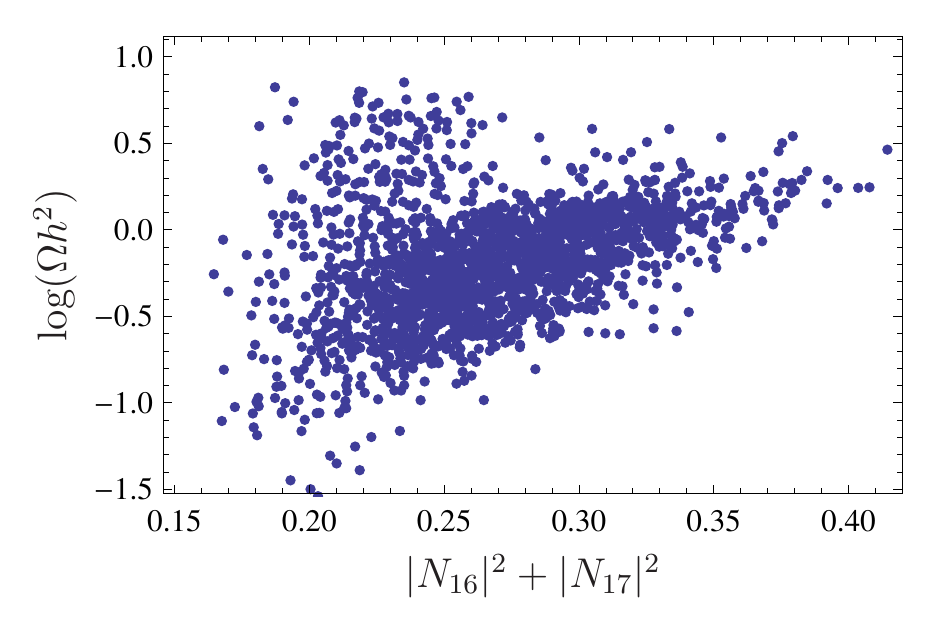} 
 \end{minipage}
 \caption{\blino t-channel annihilation. Top left: LSP mass vs.
   $\log(\Omega h^2)$; top right: \blino fraction vs. LSP mass; bottom
   left: \blino fraction vs. relic density; bottom right:
   bileptino
   fraction vs. relic density. The chosen parameter range
   were
   $m_0 = [2.7,2.8]$~TeV, $M_{1/2} = [0.7,0.8]$~TeV, $\tan\beta =
   [25,30]$, $A_0 = [1.8,2.2]$~TeV, $\tan\beta' = [1.38,1.45]$,
   $\mZp = [3.2,3.4]$~TeV, $Y_x^{33} = [0.38,0.42]$, $Y_x^{11} =
   0.42$, $Y_x^{22} = 0.43$ }
\label{fig:blino_Tchannel}
\end{figure}
The last mechanism we found to get the correct abundance of a \blino
LSP is similar to the focus point region in the MSSM and is based on a
sizable bileptino fraction of the \blino LSP. For a rather light
 LSP
with bilepton masses even lighter than the LSP mass, the
\blino/bileptino mixing leads to a strong annihilation
$\tilde{\chi}^0_1 \tilde{\chi}^0_1 \to h_B h_B$ due to a light
neutralino in the t-channel.

As can be seen in \FIG~\ref{fig:blino_Tchannel}, there is a clear
correlation between the neutralino mass and the relic density. For a
LSP mass around 200~GeV, $\Omega h^2 = 0.1$ is reached. The
 greatest 
coupling between a bilepton and the neutralino is for
 a maximally mixed
\blino/bileptino state. However, it can be seen on the
 second plot
of \FIG~\ref{fig:blino_Tchannel} that this mixing becomes smaller
with decreasing mass and the \blino clearly dominates. Therefore, the
relic density shows the counter-intuitive dependence
 on the \blino and
bileptino fractions depicted in the second row of
\FIG~\ref{fig:blino_Tchannel}.

The most important final states appearing in the annihilation for a
chosen point with $\Omega h^2 = 0.110$ are two bilepton Higgs ($h_2$),
with a contribution of $\sim$ 90\%, as well as $h_1 h_2$, $W^+ W^-$
and $h_1 h_1$ as long as the bilepton mass is lower than the LSP
 mass.
If the
 bileptons are so heavy that their presence in the final state is 
 kinematically forbidden, the
same mechanisms work if the neutralino masses are even lighter. In
that case the cross section induced by kinetic mixing can be large
enough. For instance, $\Omega h^2 = 0.103$ for a \blino mass of
158.3~GeV is obtained by
\begin{center}
\begin{tabular}{cc}
 $\tilde{\chi}^0_1 \tilde{\chi}^0_1 \, \to \, W^+ W^- $ & (52.4\%) \\
 $\tilde{\chi}^0_1 \tilde{\chi}^0_1 \, \to \,  Z Z $ & (24.9\%) \\
 $\tilde{\chi}^0_1 \tilde{\chi}^0_1 \, \to \, h_1 h_1 $ & (22.7\%)
\end{tabular}
\end{center}
Hence, especially for light neutralinos, it is very important to
include the effects of kinetic mixing. However, as soon as the LSP
mass is also below 125~GeV, \IE the mass of the light doublet Higgs,
the mechanism fails and the relic density starts to grow
 rapidly until
the Higgs resonance appears around 60~GeV, see also
\FIG~\ref{fig:blino_HiggsResonance}.

\subsubsection{Summary of \blino and bileptino dark matter} 
\paragraph*{Failed attempts} Before we summarise the different
mechanisms that work for fermionic \BL dark matter, we want to add some
remarks about scenarios which do not seem to work.
 Firstly, chargino
co-annihilation turns out to be rather difficult to achieve.
 The chargino mass is
always correlated to the mass of a MSSM neutralino and because of the
larger loop corrections the charginos are in general heavier. Hence,
before the chargino mass is close enough to the mass of a \blino or
bileptino LSP, a mass crossing between the lightest
 \BL neutralino and
the lightest MSSM-like neutralino takes place. Secondly, the bulk region
of the CMSSM is known to be ruled out by the bounds on squark masses
coming from LHC. However, in this model there are new D-term
contributions to the sfermions which are larger for the sleptons than
for the squarks because of their larger \BL charges.
 Nevertheless it was
not possible to find regions with sufficiently large annihilation of
two neutralinos into two standard model fermions through
 sleptons in
the t-channel. The reason for this is that only the \blino has a
tree-level coupling to the charged sleptons but
 obtaining a \blino LSP demands in general
even larger values of $m_0$ than needed for a bileptino LSP.

\paragraph*{Working annihilation mechanisms} A summary of possible
mechanisms to get the correct amount of dark matter for  \blino and
bileptino LSPs is given in
 Table~\ref{tab:NeutralinoSummary}. Some
scenarios are  viable for
both dark matter candidates, such as sufficient annihilation through a
 bilepton resonance. Others mechanisms like stau
co-annihilation seem to work much better for
 a bileptino LSP than for
 a \blino LSP, while a sufficient t-channel
 annihilation does not seem to be
possible at all for a bileptino LSP.

Of course, the picture changes significantly if we
allow deviations from the strict universal boundary conditions at
the GUT scale. In this framework, it should be possible to find
regions in the parameter space in which all the mechanisms
presented in table \ref{tab:NeutralinoSummary} work.
For instance, if we allow for non-unified Higgs masses
the correlation between $m_0$ and $\mu'$ due to the tadpole equations
disappears. In this case it is much easier to find co-annihilation 
regions between the squarks and sleptons and the LSP.

\begin{table}
\centering
\begin{tabular}{|c|c|c|}
\hline
\hline
Mechanism & \blino & bileptino \\
\hline 
Bilepton Higgs resonance & \checkmark
 & \checkmark    \\ 
Doublet Higgs resonance & \checkmark
 &   $\sim$   \\
CP-even sneutrino co-annihilation & \checkmark &  $\sim$   \\
CP-odd sneutrino co-annihilation &   $\sim$  & \checkmark  \\
Stau co-annihilation &   $\sim$   &  \checkmark         \\
Stop co-annihilation &   $\sim$     & $\sim$          \\
Neutralino t-channel annihilation & \checkmark &  $\times$   \\
$Z'$ resonance & $\sim$ &  \checkmark   \\
\hline
\hline
\end{tabular}
\caption{Dark matter scenarios for a $B-L$ neutralino LSP.
 The scenarios marked with a  `\checkmark'
 work very well and
 are possible in several regions in parameter
 space. Scenarios with a  `$\sim$' work in
  principle but it is much more difficult to tune the relic density to the
  correct amount or there are other drawbacks like the light Higgs mass in
  the case of stop co-annihilation. Mechanisms marked by 
  a  `$\times$' could not  be found at all. 
}
\label{tab:NeutralinoSummary} 
\end{table}

\subsection{Direct detection}
\label{sec:DD}
We have presented in the previous two subsections four new potential
dark matter candidates. However, to be a valid candidate not only
 does the
relic density have to be correct but the particle must
 also be in
agreement with all experimental limits. Therefore, here we check
against the limits on the interaction cross section with nuclei
derived by direct detection  experiments. The strongest
bounds come from the \Xenon{100} experiment \cite{Aprile:2011hi}.
The best sensitivity on the spin-independent cross section between the
LSP and a proton or neutron is roughly $10^{-44}$~$\text{cm}^2$ for a
LSP mass of around 50~GeV.

\paragraph*{\blino and bileptino} 
\BL neutralino LSPs interact with nucleons through diagrams
with s-channel squarks or t-channel neutral vector or scalar bosons.
As we have already seen in \EQS~(\ref{eq:neutqsqL}) and (\ref{eq:neutqsqR})
 only the \blino component of the LSP can couple to a
quark/squark pair. Therefore, especially for  \blino,
 the s-channel
contributions are important. 
In addition, exchange of  Higgs bosons can be important, as can also be seen
 from the corresponding couplings in \EQS~(\ref{eq:vertexNNHR}) and
 (\ref{eq:vertexNNHL}). In contrast, the $Z'$  has only an axial coupling to the
 Majorana LSPs as given in \EQS~(\ref{eq:vertexNNZpL}) and
 (\ref{eq:vertexNNZpR}). Hence, it cannot contribute 
much to the spin-independent cross section.
Furthermore, the coupling to the $Z$ boson is too small to be of
any relevance. Therefore, both \BL neutralino dark matter candidates 
interact with the
same strength with the proton as with the neutron and we don't have to
distinguish  them in the following discussion. The calculations
of the cross section were performed with \MO, with the bounds coming from
\Xenon provided by {\tt DMTools} \cite{DMTools}.

\begin{figure}[t]
 \begin{minipage}{0.99\linewidth}
\centering
\includegraphics[width=0.48\linewidth]{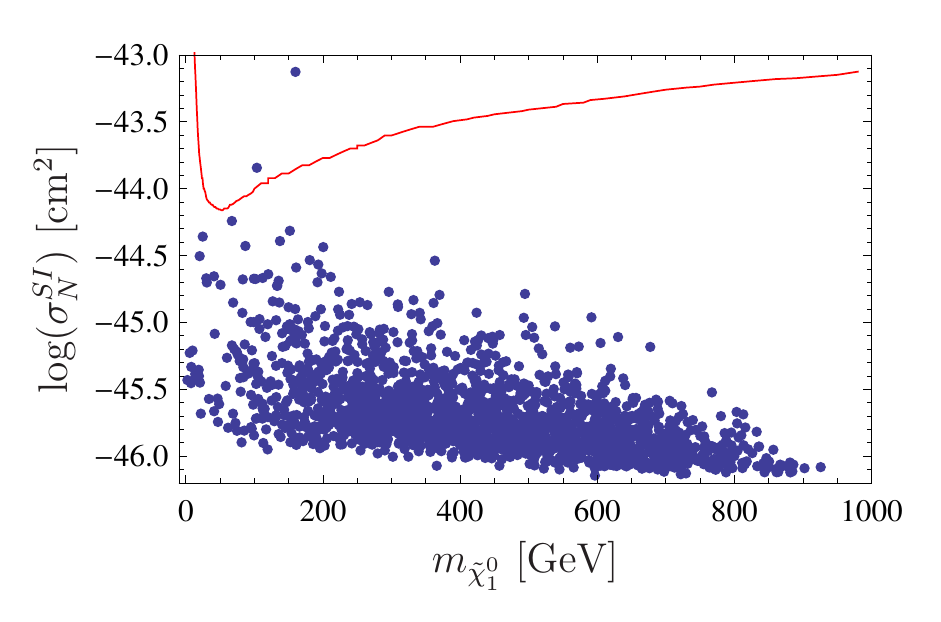}
\hfill
\includegraphics[width=0.48\linewidth]{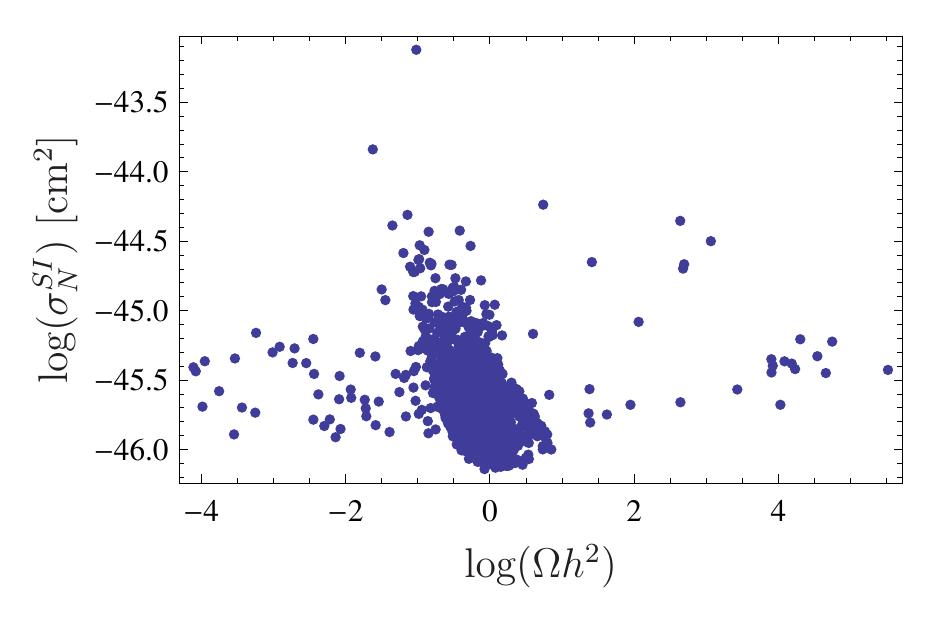} \\
\includegraphics[width=0.48\linewidth]{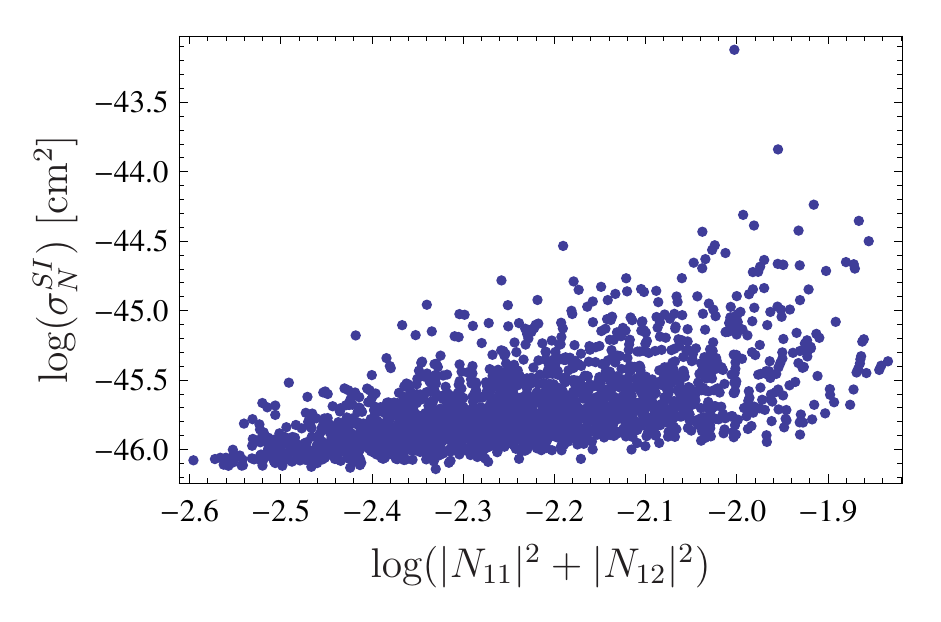}
\hfill
\includegraphics[width=0.48\linewidth]{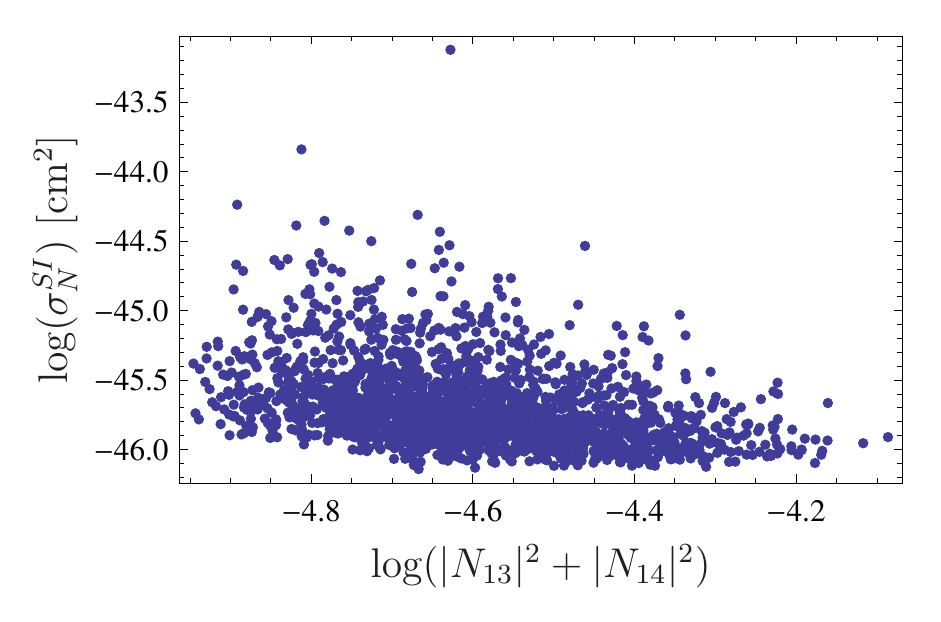} \\
\includegraphics[width=0.48\linewidth]{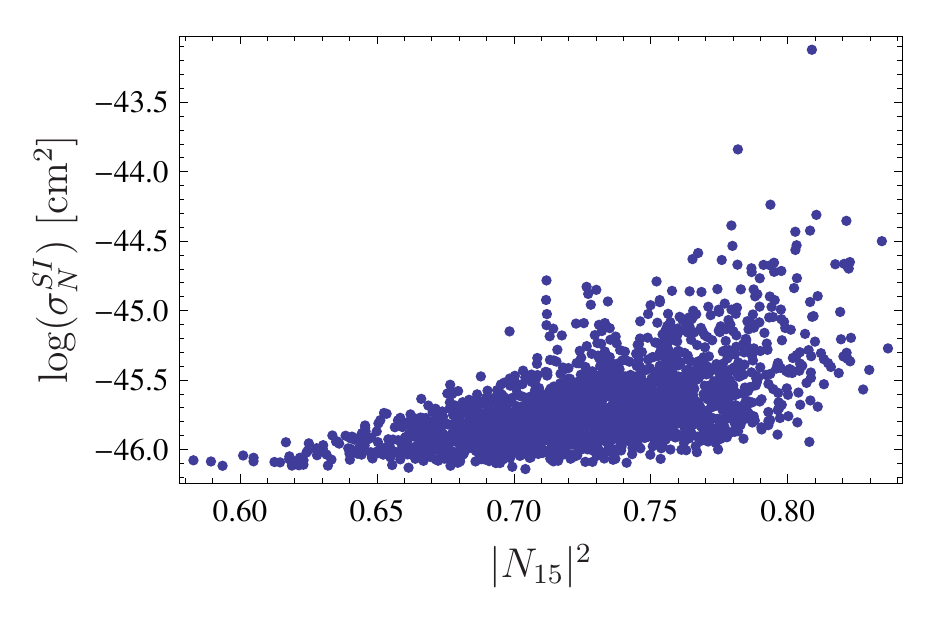}
\hfill
\includegraphics[width=0.48\linewidth]{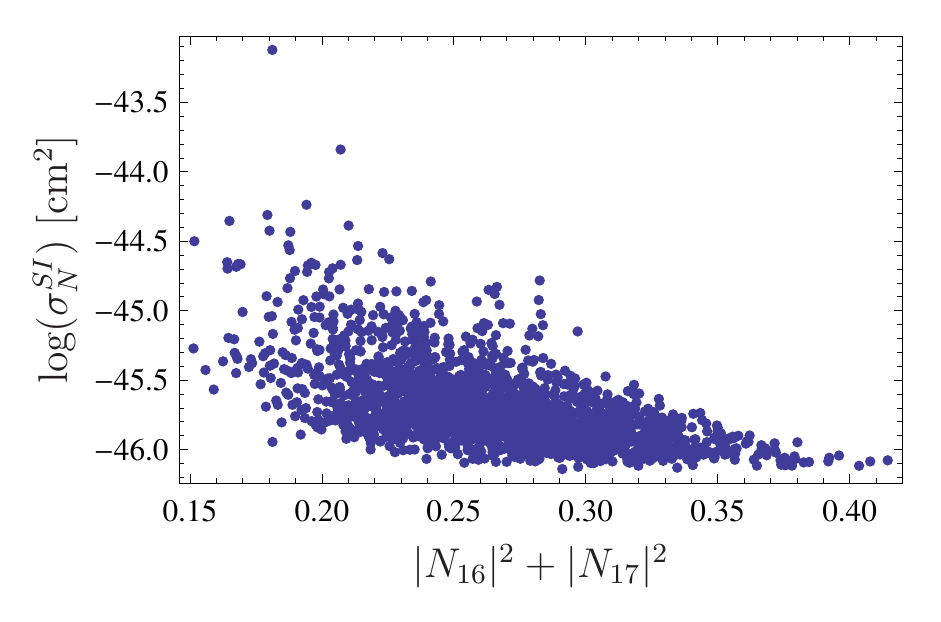} 
 \end{minipage}
 \caption{\blino direct detection. First row: direct detection cross
   section vs. LSP mass (left) and direct detection cross section vs.
   dark matter abundance (right). Second row: direct detection cross
   section as function of gaugino (left) and Higgsino (right)
   fraction. Third row: direct detection cross section as function of
   \blino (left) and bileptino fraction (right). The red line
   shows
   the exclusion limit of \Xenon{100}.  Parameters are the same as in
   \FIG~\ref{fig:blino_Tchannel}}
\label{fig:blino_DD}
\end{figure}

The coupling of a pure \blino LSP to squark/quark is comparable to
 that of a pure bino up to a factor
 $c\left(\gBL{} / g_1 \right)^2
\sim 2 c$ with a numerical coefficient of order 1 fixed by the quantum
numbers. On the other hand, a \blino LSP has 
 some bileptino
fraction which doesn't couple at all to the (s)quarks, with the
negative $\gmix$ that also reduces the coupling.
 Thus we expect the \blino interaction with a proton or neutron to be of the
same size as the bino interaction in the MSSM. This agrees
  with
the outcome of the numerical calculations shown in
\FIG~\ref{fig:blino_DD}. The large majority of the tested points are
well below the current experimental limits. Also the dependence on the
\blino and bileptino fractions is as expected:
 the cross section
increases with increasing \blino fraction and becomes very small for
 an LSP with a large bileptino component.
 The contributions of the
small admixture of MSSM states is very sub-dominant simply because of
 their smallness. The ostensible correlation between the bino fraction
 and
the cross section is based on the correlation between the \blino and
bino fraction due to kinetic mixing.

\begin{figure}[t]
 \begin{minipage}{0.99\linewidth}
\centering
\includegraphics[width=0.48\linewidth]{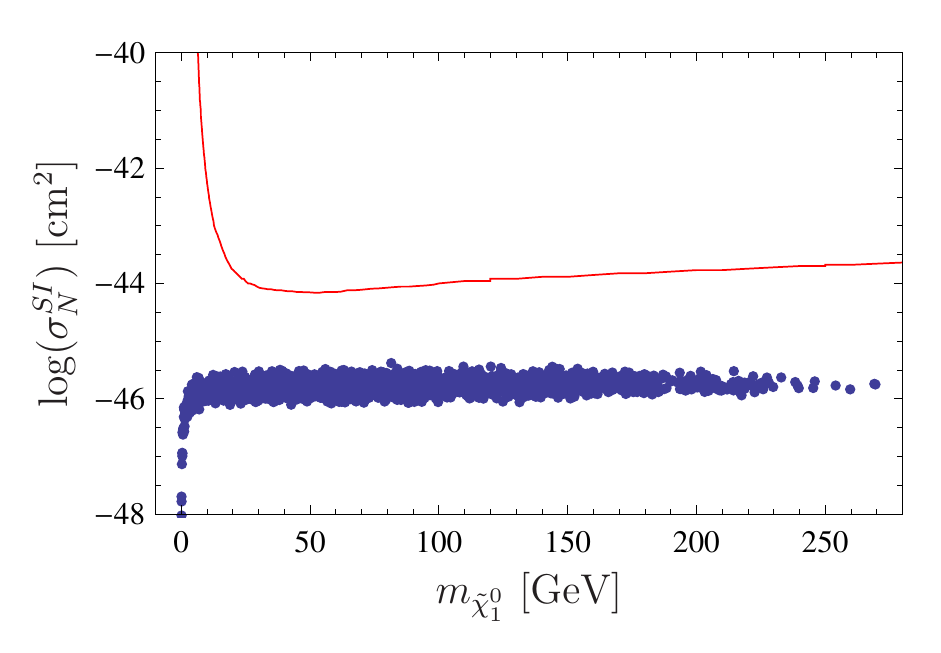}
 \end{minipage}
 \caption{\blino direct detection for a light LSP in presence of Higgs
 resonances. The red line shows the exclusion limit of \Xenon{100}.
Parameters are the same as in \FIG~\ref{fig:blino_HiggsResonance}}
\label{fig:blino_DD_HiggsResonance}
\end{figure}

The general picture doesn't change if we check regions in parameter
space in which resonances with doublet or bilepton Higgs states are
present: the coupling of both kinds of scalars to the quarks of the
first two generations is too small to increase the cross section
visibly. This is depicted in \FIG~\ref{fig:blino_DD_HiggsResonance}.
 As mentioned above, the $Z'$ couples only axially
to the bileptino LSP and the resulting contribution to the spin-independent
cross section is therefore always very small. All-in-all, the
\blino is consistent with all direct detection bounds but might be
tested with the next generation experiments, if a sensitivity of
$10^{-45} \text{cm}^2$ - $10^{-46} \text{cm}^2$ is reached.

\begin{figure}[t]
 \begin{minipage}{0.99\linewidth}
\includegraphics[width=0.48\linewidth]{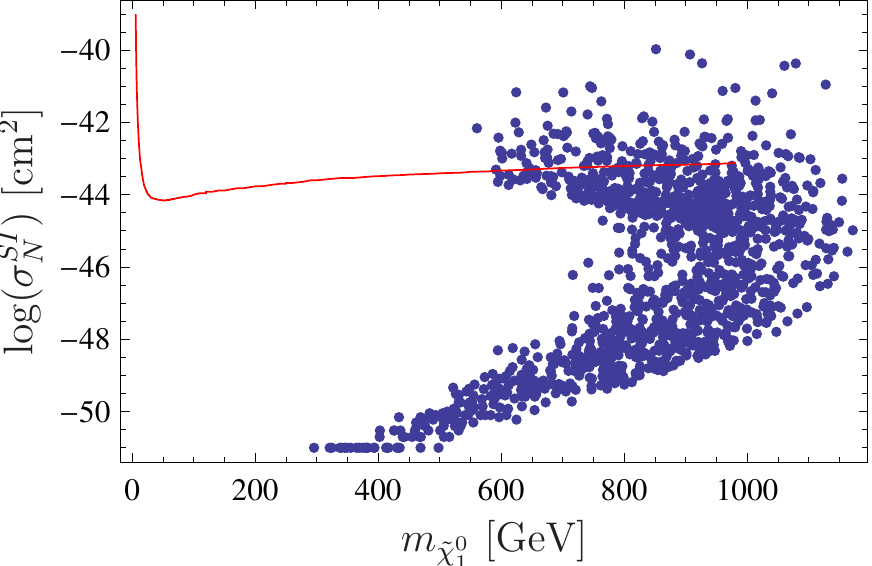}
\hfill
\includegraphics[width=0.48\linewidth]{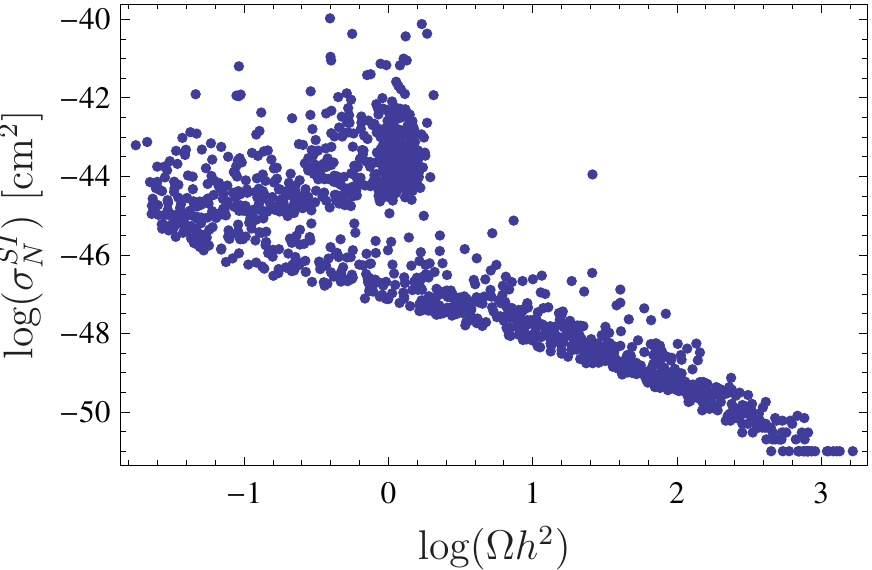} \\
\centering
\includegraphics[width=0.48\linewidth]{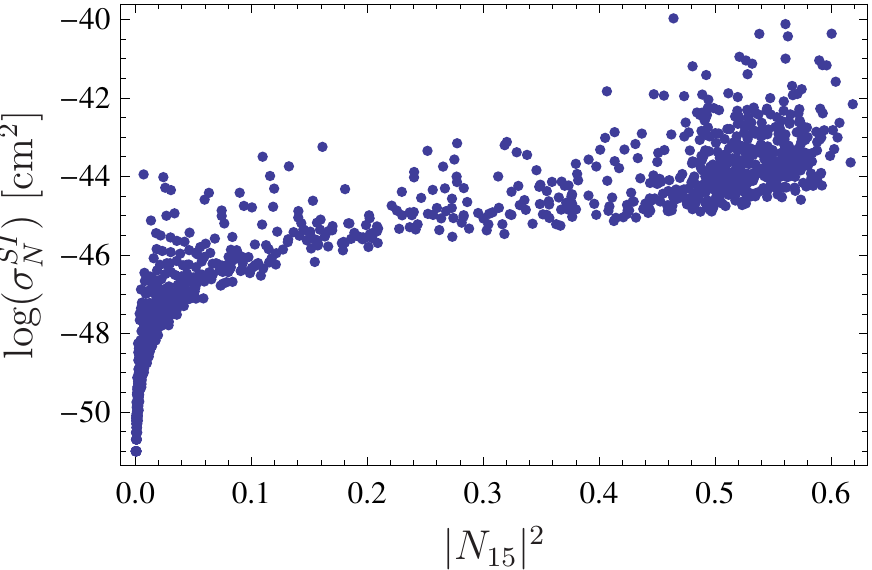}
\hfill
\includegraphics[width=0.48\linewidth]{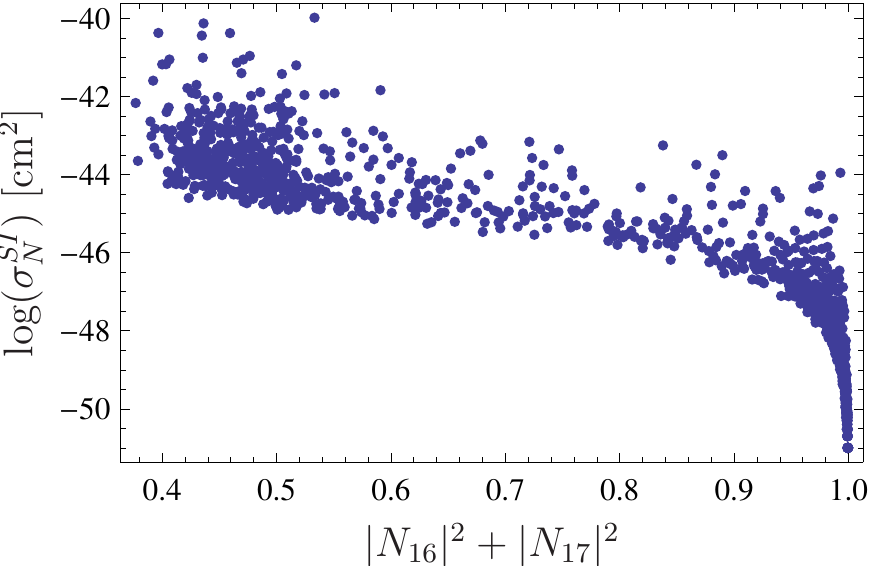} 
 \end{minipage}
 \caption{bileptino direct detection cross section near $Z'$
   resonance. First row: $m_{\tilde{\chi}^0_1}$ vs. $\sigma^{SI}_N$
   (left) and $\Omega h^2$ vs. $\sigma^{SI}_N$ (right). Second row
   $\sigma^{SI}_N$ as function of \blino (left) and bileptino
  (right)
   fraction. The red line shows the exclusion limit of \Xenon{100}.
   Parameters are the same as in \FIG~\ref{fig:bileptino_Zp}}
\label{fig:bileptino_Zp_DD}
\end{figure}

Since the spin-independent direct detection cross section of a
bileptino is in general even smaller than
 that of a \blino in the
absence of any resonance, we can immediately move on
 to the most
interesting scenario involving the $Z'$ resonance. The results of the
numerical calculation are given in \FIG~\ref{fig:bileptino_Zp_DD}. A
non-vanishing fraction of the points in the high mass regime which are
close to the resonance are already in conflict with \Xenon{100} results.
The reason for the increased cross section is that the coupling to the bilepton
 becomes large for a maximally mixed \blino-bileptino state. 
For the given parameters, the mass of the light bilepton 
is about 100~GeV for a LSP mass close to 1~TeV. Together with the large mixing,
this allows for an enhanced t-channel interaction with the quarks. 
However, there are still valid points with the correct dark matter
relic density and a cross section below $10^{-44}$~$\text{cm}^2$.
With the expected sensitivity of
 $10^{-46}$~$\text{cm}^2$ of
Xenon1t \cite{xenon1t}, all of these points could
 be tested. The
surprising dependence of the cross section on the \blino fraction is
caused by the dependence of the LSP mass on the \blino fraction:
while the $Z'$ boson couples to the bileptino component, it is only possible to have an
LSP which is mainly bileptino if it is also relatively light and thus far
from the resonance.

\begin{figure}[t]
 \begin{minipage}{0.99\linewidth}
\centering
\includegraphics[width=0.48\linewidth]{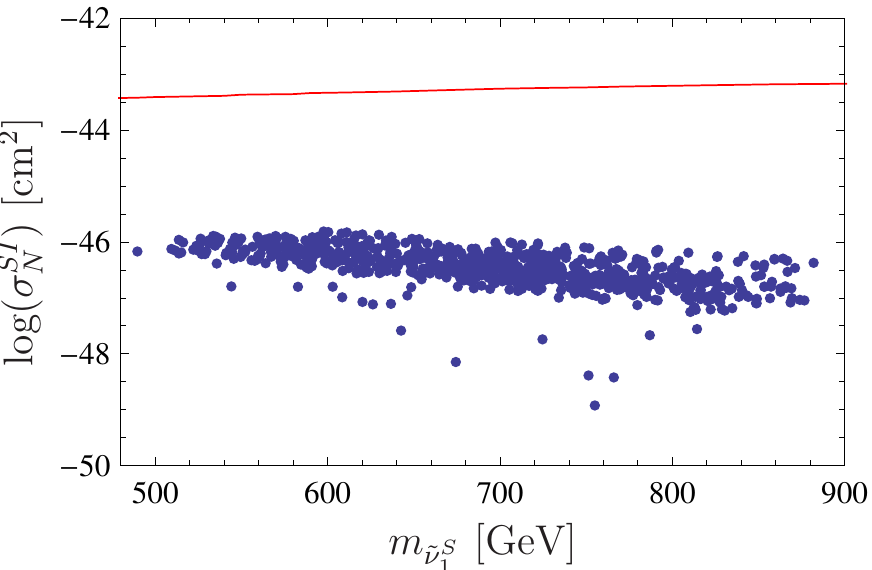} 
\hfill
\includegraphics[width=0.48\linewidth]{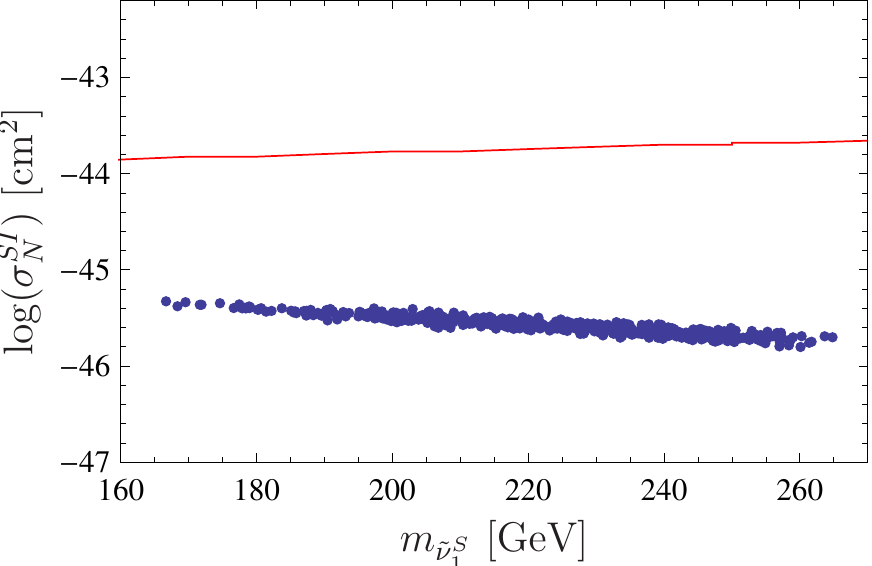} \\
\includegraphics[width=0.48\linewidth]{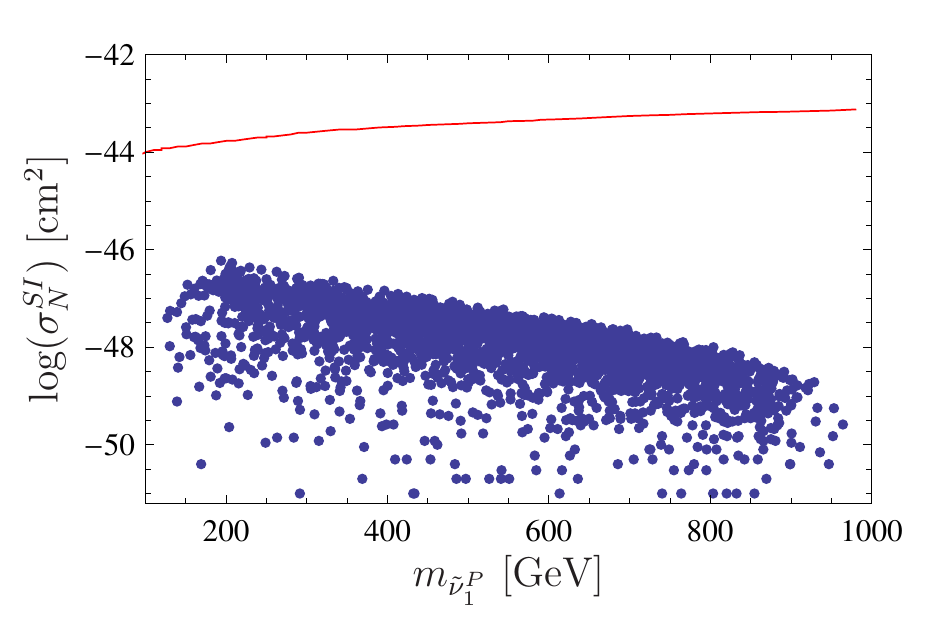}
 \end{minipage}
 \caption{First row: direct detection cross section for CP-even
   sneutrinos: $\sigma^{SI}_N$ as function of the mass of the lightest
   scalar sneutrino. The parameters were those of
   \FIG~\ref{fig:CPeven} (left) and \FIG~\ref{fig:lightCPeven}
   (right). Second row: direct detection cross section for CP-odd
   sneutrinos using the parameters of
   \FIG~\ref{fig:PSsneutrino_KM_NKM}. The red line shows the exclusion
   limit of \Xenon{100}. }
\label{fig:Sneutrino_DD}
\end{figure}

\paragraph*{Sneutrinos} Sneutrino dark matter in the MSSM is already
ruled out because of the t-channel contributions
 of the $Z$ boson.
However, \BL sneutrinos as dark matter candidates are CP
 eigenstates with different masses,
hence neither the $Z$ nor the $Z'$ boson can
 contribute. In addition, there
is no tree-level coupling between the sneutrinos and 
 squarks. The
only possible contributions are due to t-channel diagrams
 involving
Higgs bosons which couple only weakly to the quarks.
 All-in-all, the
cross section is very small and always more than one order of magnitude
 below the
current experimental limits as shown in \FIG~\ref{fig:Sneutrino_DD}.

\subsection{Impact on MSSM dark matter candidates}
\label{sec:impactMSSM}
In addition to the new candidates for
 dark matter, the \BLSSM also retains the candidates of the MSSM.
There are, however, some differences in the properties of
 these shared candidates in how their relic densities are in the context of the
 \BLSSM.  We
give here a short list of possible effects, since a detailed study of
these cases goes beyond the scope of this paper:
\begin{itemize}
\item {\bf Stau/squark co-annihilation}: the masses of the sfermions
  receive new contributions from the D-terms involving bileptons.
  The bino mass is also slightly altered due to
  the effect of kinetic
  mixing. Therefore it can be expected that the co-annihilation
  regions of the MSSM get shifted.
\item {\bf Focus point}: it has already be pointed out in
  Ref.~\cite{O'Leary:2011yq} that a change in $\mZp$ or $\tan\beta'$
  can also alter the Higgsino fraction of an MSSM neutralino LSP.
  Hence the new parameter will have an impact on the focus point
  region.
\item {\bf Higgs resonances}: the MSSM and the \BL sector are already
  coupled at tree level due to kinetic mixing.
  Therefore 
  resonances between a MSSM neutralino and a bilepton are also
 possible. In
  contrast to the Higgs funnel in the CMSSM, these resonances don't
  demand a very large value of $\tan\beta$, but are rather easy to
  find because the mass of a bino LSP and the light
 bilepton are
  sensitive to different parameters.
\item {\bf $Z'$ resonances}: the coupling of a MSSM neutralino to the
  $Z'$ boson is very weak and the corresponding resonance won't reduce
  the abundance to an acceptable amount. Nevertheless, there exists
  the possibility of indirect resonances similar to the one discussed
  in \SEC~\ref{sec:HiggsResonance} for the bileptino with the
  MSSM
  Higgs. If the second neutralino is a bileptino and also very
  close
  in mass to the LSP and to the resonant point with the $Z'$ boson, it
  annihilates very efficiently and can also
  co-annihilate with the
  LSP to reduce its relic density.
\end{itemize}

\section{Conclusion}
\label{sec:conclusion}
We have discussed here the  additional possibilities for dark matter
candidates arising in a constrained version of the
 minimal $R$-parity-conserving
supersymmetric model with a gauged \UBL.
 In addition to the candidates known from the
MSSM, the LSP can be either a \blino or bileptino
 neutralino, or a
 CP-even or -odd  bosonic partner of a right-handed neutrino.
 In the case of a sneutrino LSP, the
dark matter relic density is often of the correct order of magnitude
 or even well below the measured value because of the strong interaction between the sneutrinos and the light
 bilepton. However,  kinetic mixing is crucial for many annihilation channels
 and light sneutrinos below the bilepton mass
threshold would be ruled out without kinetic mixing. Since the
sneutrino LSP is a CP eigenstate with a sizable mass 
difference to the eigenstate with opposite CP quantum number,
there is no tree-level coupling
to the $Z$ boson and the spin-independent cross section to nuclei is
much smaller than that of
 the scalar partner of the left-handed neutrino
 in the MSSM.
Therefore, the sneutrinos clearly fulfill all bounds coming from
direct detection experiments like \Xenon{100}.

In contrast, for a \blino or bileptino LSP, specific mass
configurations are needed to reduce the abundance to a level
consistent with the measured dark matter abundance in the Universe.
 Possible mechanisms that we have discussed
 are resonances with Higgs fields and
co-annihilations with $SU(2)_{L}$-singlet
 sneutrinos, which work very well
for both the \blino and the bileptino. A co-annihilation with staus is
also possible but easier to realize for a bileptino.
 Stop
co-annihilation would also work but leads in general to a mass of the
light doublet Higgs which is in conflict with recent indications
 coming from
the LHC. In the case of a rather light \blino LSP with a
 non-negligible bileptino fraction, the t-channel annihilation into two
 bileptons can also be
sufficient to get the correct relic density. None
 of these scenarios
are in conflict with the bounds coming from \Xenon{100}.
 However, the
$Z'$ resonance, which is only relevant for a
 bileptino LSP, can
increase the spin-independent cross section of the LSP with nuclei to
a level excluded by \Xenon{100} for smaller values of $\mZp$.

\section*{Acknowledgments}
LB is supported by the Deutsche Forschungsgemeinschaft through the
Research Training Group grant GRK\,1102 \textit{Physics of Hadron
  Accelerators}. BOL and WP have been supported by the
German Ministry of Education and Research (BMBF) under contract
no.\ 05H09WWEF.

\bibliographystyle{h-physrev}

\end{document}